\documentclass[a4paper,12pt]{article}
\usepackage[utf8]{inputenc} 
\usepackage{mymacros}

\newif\ifdraft
\draftfalse 

\ifdraft
  \usepackage[right]{showlabels}
  
\fi

\title{Flow-geometry microstates}
\author[a]{Ricardo Espíndola}
\author[b]{and~Shoichiro Miyashita}
\affiliation[a]{Institute for Advanced Study, Tsinghua University, Beijing 100084, China}
\affiliation[b]{Department of Physics, National Dong Hwa University, Hualien, Taiwan, R.O.C.}
\emailAdd{ricardo.esro1@gmail.com, s-miyashita@gms.ndhu.edu.tw}

\abstract{We construct geometric microstates for a class of two-dimensional flow geometries—spacetimes that interpolate from an asymptotic AdS$_2$ boundary to a dS$_2$ static patch in the interior—by inserting particles behind the horizon. We show that this mechanism produces dS microstates with an Einstein–Rosen bridge of infinite length behind the horizon. The state-counting of these microstates, including wormhole contributions, reproduces the Gibbons-Hawking entropy, $S_{\rm dS}=A^{\rm dS}_{\rm horizon}/4G$. Furthermore, we extend the microstate-counting method to the case of a finite-length Einstein–Rosen bridge. As a result, the Hilbert space of the dS horizon in the flow geometry can be spanned by states with a purely dS Einstein–Rosen bridge, containing no AdS portion on the time-symmetric slice. This provides a concrete realization of dS microstates within a controlled holographic framework.
}

\begin{document}

\maketitle
\newpage

\section{Introduction} \label{sec:intro}
The Bekenstein-Hawking entropy formula \cite{Bekenstein:1972tm, Bekenstein:1973ur, Hawking:1975vcx}, a cornerstone of black hole thermodynamics \cite{Bardeen:1973gs, Hawking:1976de}, has inspired many modern approaches to quantum gravity—including a novel resolution to the black hole information paradox \cite{Penington:2019npb, Almheiri:2019psf,
Almheiri:2019hni, Penington:2019kki, Almheiri:2019qdq}. Shortly after its discovery, Gibbons and Hawking demonstrated that cosmological horizons obey a similar, yet even more profound, area relation. Specifically, an observer in the static patch of de Sitter (dS) space will detect thermal radiation emanating from the cosmological horizon, with an entropy given by \cite{Gibbons:1977mu,Gibbons:1976ue},
\beq 
\label{eq:GH}
S_{\text{GH}}=\frac{A}{4G_N}~,
\eeq
where $A$ is the cross-sectional area of the observer's horizon, and $G_N$ is Newton constant. Unlike black hole horizons, the entropy of a cosmological horizon is intrinsically tied to the observer measuring it leading to a radical departure from the black hole case: distinct observers in cosmology will perceive different horizons. Furthermore, it has been argued that empty de Sitter space corresponds to a maximally mixed state, suggesting that a Hilbert space of dimension of order $\exp (A/4G_N)$ suffices to describe all possible states, including excitations, in the static patch \cite{Banks:2003cg, 
Banks:2006rx, Fischler:2000, Dong:2018cuv}.

In the modern approach, the observer-dependent nature of de Sitter space manifests in its entanglement structure. This can be understood by dressing the algebra of observables for a static patch with the worldline of an observer, which promotes the original type III algebra of quantum fields to the Murray-von Neumann type II$_1$ factor. Crucially, unlike type III algebras, the type II$_1$ algebra admits a well-defined trace and a state of maximum entropy—naturally identified with the de Sitter vacuum \cite{Chandrasekaran:2022cip, Witten:2023qsv, Witten:2023xze}.

The origin of de Sitter entropy is even more enigmatic than its black hole counterpart, posing a fundamental challenge to our understanding of quantum gravity. While the Gibbons-Hawking formula (\ref{eq:GH}) suggests a statistical interpretation, a key question remains: \emph{What are the microstates accessible to an observer in the static patch?} In this work, we advance towards resolving this question by systematically constructing microstates in a class of geometries containing de Sitter bubbles by employing techniques from wormhole statistics \cite{Balasubramanian:2022gmo, Balasubramanian:2022lnw, Climent:2024trz, Balasubramanian:2024rek}.
 
Recent progress has been achieved in the context of black holes in anti-de Sitter space (AdS). By means of a modern statistical approach that accounts for wormhole contributions to the gravitational path integral. For example such wormhole contributions are responsible for recovering the Page curve for evaporating black holes \cite{Penington:2019kki, Almheiri:2019qdq}. More recently, similar inspired techniques have been used to count the number of microstates of black holes with different semiclassical interiors \cite{Balasubramanian:2022gmo, Balasubramanian:2022lnw, Climent:2024trz, Balasubramanian:2024rek}. 

A key element in the microstate counting is the use of matter shells to produce different black hole interiors, generating a large number of gravitational `bag of golds' \cite{Wheeler:1964qna} 
\footnote{
For more details on the name `bag of gold', see footnote~1 in \cite{Fu:2019oyc}.
},
\ie, states featuring a long Einstein-Rosen (ER) bridge behind the horizon. However, establishing a controlled framework for these bag of golds states in dS space is challenging for two related reasons: ({\textbf{i}}) the absence of a boundary dual, which allows a precise definition of the dual local operators $\mathcal{O}(x)$ that create matter shells, and (\textbf{ii}) the intrinsic property of dS space where infalling matter produces a decrease in the size of the cosmological horizon
\footnote{\label{foot:GW}In a recent paper \cite{Wang:2025jfd}, they attempted to construct dS microstates with long ER bridges by injecting branes from a Dirichlet boundary placed near $r=0$ (in Euclidean signature) and showed that the dimension of the Hilbert space of the dS horizon is the exponential of the Gibbons–Hawking entropy by applying the method of \cite{Balasubramanian:2022gmo}. However, we believe that they used branes of {\it negative} tension, although they explained that the microstates are created by branes with the tension parameter $T>0$. It seems that the source of this discrepancy (in version 2) may be the abnormal sign convention of the brane action in (3.1) of \cite{Wang:2025jfd}, meaning that the sign of the tension parameter $T$ is opposite to that of the actual brane tension. (In version 1, the sign convention itself was normal, but the Darmois–Israel junction condition (3.8) in \cite{Wang:2025jfd} seems to contain an error.)  
}
; in contrast to the growth of the ER bridge in AdS. The latter property was recently exploited for teleportation protocols in dS \cite{Aguilar-Gutierrez:2023ymx}.

In this work, we construct semiclassical microstates that overcome these challenges, which we dub \emph{centaur microstates}. These are realized within a class of solutions known as centaur geometry \cite{Anninos:2017hhn, Anninos:2018svg} (or more generally flow geometries \cite{Anninos:2020cwo}).  These are solutions to dilaton gravity theories that interpolate between an AdS$_2$ boundary region and a dS$_2$ static patch in the interior. From the boundary perspective, the dS$_2$ physics arises from turning on a relevant deformation in the quantum system, which flows to interesting infrared physics. These geometries feature a horizon (whose properties depend on the specific dilaton potential $V(\Phi)$) and can be interpreted as thermal states. Crucially, they allow us to anchor the static patch worldline to an AdS$_2$ boundary, thereby defining a holographic dual quantum mechanical system where standard AdS/CFT techniques can be applied to study dS physics.

\begin{figure}[t]
\begin{center}
\includegraphics[width=5.cm]{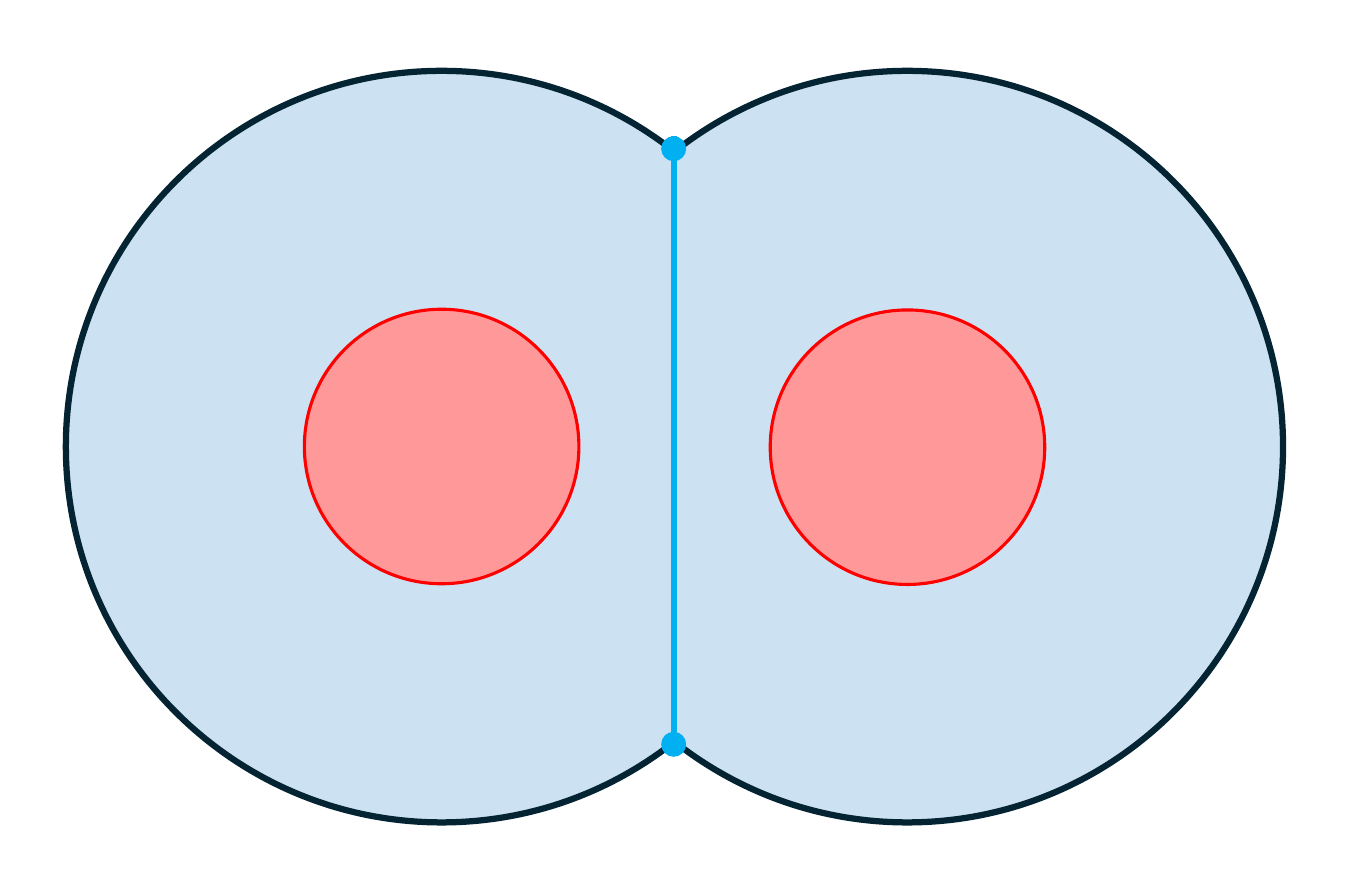} ~ \includegraphics[width=5.cm]{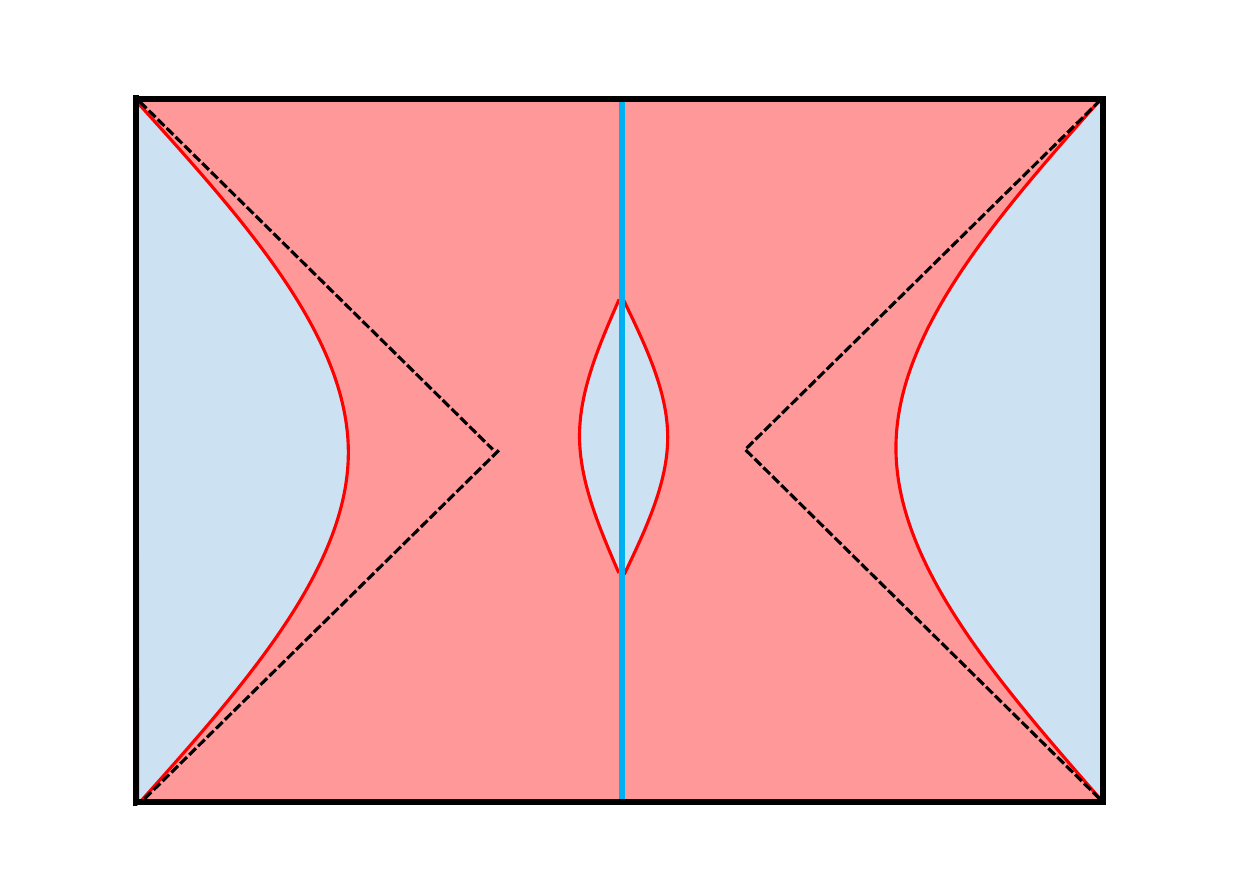} 
\caption{
(\textbf{Left}) Euclidean flow PETS geometry. By acting with an operator ${\cal O}(x)$, a heavy particle is created, which backreacts on the geometry. When the particle mass is sufficiently large, two disjoint dS regions are formed.  
(\textbf{Right}) Lorentzian flow PETS geometry. The particle (blue line) elongates the ER bridge. When the mass is large, an AdS region appears in the middle as in the figure; we call this the \textit{flow ER bridge}. Conversely, if the mass is sufficiently small, there is no AdS region in the interior; we call this the \textit{dS ER bridge}.}
\label{fig1}
\end{center}
\end{figure}

This framework directly addresses the challenges mentioned above. First, the presence of an AdS boundary allows to use the Hartle-Hawking construction to prepare global states from an Euclidean saddle. Second, the centaur geometry exhibits a Shapiro time delay for matter obeying the null energy condition. Together, these features enable us to prepare novel `partially entangled thermal states' (PETS) \cite{Goel:2018ubv} with non-trivial dS$_2$ interiors by acting with local operators ${\cal{O}}(x)$ on the Euclidean boundary (see Fig. \ref{fig1}). These PETS give rise to distinct Lorentzian interiors behind the horizon, which we identify in two classes depending on the mass of the matter particles: \emph{flow Einstein-Rosen bridges} and \emph{dS Einstein-Rosen bridges}. Furthermore, we compute the Hilbert space dimension spanned by these centaur microstates, the \emph{flow Hilbert space} denoted $\mathcal{H}_{\mathbf{Flow}}$, by including wormhole contributions to the gravitational path integral (GPI). Importantly, these microstates correspond to the same asymptotic geometry.

Although our explicit calculations focus on centaur geometries that flow from an asymptotic AdS$_2$ to a dS$_2$ spacetime in the interior, our core results are expected to hold for a broader class of flow geometries. The key ingredient is the asymptotic AdS region, which guarantees the existence of a long wormhole in a suitable parameter regime. This AdS region is what renders the microstate construction and subsequent state counting tractable.

\emph{dS microstates}: This work provides a concrete realization, within a controlled holographic framework, of states that can be interpreted as microstates of the dS horizon from the perspective of a static patch observer. The construction relies on the two-dimensional nature of the theory, which allows matter inserted from $\mathcal{I}^-$ (in the Lorentzian picture) outside the static patch to generate a long Einstein-Rosen bridge behind the cosmological horizon, in contrast to the higher-dimensional expectation due to the Gao-Wald theorem (see Section \ref{subsec:dSER} for details). In the finite-mass regime, we construct states whose time-symmetric slice consists purely of a dS static patch with no AdS region: a \emph{dS Einstein-Rosen bridge}. From the perspective of a static patch observer, these states represent distinct microstates of the dS horizon, with the interior geometry indistinguishable from empty de Sitter\footnote{We thank the JHEP referee for his interesting comments and suggestions on this point.}.

We now present a summary of the main results of our paper.

In Section \ref{sec:PETS}, we construct a new family of PETS for flow geometries. Starting from a time-symmetric slice of the centaur geometry, we prepare a global thermal state via the Euclidean path integral. Different microstates are modeled by acting with boundary operators $\mathcal{O}_i$, followed by Euclidean time evolution. In the bulk, these operators correspond to matter particles of mass $m_i$ (the two-dimensional analogue of infalling matter shells for higher-dimensional black holes), where the index $i$ 
labels different operator species. These particles follow different worldline trajectories. 

We begin by reviewing the basic elements of flow geometries, deferring a detailed thermodynamic discussion of both the canonical and microcanonical ensembles to Appendix \ref{app:thermodynamics}. For concreteness, we focus on potentials of the form $V(\Phi)=2 |\Phi|/\ell^2$, with $\ell$ a characteristic length scale of the system. This family of potentials leads to sharp centaur solutions (\ref{eq:sharp}).

We then study single-particle states and analyze the particle dynamics by coupling the dilaton-gravity action to that of a massive particle. In this framework, a particle worldline manifests as the co-dimension-one locus where two different dilaton solutions (e.g., $\Phi_1$ and $\Phi_2$) are glued. Since the dilaton determines the properties of the geometry, the particle geodesic can be obtained by solving the Darmois-Israel junction conditions. Our analysis reveals geodesics, parameterized by $m_i \beta_i$, that smoothly interpolate between AdS and dS regions. Furthermore, we compute the renormalized distance, $\Delta \cal{T}$, between the endpoints of these worldlines  (see Eq. (\ref{eq:TimeDifference})). This calculation reveals a new class of microstates, absent in standard black holes, which are associated with trajectories that probe the dS interior.  

This mass-dependent behavior leads to a new class of \emph{flow PETS geometries} in the Lorentzian picture. These geometries exhibit distinct interiors: a \emph{flow Einstein-Rosen bridge} and a \emph{dS Einstein-Rosen bridge}.  Specifically, for a sufficiently large particle mass (see condition (\ref{eq:conditionAdS})), the flow ER bridge contains both AdS and dS portions. Conversely, when the mass is sufficiently small compared to the system temperature (see condition (\ref{eq:conditiondS})), the corresponding Lorentzian geometry lacks an AdS region behind the horizon (see Fig. \ref{fig:ERbridges}). 

In Section \ref{sec:4}, we develop the mathematical framework for computing the dimensionality of the Hilbert space spanned by flow PETS. This begins with the Gram matrix of overlaps between flow PETS in the canonical ensemble. We then perform an inverse Laplace transform (ILT) to obtain the Gram matrix, and its $n$-th moments generalization, in the microcanonical ensemble. Subsequently, we introduce the resolvent for the normalized Gram matrix, and compute its trace by solving a Schwinger-Dyson equation. Finally, we analyze the discontinuity of the resolvent, which determines the Hilbert space dimension.

Section \ref{sec:cGm} presents a detailed analysis of the canonical Gram matrix using the GPI. We compute the on-shell gravity action for flow PETS geometries, including contributions from Euclidean wormholes. Notably, these saddles become dominant after transitioning to the microcanonical ensemble (see Appendix \ref{app:thermodynamics}), which can be computed by performing an inverse Laplace transform. 

The on-shell action simplifies considerably in the case of very large particle mass, which we analyze in detail in Section \ref{sec:heavymass}. There, we compute the $n$-th moments of the normalized Gram matrix, from which we derive the dimensionality of the Hilbert space:
\be
\label{eq:dimH}
{\rm dim}(\mathcal{H}_{\mathbf{Flow}}) \simeq e^{{2S_{\rm flow}}} ~ ,
\ee
a key result of our work, where $S_{\rm flow}$ is the horizon entropy of the flow geometry (see Eq. (\ref{S_flow})).
This is the Hilbert space dimension perceived by static patch observers whose worldlines have been pushed all the way to AdS$_2$ boundaries. We also discuss the connection between this result and previous work in the large mass limit \cite{Climent:2024trz}.

Finally, in Section \ref{sec:microcanonical}, we analyze the microcanonical ensemble directly via the inverse Laplace transform of the Gram matrix. We pay particular attention to the small-mass 
case, where the semiclassical geometry features a dS ER bridge. This analysis demonstrates that the flow  
Hilbert space—with the same dimension $\exp(2S_{\bf Flow})$—can be spanned entirely by microstates with a dS ER bridge, opening a new avenue for understanding the microscopic nature of de Sitter space.

\paragraph{Organization}  This paper is organized as follows. In Section \ref{sec:PETS},  we review the necessary background on flow geometries and construct novel PETS for flow geometries, which serve as semiclassical microstates of the flow geometry horizon; the \emph{centaur microstates}. In Section \ref{sec:4}, we introduce the Gram matrix for flow PETS and compute the dimensionality of the Hilbert space spanned by centaur microstates. Section \ref{sec:cGm} details the computation of Euclidean wormhole contributions to the GPI in the canonical ensemble. We then analyze the heavy-mass limit in Section \ref{sec:heavymass}, where the microstates feature an AdS Einstein-Rosen bridge. In Section \ref{sec:microcanonical}, we address the finite mass case, detailing the inverse Laplace transform from the canonical to the microcanonical ensemble. In this case, the corresponding centaur microstates are associated with a de Sitter ER bridge. We conclude with a summary and discussion of future directions in Section \ref{sec:Discuss}.
\\

\section{Flow PETS Geometries}\label{sec:PETS}
Flow geometries are solutions within a class of dilaton gravity theories that interpolate between AdS$_2$ at the boundary and a dS$_2$ static patch in the bulk interior. They represent the semiclassical geometries of a (microcanonical) thermofield double (TFD) state in our model. In this section, we construct a family of quantum states that can be interpreted as microstates of the horizon of the flow geometry. Following the method developed in \cite{Balasubramanian:2022gmo, Climent:2024trz}, we identify these microstates with (a microcanonical version of) partially entangled thermal states (PETS) \cite{Goel:2018ubv}. The semiclassical geometry of each PETS is a flow geometry containing a long ER bridge, created by the gravitational backreaction of a particle inserted at the Euclidean boundary. We refer to these configurations as \emph{flow PETS geometries}.

The rest of this section is organized as follows. In Section \ref{sec:flowgeometries}, we review the flow geometries introduced in \cite{Anninos:2017hhn}. In Section \ref{subsec:flowPETSgeometry}, we explicitly construct Euclidean flow PETS geometries. The construction begins by studying the dynamics of a particle with mass $m_i$ dual to a boundary operator $\mathcal{O}_i$, propagating on a Euclidean flow geometry. We find that different particle trajectories depend critically in its mass. For large masses, the worldlines are confined to the AdS region near the boundary, while for sufficiently small masses, the trajectories probe the dS region. This behavior induces a change in the renormalized boundary length, which is quantified by the renormalized worldline distance $\Delta \mathcal{T}$ derived in Eq. (\ref{eq:TimeDifference}). 

Finally, in Section \ref{sec:ERstates}, we analyze the Lorentzian continuation of these geometries, which describe novel classes of wormholes. The backreaction of the particle mass $m_i$ leads to two distinct types of interior structures: 
\begin{itemize}
    \item {\bf Flow ER Bridge:} For large masses, the backreaction generates a long ER bridge that contains an AdS portion. In the heavy-mass limit, this bridge becomes almost entirely AdS-like. 
    \item {\bf dS ER Bridge:} For small masses, the backreaction results in an ER bridge that consists solely of a dS interior at the time-symmetric slice, with no connected AdS portion. 
\end{itemize}
This classification of wormholes based on their interior structure is a unique feature of our centaur geometry construction and provides the foundation for our subsequent microstate counting.

\subsection{Flow geometries}
\label{sec:flowgeometries}
The Euclidean action of the dilaton-gravity system we analyze has the form 
\begin{equation}\label{eq:I centaur}
    I_E=-\frac{1}{16\pi G_2}\int_{\mathcal{M}} \rmd^2x\sqrt{g}(R\Phi+V(\Phi))-\frac{1}{8\pi G_2}\int_{\partial\mathcal{M}}\rmd \tau\sqrt{h}\,\Phi K+I_{\text{ct}}~,
\end{equation}
where $\cal{M}$ is the spacetime manifold, $h$ is the induced metric at $\partial \cal{M}$, and the counterterm in the action is 
\beq 
\label{eq:IctJT}
I_{\text{ct}}
=\frac{1}{8\pi G_{2}}\int_{\partial\mathcal{M}}\rmd \tau\sqrt{h}\frac{\Phi}{\ell}~.
\eeq
Here $\ell$ is a bulk curvature scale determined by the dilaton potential $V(\Phi)$. It is also possible to add a topological term to the action
\begin{align}
    I_{\text{top}}&=-\frac{\Phi_{0}}{16\pi G_{2}}\int_{\mathcal{M}}\rmd^{2}x\sqrt{g}R-\frac{\Phi_{0}}{8\pi G_{2}}\int_{\partial\mathcal{M}}\rmd\tau\sqrt{h}K=-\frac{2\pi}{8\pi G_{2}}\Phi_{0}\chi\;,\label{eq:topterm}
\end{align}
where $\chi$ is the Euler characteristic of $\mathcal{M}$.

The equations of motion for the dilaton field and metric are
\be
\label{eq:EOM}
\nabla_\mu\nabla_\nu\Phi-g_{\mu\nu}\nabla^2\Phi-\frac{1}{2}g_{\mu\nu}V(\Phi)=0~, \quad R=-V'(\Phi)~.
\ee
The dynamics of the dilaton is governed by the gravity equation of motion, and the dilaton equation, in turn, determines the background $\mathcal{M}$. A class of solutions written in the Schwarzschild gauge are given by
\begin{equation}
\label{eq:static patch sol}
    \rmd s^2=f(r)\rmd\tau^2+\frac{\rmd r^2}{f(r)}~,\quad \Phi=\widetilde{\Phi}_{b} \frac{r}{\ell}~,
\end{equation}
with a positive, dimensionless constant $\widetilde{\Phi}_{b}$.  We will assume that $\Phi_{0}\gg \widetilde{\Phi}_{b}$. From the higher dimensional point of view, this condition is a consequence of near black hole extremality. The Ricci scalar of (\ref{eq:static patch sol}) is non-constant, $R=-\partial^{2}_{r}f(r)$, and given in terms of the function
\begin{equation}\label{eq:N(r) factor}
    f(r)=\frac{\ell}{\widetilde{\Phi}_{b}}\int_{r_h}^r d r'\,V(r')~. 
\end{equation}
In this gauge, the horizon geometry is located at $r = r_h$, where $f(r)$ vanishes. Expanding the geometry (\ref{eq:static patch sol}) near $r=r_h$, we can periodically identify the Euclidean time with period
\beq
\beta:= \frac{1}{T} = \frac{4\pi \widetilde{\Phi}_{b}}{\ell|V(r_{h})|}~, \label{eq:flowtemp}
\eeq 
by requiring smoothness of the geometry near the tip. For concreteness, we focus on a particular class of geometry known as \emph{sharp centaur} \cite{Anninos:2017hhn}
\be
\label{eq:sharp}
\begin{split}
ds^2 =  f_{\rm flow}(r) d\tau^2 + \frac{dr^2}{ f_{\rm flow}(r)}~ , \quad -(2\pi \ell^2 T) \leq r < \infty~ , \\ f_{\rm flow}(r)=(2\pi \ell T)^2+ \frac{1}{\ell^2}\frac{r^3}{\lvert r \lvert}
\end{split}
\ee
This is a solution to the equations of motion (\ref{eq:EOM}) with the potential 
\be
V(\Phi)=\frac{2|\Phi|}{\ell^2} ~,
\ee
and the horizon radius is given by $r_{h} = -2 \pi \ell^2 T$. Therefore, the horizon entropy of the flow geometry, which is given by the “total” dilaton value at the horizon divided by $4G_{2}$, is written as
\be\label{S_flow}
S_{\rm flow} = -\frac{\pi \ell}{2G_{2}} \widetilde{\Phi}_{b} T + \frac{\Phi_{0}}{4G_{2}} ~ .
\ee
The solution is continuous and (twice-)differentiable. 
The parameter $T$ represents the temperature, consistent with Eq. (\ref{eq:flowtemp}). The dilaton value at the horizon, $\Phi_{h}$, is given by $\Phi_{h} = \Phi(r_{h}) = -2\pi \ell T \widetilde{\Phi}_{b}$. Therefore, the absolute value $|\Phi_{h}|$ can be interpreted as the temperature of the system. For a given temperature, there exist two solutions in centaur gravity, corresponding to opposite signs of $\Phi_h$ and associated with dS$_2$ horizons or AdS$_2$ black holes. The former corresponds to (\ref{eq:sharp}), while the latter corresponds to the usual AdS$_{2}$ solution (see Eq. \eqref{eq:AppendixBH}).

The sharp centaur solution is a limiting case of a larger family of smooth potentials parametrized by $\epsilon$, which is the scale size that controls the transition region between AdS$_2$ and dS$_2$. 
\footnote{
An example of a family of smooth potentials is 
\ben
V(\Phi) = \frac{2}{\ell^2} \Phi \tanh \left( \frac{\Phi}{\epsilon} \right) ~ .
\een
}
Our results can be easily extended for a more general class of potentials.

\subsection{Euclidean Flow PETS geometry}\label{subsec:flowPETSgeometry}
The Euclidean action for the PETS geometries in our model is characterized by two parts: the action of the centaur geometry (\ref{eq:I centaur}), and the action for a particle with mass $m_{i}$
\be
I_{E}^{i}[\boldsymbol{g},\Phi,\gamma_{1}]= I_{E}^{cent}[\boldsymbol{g},\Phi]+I^{m_{i}}_{E}[\boldsymbol{g},\Phi, \gamma_{1}]~.
\ee
The particle action is given by 
\be
\label{eq:particle}
I^{m_{i}}_{E}[\boldsymbol{g},\gamma_{1}]=m_{i} \int_{\gamma_{1}} ds \sqrt{h} + \frac{-1}{8\pi G_{2}} \int_{\gamma_{1}} ds \sqrt{h} \Phi (K_{L}+K_{R}) - \left. m_{i} \ell \log \left(\frac{\Phi}{2 \pi G_{2} \ell m_{i}}\right) \right|_{\partial \gamma_{1}}~.
\ee
Here, $\gamma_{1}$ represents the particle worldline on the manifold $\mathcal{M}$, with its endpoints $\partial \gamma_{1}=\{ \gamma_{1}(\infty)$, $\gamma_{1}(-\infty) \}$ located on the boundary $\partial\mathcal{M}$. The induced metric along $\gamma_1$ is denoted by $h$. $K_L$ and $K_R$ represent the extrinsic curvatures from the Gibbons-Hawking-York terms. Importantly, the final term in the action (\ref{eq:particle}) is a particle counterterm. It consists of the intrinsic quantities at $\partial \gamma_{1}$, similar to the case with usual counterterm for pure gravity.

This particle counterterm, like its pure gravity counterpart, is not unique. We fix this ambiguity by requiring that the on-shell particle action vanishes in the limit $m_{i} \to \infty$. This choice is physically motivated: in the infinite mass limit, the particle trajectory effectively becomes point-like. Its contribution to the action should vanish. This specific choice of counterterm allows us to safely take the $m_{i} \to \infty$ limit, as required by the method in \cite{Balasubramanian:2022gmo, Climent:2024trz}. whereas other choices would lead to divergences. For example, if we choose the counterterm to be $- \left. m_{i} \ell \log \left(\Phi\right) \right|_{\partial \gamma_{1}}$ instead of the counterterm in (\ref{eq:particle}), we can renormalize the divergence related to the AdS volume for finite mass $m_{i}$, but cause the action to diverge negatively as $m_i$ is increased. 

\begin{figure}[t]
\begin{center}
\includegraphics[width=12.cm]{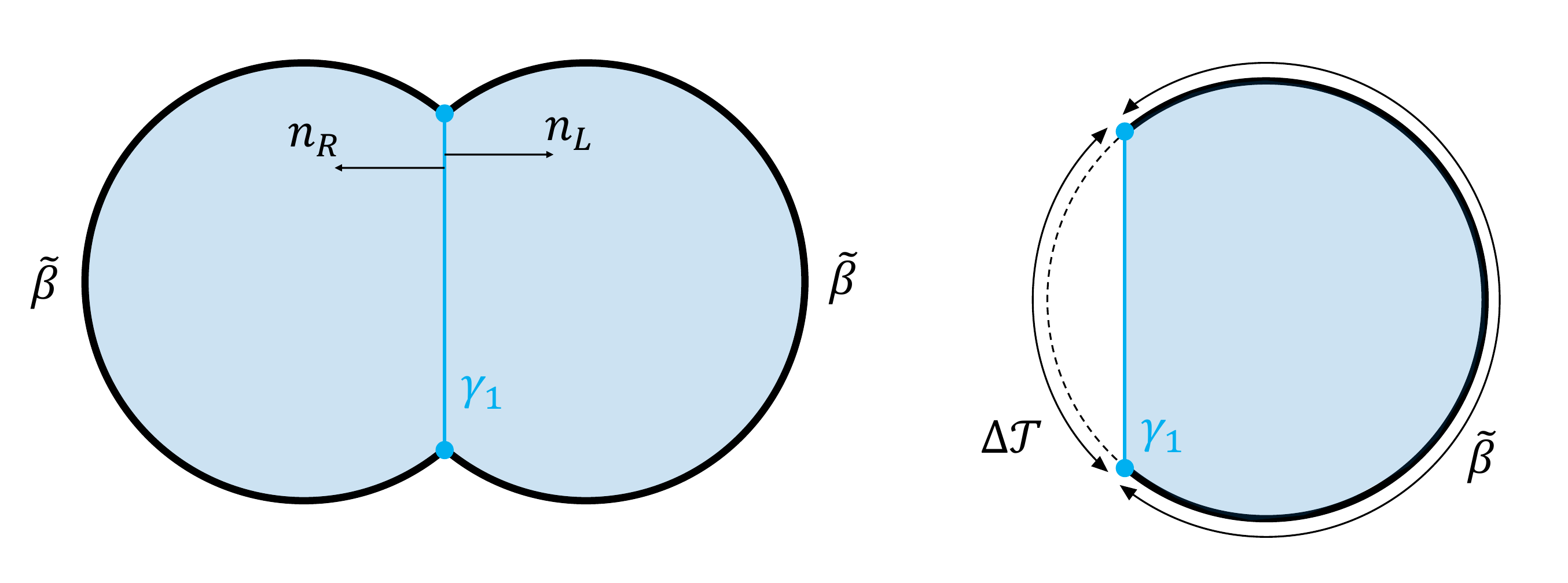} 
	\caption{(\textbf{Left}) Euclidean flow PETS geometry. The worldline $\gamma_{1}$ with particle mass $m_{i}$ separates the geometry into left and right regions. The renormalized boundary lengths of the left and the right boundaries are, respectively, equal to $\widetilde{\beta}$. Therefore the total renormalized boundary length is $2\widetilde{\beta}$. $n_{L} (n_{R})$ represents the normal vector at $\gamma_{1}$ for the left (right) region. (\textbf{Right}) Right half of the geometry. Let $\beta_{i}$ be the full disc renormalized length, and let $\Delta \mathcal{T}(m_{i}, \beta_{i})$ be the time difference between the two endpoints of $\gamma_{1}$. They satisfy $\beta_{i} = \widetilde{\beta} + \Delta \mathcal{T}(m_{i}, \beta_{i})$.   }
\label{fig:SecGramMatrixPETS}
\end{center}
\end{figure}

In dilaton gravity theories, the gluing of geometries across a matter source is controlled by the dilaton profile. In fact, the equation of motion for the particle, yields the Darmois-Israel junction conditions. For two dilaton solutions $\Phi_L$  and $\Phi_R$ joined at the particle worldline $\gamma_{1}$, these conditions are
\be
\label{eq:junction}
(n^{\mu}_{L} \nabla_{\mu} \Phi_{L} + n^{\mu}_{R} \nabla_{\mu} \Phi_{R})- 8 \pi G_{2} m_{i} =0~, ~~~~~ K_{L} + K_{R}=0~.
\ee
In higher dimensions, these conditions  imply a discontinuity in the first derivative of the metric at $\gamma_{1}$. Since our boundary conditions respect a $\mathbb{Z}_{2}$ symmetry, we assume the on-shell geometry shares this symmetry. This implies the normal derivatives are not independent, $n^{\mu}_{L} \nabla_{\mu} \Phi_{L} = n^{\mu}_{R} \nabla_{\mu} \Phi_{R}$. Thus, the junction conditions (\ref{eq:junction}) simplify to
\be
\label{eq:eomsimple}
n^{\mu} \nabla_{\mu} \Phi=4\pi G_{2} m_{i}~~  {\rm and}~~  K=0~. 
\ee

We are interested in solutions to the equation of motion (\ref{eq:eomsimple}) for a given value of the renormalized boundary length $2\widetilde{\beta}$, a parameter of the canonical PETS (details in Section \ref{sec:4}). As a consequence of the $\mathbb{Z}_{2}$ symmetry, we can focus on the one-sided geometry with renormalized boundary length $\beta_{i}$ and worldline trajectory $\gamma_{1}$. The relation between these renormalized lengths is given by 
\be
\label{eq:lengths}
\beta_{i}(\widetilde{\beta}, m_{i}) = \widetilde{\beta} + \Delta \mathcal{T}(m_{i}, \beta_{i})~.
\ee
In this equation, $\Delta \mathcal{T}$ is the renormalized distance between the endpoints of the worldline $\gamma_{1}$  (see Fig. \ref{fig:SecGramMatrixPETS}), whose explicit form will be determined shortly. 

Let the particle trajectory be $\gamma_1:\mathbb{R}\to \mathcal{M}$. We use embedding functions to parametrize the trajectory $C(s) = (\mathcal{T}(s), \mathcal{R}(s))$, with parameter $s\in (-\infty , \infty)$. The projection tensor $e^{\mu}_{s}$ and the normal vector $n_{\mu}$ are given respectively by the expressions
\be
(e^{\tau}_{s}, e^{r}_{s}) = (\dot{\mathcal{T}}~, \dot{\mathcal{R}})~, ~~~~ (n_{\tau}, n_{r}) = (-\dot{\mathcal{R}}, \dot{\mathcal{T}})~,
\ee
where we have defined $\cdot \equiv \partial_s $. We can relate these geometric quantities using the normalization condition for $e^{\mu}_{s}$
\be
f_{\rm flow}^2 \dot{\mathcal{T}}^2 = f_{\rm flow} - \dot{\mathcal{R}}^2 ~.
\label{eq:ParticleConstraint}
\ee
We then find that the junction condition (\ref{eq:junction}) for the metric (\ref{eq:sharp}) becomes
\be
\label{eq:geodesicf}
f_{\rm flow}(\mathcal{R}) - \dot{\mathcal{R}}^2  = \frac{16\pi^2 G_{2}^{ ~2} \ell^2 m_{i}^2}{\widetilde{\Phi}_{b}^2} 
 ~. 
\ee
\begin{figure}[t]
\begin{center}
	\includegraphics[width=6.cm]{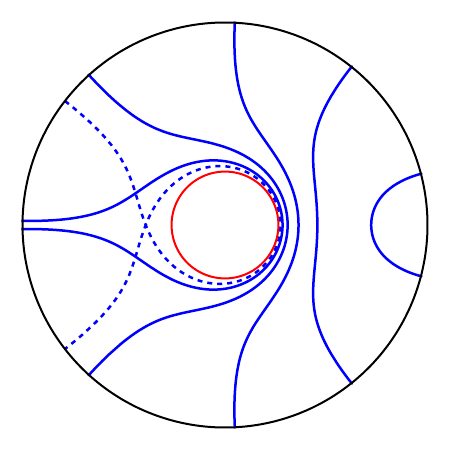}
	~~~~ \includegraphics[width=6.cm]{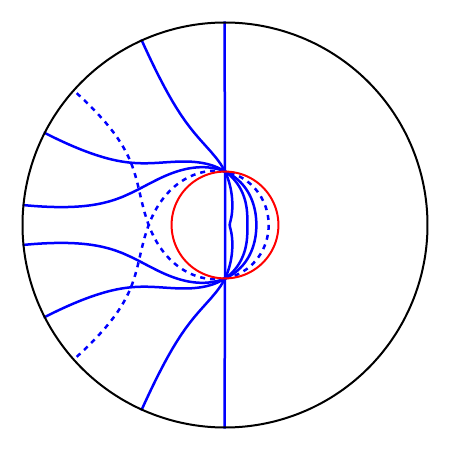}
	\caption{Qualitative behaviors of geodesics of various masses. The red circle represents the boundary between the AdS and dS regions. (\textbf{Left}) $m_{i}>\frac{\widetilde{\Phi}_{b} }{2 G_{2}\beta_{i}}$. When the mass is large, the rightmost geodesic has a small value of $\Delta \mathcal{T}$.  As the mass decreases, the geodesics tend to wrap around the dS region and the value of $\Delta \mathcal{T}$ becomes large. When the mass is lower than a critical value, $\widetilde{\beta}$ is no longer positive. This case is depicted by the dashed curve.  (\textbf{Right}) $m_{i}<\frac{\widetilde{\Phi}_{b} }{2 G_{2}\beta_{i}}$. In this mass range, every geodesic must pass through the `north pole' and the `south pole' of the dS region. When the mass is small, the geodesic is nearly a straight line, as in the figure, \ie $\Delta \mathcal{T} \simeq \frac{\beta_{i}}{2}$. Increasing the mass also increases the value of $\Delta \mathcal{T} $. As in the previous case, when the mass exceeds another critical value, $\widetilde{\beta}$ is no longer positive. This case is depicted by the dashed curve.  }
\label{fig:geodesics}
\end{center}
\end{figure}
The solutions to Eq. (\ref{eq:geodesicf}) yield different classes of trajectories, depending on the ratio of the particle mass to the background temperature (controlled by $\beta_i$), which smoothly interpolate between dS and AdS regions. For a large mass-to-temperature ratio, the particle remains confined to the AdS region close to the boundary. On the other hand, for a small ratio, the particle probes the dS region deep in the IR. The qualitative behavior of the worldline $\gamma_1$ is shown in Fig. \ref{fig:geodesics}. (As we will explain shortly, there is an intermediate mass range where the particle trajectory self-intersects.) We refer to the reader to the Appendix \ref{appedix:B}, where we write explicit expressions for the trajectories $C(s)$ in the different geometric regions.

Every trajectory asymptotically reaches the AdS boundary in the limit $s \to \pm \infty$. We can compute the renormalized distance between endpoints, $\Delta \mathcal{T} := \mathcal{T}(\infty) - \mathcal{T}(-\infty)$, using the solutions (\ref{eq:centaurTlargemass}) or (\ref{eq:centaurTsmallmass}). In the different regions of the centaur geometry we find the following expressions 
\be \label{eq:TimeDifference}
\Delta \mathcal{T} =  \left\{
\begin{array}{ll}
\displaystyle \frac{ \beta_{i}}{\pi} {\rm arctanh} \left[  \frac{\widetilde{\Phi}_{b} }{2 G_{2} \beta_{i} m_{i} } \right]~ & ~~~ \displaystyle ( m_{i} > \frac{\widetilde{\Phi}_{b} }{2 G_{2} \beta_{i}} ) ~  \\
 ~ & ~ \\
\displaystyle \frac{ \beta_{i}}{\pi} {\rm arctanh} \left[  \frac{2 G_{2} \beta_{i} m_{i} }{\widetilde{\Phi}_{b} } \right] + \frac{\beta_{i}}{2} ~ & ~~~ \displaystyle (m_{i} < \frac{\widetilde{\Phi}_{b} }{2 G_{2} \beta_{i}}) ~ .
\end{array}
\right.
\ee

For large masses, the particle is very close to the boundary. In consequence, the value of $\Delta \mathcal{T}$ is very small at fixed temperature $\beta_i$. As the mass decreases, the geodesic begins to wrap around the dS region. For sufficiently small masses, the geodesic goes into the dS region. Around this mass range, all geodesics go through the north and south pole of the dS region.
Notably, in an intermediate mass range, $\Delta \mathcal{T}$ can exceed $\beta_{i}$, meaning the particle trajectory self-intersects, as shown by the dashed curves in Fig. \ref{fig:geodesics}. In this range, constructing a real Euclidean geometry by gluing two copies at the trajectory is not allowed. From Eq. (\ref{eq:TimeDifference}), the forbidden mass range is given by 
\be
 \tanh \left( \frac{\pi}{2} \right)\frac{\widetilde{\Phi}_{b} }{2 G_{2} \beta_{i}} \leq m_{i} \leq \frac{1}{\tanh(\pi)}\frac{\widetilde{\Phi}_{b} }{2 G_{2} \beta_{i}} ~ . \label{eq:massrange}
\ee

\begin{figure}[t]
\begin{center}
	\hspace{-.2cm} \includegraphics[width=7.cm]{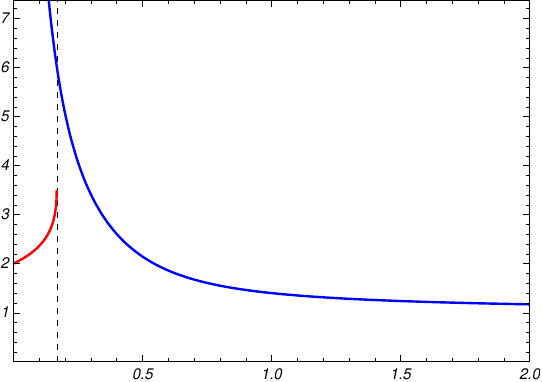} \hspace{.8cm} \includegraphics[width=7.cm]{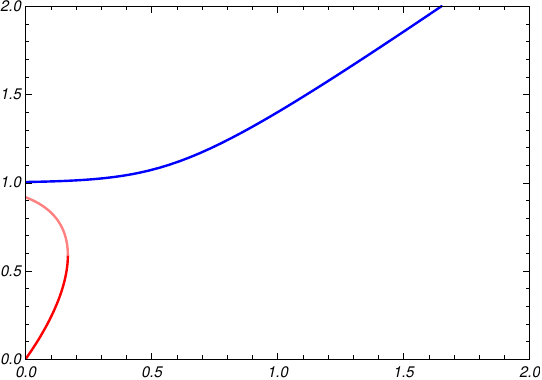}
    \put(-465,65){$\beta_i/\widetilde\beta$}
     \put(-225,65){$m_i\beta$}
    \put(-340,-15){$m_i\widetilde\beta$}
    \put(-110,-15){$ m_i\widetilde\beta $}
    \vspace{.2cm}
	\caption{ The behavior of $\beta_{i}(\widetilde{\beta}, m_{i})$. The blue curve corresponds to the $ m_{i} > \frac{\widetilde{\Phi}_{b} }{2G_{2} \beta_{i}} $ case, and the red and pink curves correspond to the $m_{i} < \frac{\widetilde{\Phi}_{b} }{2G_{2} \beta_{i}} $ case. 
   For both figures, the horizontal axes and the vertical axis of the right figure are divided by $2G_{2}\widetilde{\Phi}_{b}$.
   (\textbf{Left}) Relationship between $\beta_{i}$ and $\widetilde{\beta}$, normalized by $1/m_{i}$. The red curve terminates around $(\beta_{i}/\widetilde{\beta}, \frac{2G_{2}}{\widetilde{\Phi}_{b}}\widetilde{\beta} m_{i}) \simeq (3.5023, 0.1677)$.  (\textbf{Right}) Relationship between $\widetilde{\beta}$ and $E$, normalized by $\beta_{i}$. There are no real Euclidean saddles in the energy range given by Eq.~\eqref{eq:EnergyRange}.
    }
\label{fig:betafunction}
\end{center}
\end{figure}

By solving the Eq. (\ref{eq:lengths}), we can obtain the relation between temperatures $\beta_{i}=\beta_{i}(\widetilde{\beta}, m_{i})$. In general, this can be solved numerically as shown in Fig. \ref{fig:betafunction}. The left panel shows the relationship between $\beta_{i}$ and $ m_{i}$ normalized by $\widetilde{\beta}$. Thus, for a given $\widetilde{\beta}$, the first type of solution (Fig. \ref{fig:geodesics} (\textbf{Left})) exists for any mass $m_{i}$ 
in the range $(0, \infty)$. The second type of solution (Fig. \ref{fig:geodesics} (\textbf{Right})) exists only for small masses, with an upper bound of $m_{i} \lesssim 0.1677 \frac{\widetilde{\Phi}_{b} }{2G_{2} \widetilde{\beta}} $.

The right panel of Fig. \ref{fig:betafunction} shows the relationship between $\beta_{i}$ and $\widetilde{\beta}$ normalized by $1/m_{i}$. This perspective is useful for performing an inverse Laplace transform to the microcanonical ensemble later. The energy $E$ in the microcanonical ensemble and $\beta_{i}$ are related via \cite{Anninos:2017hhn}
\ben
E = - \frac{\pi \ell \widetilde{\Phi}_{b}}{4 G_{2} \beta_{i}^2} ~ .
\een
Thus, the right panel also illustrates the relationship between the variable $\widetilde{\beta}$ in the canonical ensemble and the variable $E$ in the microcanonical ensemble. For a fixed $m_{i}$, when $\widetilde{\beta}$ is sufficiently large there is only one saddle. For $\widetilde{\beta} \lesssim 0.1677 \, \frac{\widetilde{\Phi}_{b} }{2G_{2} m_{i}}$, there are three saddles in the canonical ensemble. In the microcanonical ensemble, the number of real saddles is always one except in the energy range
\be
 \tanh^2 (\pi) \, \frac{\pi \ell G_{2} m_{i}^2}{ \widetilde{\Phi}_{b}} \ \leq \ (-E) \ \leq \ \frac{1}{\tanh^2 \left( \frac{\pi}{2} \right)} \, \frac{\pi \ell G_{2} m_{i}^2}{ \widetilde{\Phi}_{b}} ~ , \label{eq:EnergyRange}
\ee
which originates from the forbidden mass range in Eq. (\ref{eq:massrange}).

\subsection{Flow ER bridge and dS ER bridge in Flow PETS Geometry}
\label{sec:ERstates}

As we have seen in the previous section, the behavior of the infalling particle in the Euclidean geometry is qualitatively different depending on its mass. This difference is also reflected in the structure of the Lorentzian flow PETS geometries.

\begin{figure}[t]
\begin{center}
\includegraphics[width=14.cm]{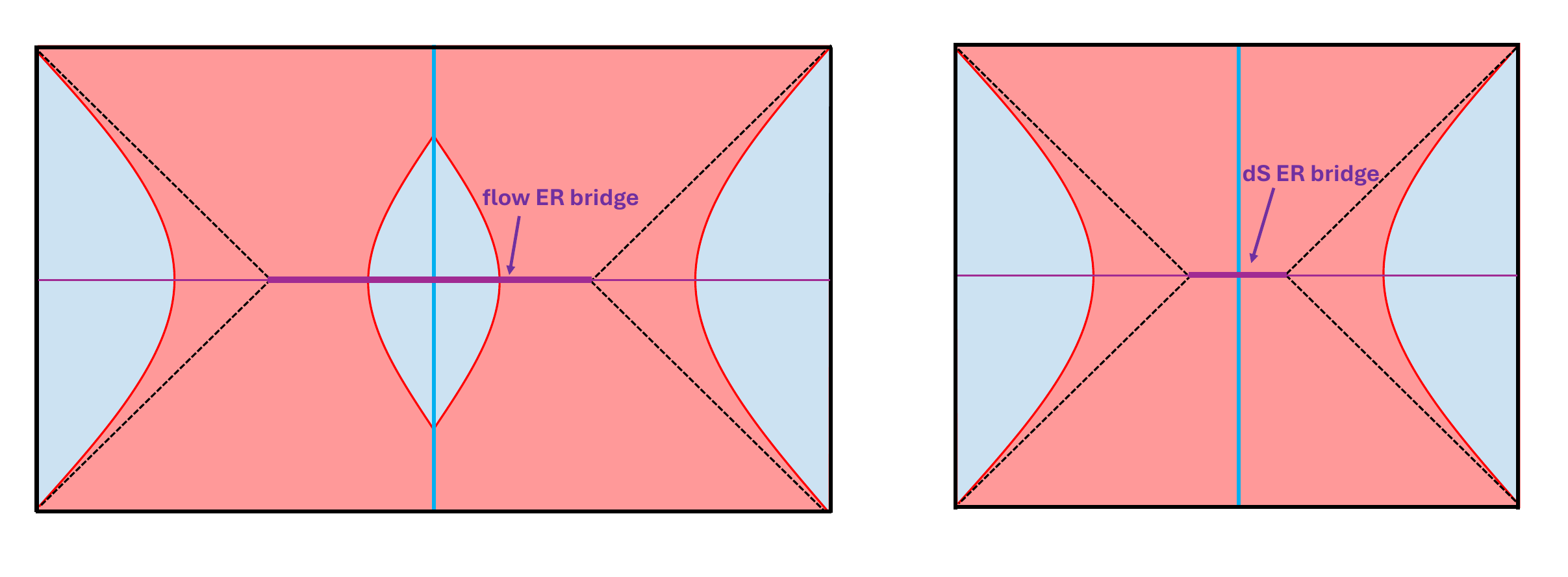} 
	\caption{(\textbf{Left}) Flow PETS geometry with flow ER bridge. (\textbf{Right}) Flow PETS geometry with dS ER bridge. The blue line represents the particle trajectory used to prepare the state.}
\label{fig:ERbridges}
\end{center}
\end{figure}

When the particle mass is sufficiently large compared to the temperature or the absolute value of energy 
\be
\label{eq:conditionAdS}
m_{i} > \frac{1}{\tanh(\pi)} \frac{\widetilde{\Phi}_{b}}{2 G_{2} \beta_{i}} ~~~~ {\rm or} ~~~~ m_{i} > \frac{1}{\tanh(\pi)}  \sqrt{ \frac{- E \widetilde{\Phi}_{b}}{\pi \ell G_{2}} } ~ ,
\ee
the ER bridge at the time-symmetric slice in the flow PETS geometry contains \emph{both AdS and dS portions}, as shown in the left panel of Fig.~\ref{fig:ERbridges}.

As illustrated in Fig.~\ref{fig:ERbridges}, the AdS portion has a finite lifetime and does not extend to future infinity. However, in the large-mass limit, the AdS portion can be made arbitrarily long-lived. Since, in the method \cite{Balasubramanian:2022gmo, Climent:2024trz}, we always take the large-mass limit and the AdS portion has infinite extent to both the future and the past, we refer to this type of Lorentzian wormhole as a \textit{flow Einstein–Rosen bridge}.

Conversely, if the mass is sufficiently small compared to the temperature, 
\be
\label{eq:conditiondS}
m_{i} < \tanh\left( \frac{\pi}{2} \right) \frac{\widetilde{\Phi}_{b}}{2 G_{2} \beta_{i}}   ~~~~ {\rm or} ~~~~  m_{i} < \tanh\left( \frac{\pi}{2} \right) \sqrt{ \frac{- E \widetilde{\Phi}_{b}}{\pi \ell G_{2}} } ~ ,
\ee
the corresponding Lorentzian geometry does not contain an AdS portion behind the horizon, as shown in the right panel of Fig.~\ref{fig:ERbridges}. We call this a \textit{dS Einstein–Rosen bridge}, as it more closely resembles the would-be ER bridge of pure dS spacetime.
\footnote{
Note that such a long wormhole cannot be constructed in higher-dimensional dS spacetimes with positive-energy matter, as it would contradict the Gao–Wald theorem for dS spacetimes~\cite{Gao:2000ga}. See footnote~\ref{foot:GW}, or Subsection~\ref{subsec:dSER}.
}

When the mass lies in the range (\ref{eq:massrange}), it is unclear how to interpret the states and the corresponding Lorentzian geometries. As explained in Subsection~\ref{subsec:flowPETSgeometry}, the gluing process to produce a smooth, real Euclidean geometry does not work for masses in this range. However, Lorentzian geometries obtained by Wick rotating the Euclidean solutions (\ref{eq:SolutionLargeMass}) and (\ref{eq:SolutionSmallMass}) do exist for masses in this range.

One possible interpretation is that the Euclidean configurations are in fact saddle points of the GPI, even if the gluing procedure does not yield a real Euclidean geometry (a similar assumption will be used for the GPI of Euclidean wormholes in Subsection~\ref{subsec:microUniOver}.) Regardless of the interpretation, we will exclude states with particle masses in this intermediate range from the remainder of our analysis.

Therefore, in centaur gravity, the magnitude of the mass not only modulates the length of the ER bridge but also determines its fundamental type. Following \cite{Balasubramanian:2022gmo, Climent:2024trz}, we will primarily consider the large-mass limit, where the microstates correspond to geometries with a flow ER bridge. We will show that these states can account for the horizon entropy within that framework, which is a valuable result in its own right.  

However, in the specific context of investigating dS horizon entropy, it is also interesting to ask whether the entropy can be explained by states that are intrinsically de Sitter-like, \ie, those containing only a dS ER bridge without an AdS interior region. 
Therefore, in the following sections, we will also extend the analysis to the finite-mass regime. We will show that, under certain assumptions, the dS horizon entropy can be explained solely by states with a dS ER bridge, generalizing the state-counting method beyond the strict large-mass limit.

\section{Microstate counting}
\label{sec:4}
We now identify the flow PETS geometries, introduced previously, as the semiclassical counterparts of \emph{flow geometry microstates} or \emph{centaur microstates,} and define their canonical Gram matrix. The central objective of this section is to calculate the dimensionality of the Hilbert space spanned by these microstates for a flow geometry with fixed energy $E$. We refer to this space as the {\it flow geometry Hilbert space}, denoted as $\mathcal{H}_{\rm \mathbf{Flow}}(E)$.

We propose that the Hilbert space $\mathcal{H}_{\rm \mathbf{Flow}}(E)$ is spanned by a family of microcanonical flow geometry PETS states, which have an associated Hilbert space denoted as
\begin{align}
\mathbf{Flow}(E)  = \left\{ \left|\Psi_{i}^{(mc)}(E) \right\rangle \in \mathcal{H} ~ | ~ i=1,\cdots, \Omega \right\} \nonumber \\
\mathcal{H}_{\mathbf{Flow}}(E) = {\rm Span}(\mathbf{Flow}(E))~, \hspace{1.9cm}
\end{align}
where $\Omega$ is the total number of flow states. 

This is a natural choice because these PETS geometries share the same exterior as the flow geometry but contain different interiors. We employ the microcanonical ensemble for two crucial reasons: (i) Standard canonical PETS have non-zero coefficients for all energy eigenstates, and a generic family of such states cannot be confined to a finite-dimensional subspace (\ie much smaller than $\Omega$). (ii) The flow geometry and its related long-ER-bridge geometries correspond to semiclassical geometries for microcanonical PETS, not their canonical counterparts. For further details on (i), see \cite{Balasubramanian:2022gmo, Climent:2024trz}. 

\begin{figure}[t!]
\begin{center}
\includegraphics[width=12.cm]{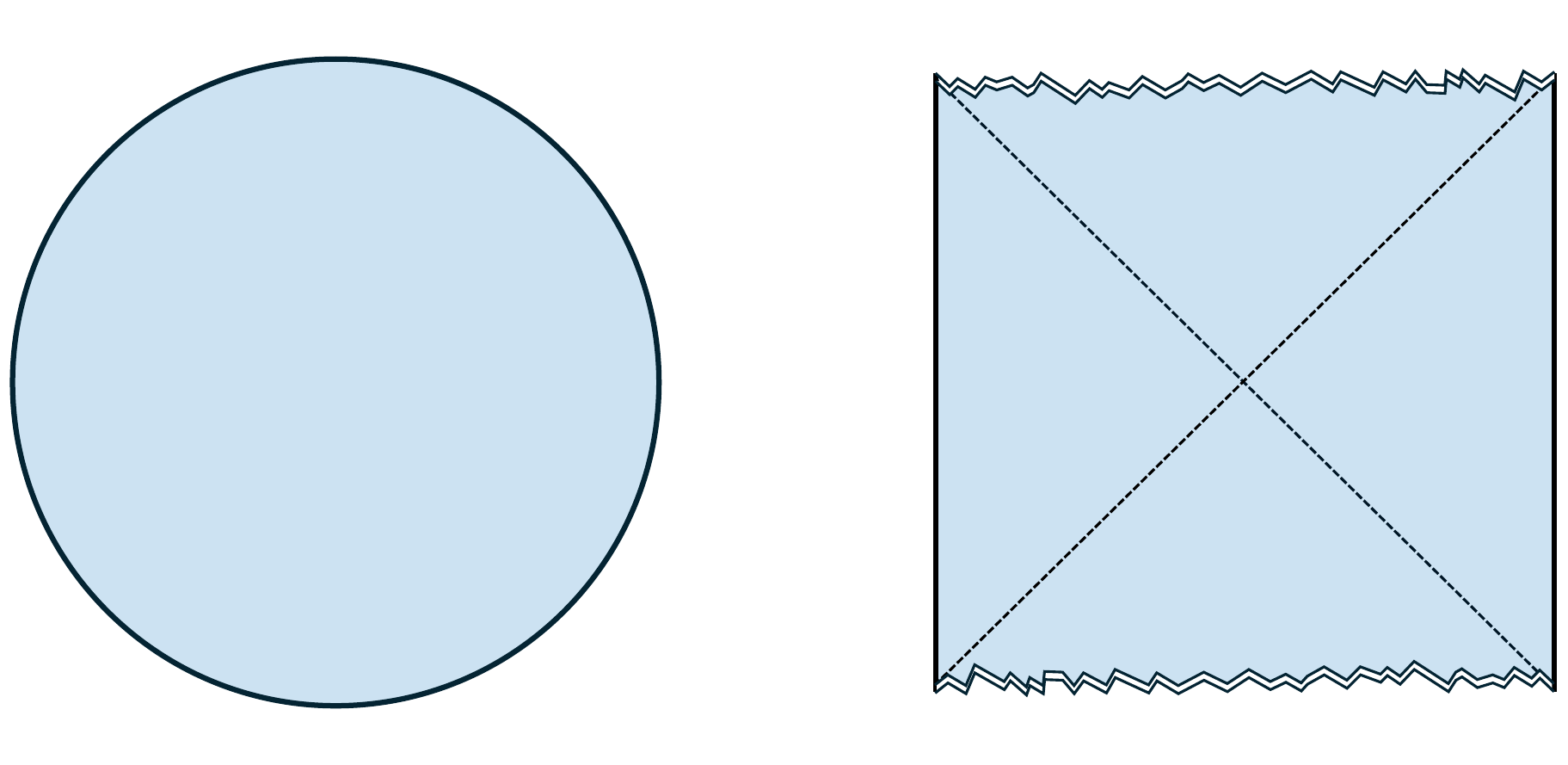} 
	\caption{(\textbf{Left}) Euclidean AdS BH. This is the dominant saddle of the GPI for ${\rm Tr ~} e^{-\beta H}$. (\textbf{Right}) (Lorentzian) two-sided AdS BH.  }
\label{fig:TFD}
\end{center}
\end{figure}

We will not use the explicit form of the microcanonical states themselves. Instead, our strategy will be to compute their overlaps using the GPI. Specifically, we avoid calculating the overlaps of microcanonical PETS directly. We first calculate the Gram matrix and its higher moments for PETS in the canonical ensemble. We then perform an inverse Laplace transform to obtain the moments of the \emph{microcanonical} Gram matrix. 

Crucially, because these quantities are defined through the GPI, they can be regarded as ensemble averaged quantities. In particular, $G_{ij}$ can be treated as a random matrix \cite{Saad:2019lba}. Its resolvent, defined by an infinite sum over its moments, ultimately allows us to extract the Hilbert space dimension. Therefore, this approach modifies the standard prescription for usual Gram matrix, outlined in Appendix \ref{AppGram}, by incorporating an ensemble average. The details of this framework will be elaborated later in this section. 

In summary, our computational strategy is organized as follows:
\begin{itemize}
\item Calculate statistical moments of the \emph{canonical} Gram matrix $ \boldsymbol{G}^{(c)}$ by GPIs. (Section \ref{subsec:cGram})
\item Perform inverse Laplace transforms to obtain the microcanonical Gram matrix $ \boldsymbol{G}^{(mc)}$ and their statistical moments. (Section \ref{subsec:mcGram})
\item Calculate the trace of the resolvent matrix $\overline{R(\lambda)}$ for $ \boldsymbol{G}^{(mc)}$ and calculate ${\rm dim}(\mathcal{H}_{\mathbf{Flow}}(E))$ (Section \ref{subsec:dimensionality})
\end{itemize}
In the remainder of this section, we briefly explain the rationale behind each step. Detailed calculations are presented in subsequent sections.
\subsection{Canonical PETSs and Their Gram Matrix} \label{subsec:cGram}
\begin{figure}[t]
\begin{center}
\includegraphics[width=13.cm]{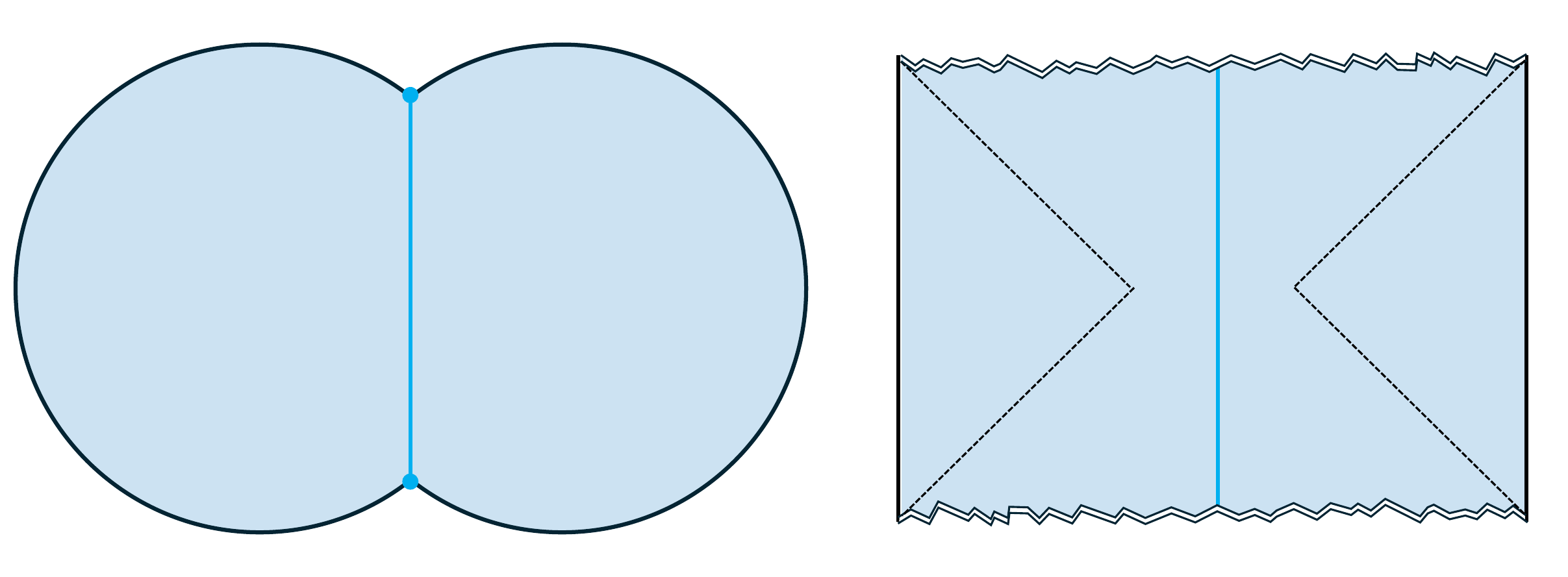} 
	\caption{
(\textbf{Left}) The dominant saddle of the GPI for the normalization of the canonical PETS. The inserted operators, $\mathcal{O}_{i}$, create a heavy particle, which backreacts on the Euclidean BH geometry.  
(\textbf{Right}) The Lorentzian continuation of the Euclidean saddle. Due to the backreaction of the heavy particle, the ER bridge between the two sides increases.}
\label{fig:PETS}
\end{center}
\end{figure}

The discussion on canonical PETS states in this section is similar to the black hole case (see Fig. {\ref{fig:PETS}}). We set notation used in the rest of the paper. Readers familiar with this method can proceed directly to Section \ref{subsec:mcGram}.

A partially entangled thermal state is a one-parameter generalization of the thermofield double state \cite{Goel:2018ubv}. The TFD state, $\left|\Psi_{\rm TFD}^{(c)} (\beta) \right\rangle$, is canonical purification of the thermal density matrix 
\be
e^{-\beta H} ~ \to ~ \left|\Psi^{(c)}_{\rm TFD}(\beta) 
 \right\rangle = \sum_{a} e^{- \frac{\beta}{2} H} |E_{a}\rangle_{L} | E_{a}\rangle_{R}~.
\ee
The trace of the thermal density matrix and the TFD state admit representations via Euclidean and Lorentzian GPIs, respectively, with dominant saddles being the Euclidean AdS black hole and the two-sided AdS black hole (see Fig. \ref{fig:TFD}). 

A canonical PETS is defined by the canonical purification of a particular generalization of the thermal density matrix, obtained by inserting an operator $\mathcal{O}_{i}$;
\be
e^{-\frac{\widetilde{\beta}_{R}}{2}H} \mathcal{O}_{i} e^{-\widetilde{\beta}_{L}H }  \mathcal{O}_{i} e^{-\frac{\widetilde{\beta}_{R}}{2}H} ~ \to ~ \left|\Psi^{(c)}_{i} (\widetilde{\beta}_{L}, \widetilde{\beta}_{R}) \right\rangle = \sum_{a, b} \left( e^{-\frac{\widetilde{\beta}_{R}}{2}H} \mathcal{O}_{i} e^{-\frac{\widetilde{\beta}_{L}}{2} H } \right)_{ab}  |E_{a}\rangle_{L} | E_{b}\rangle_{R}~.
\ee
These states also have GPI representations; their dominant saddles are black holes with a heavy particle insertion as illustrated in Fig. \ref{fig:PETS}. 

\begin{figure}[t]
\begin{center}
\includegraphics[width=3.cm]{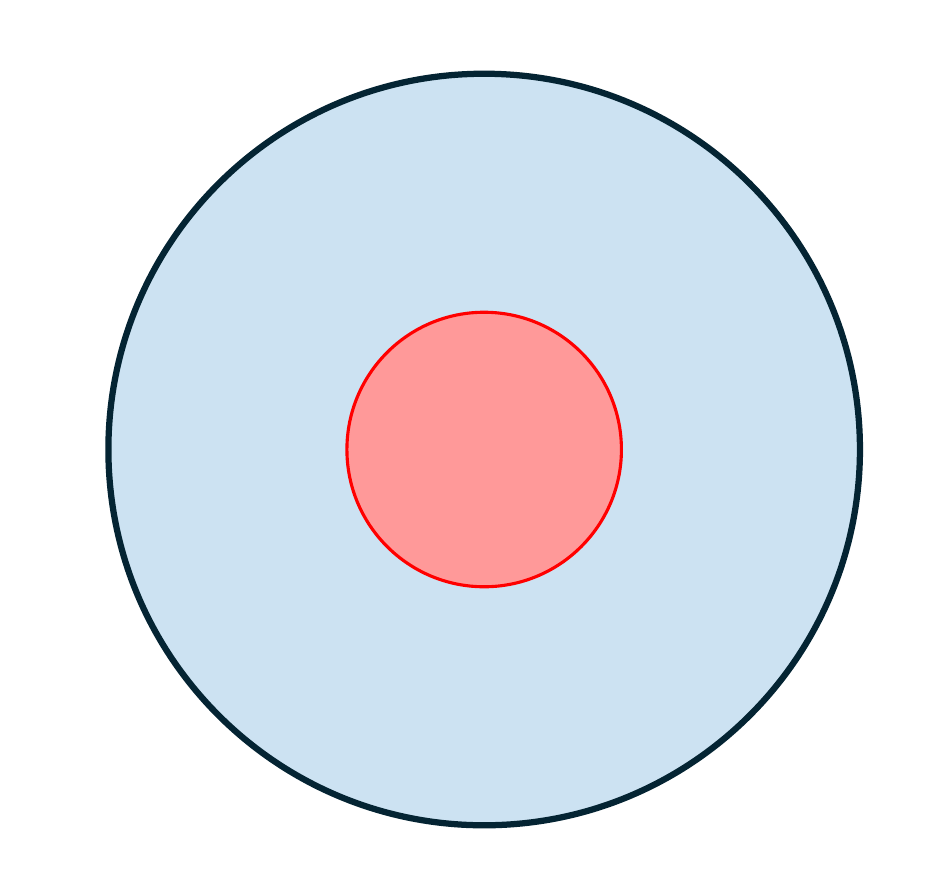} ~
\includegraphics[width=3.cm]{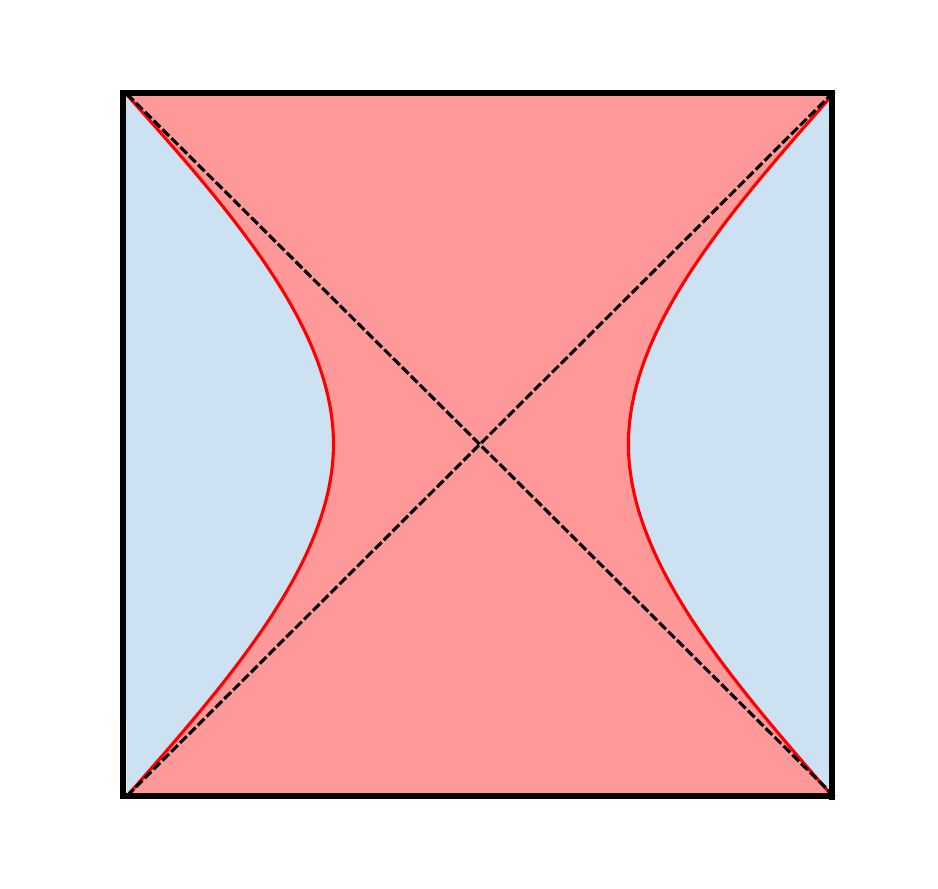} ~
\includegraphics[width=4cm]{EflowPETS.pdf} ~
\includegraphics[width=4cm]{LflowPETS.pdf} \\
(\textbf{a}) \hspace{2.5cm} (\textbf{b})  \hspace{3cm} (\textbf{c})  \hspace{3.6cm} (\textbf{d})  \hspace{0.1cm} 

\caption{
(\textbf{a}) Euclidean version of the flow geometry introduced in Subsection~\ref{sec:flowgeometries}. A dS region appears at the center.  
(\textbf{b}) The flow geometry.  
(\textbf{c}) The Euclidean flow PETS geometry introduced in Subsection~\ref{subsec:flowPETSgeometry}.  
(\textbf{d}) The Lorentzian continuation of the flow PETS geometry.}
\label{fig:flowPETS}
\end{center}
\end{figure}

Since the PETS geometries share the same exterior as the black hole geometry, they can be interpreted as black hole microstates. In our centaur gravity setup with $G_{2}>0$, flow geometries also exist alongside AdS black holes (Fig. \ref{fig:flowPETS}), but they are subdominant saddles of the GPIs in the canonical ensemble.
However, they become the dominant saddles in the \textit{microcanonical} ensemble, for microcanonical versions of TFD and PETS, where we specify energy instead of temperature. This is why we will consider microcanonical versions to construct a family of microstates for flow geometry later.

In the following, we restrict our attention to the symmetric case  $\widetilde{\beta}_{L}=\widetilde{\beta}_{R}=\widetilde{\beta}$;
\be
 \left|\Psi^{(c)}_{i} (\widetilde{\beta}) \right\rangle = \sum_{a, b} e^{-\frac{\widetilde{\beta}}{2} H} \mathcal{O}_{i} e^{-\frac{\widetilde{\beta}}{2} H }  |E_{a}\rangle_{L} | E_{b}\rangle_{R}~.
\ee
The unnormalized canonical Gram matrix of the PETSs is defined by the overlaps
\be
\mathcal{G}_{ij}^{(c)}(\widetilde{\beta}) := \left\langle \Psi^{(c)}_{i} (\widetilde{\beta}) \right| \left. \Psi^{(c)}_{j} (\widetilde{\beta}) \right\rangle~.
\ee
As we will explain in the next section, this matrix and its higher $n$-th moments have explicit GPI expressions.

\subsection{Inverse Laplace Transform and the Microcanonical Gram Matrix} \label{subsec:mcGram}
To construct the flow geometry microstates, $\textbf{Flow}(E)$, and the flow Hilbert space, $\mathcal{H}_{\textbf{Flow}}(E)$, we make use of microcanonical PETSs. The preparation of the microcanonical PETS is similar to its canonical analogue. It is a generalization of the microcanonical TFD state given by
\be
\left|\Psi^{(mc)}_{\rm TFD}(E) \right\rangle = \sum_{a} f_{E}(E_{a}) |E_{a}\rangle_{L} | E_{a}\rangle_{R}~,
\ee
where $f_{E}(E_{a})$ is a window function sharply peaked around $E_{a}=E$. For the purpose of building a finite-dimensional $\mathcal{H}_{\textbf{Flow}}$, it is crucial to choose $f_{E}(E_{a})$ so that it does not have tails, that is, it has a sufficiently small compact support. In this work, we simply take it as a step function
\be
f_{E}(E_{a})= \Theta(E_{a}-E+\frac{\varepsilon}{2}) - \Theta(E_{a}-E - \frac{\varepsilon}{2})~.
\ee
We define the symmetric microcanonical PETS by 
\footnote{
This is slightly different from the one where we take $E_{L}=E_{R}=E$ for the non-symmetric microcanonical PETS, defined by
\ben
\left|\Psi_{i}^{(mc)}(E_{L}, E_{R}) \right\rangle = \sum_{a, b} f_{E_{L}}(E_{a}) f_{E_{R}}(E_{b}) \big( \mathcal{O}_{i} \big)_{ab} |E_{a} \rangle_{L} | E_{b} \rangle_{R} 
\een 
When setting $E_{L}=E_{R}=E$, it apparently includes non-diagonal elements, whereas our definition does not. In both definitions, when we set $\mathcal{O}_{i}=\mathbf{1}$, they reduce to the microcanonical TFD state.
}
\be
\left|\Psi_{i}^{(mc)}(E) \right\rangle = \sum_{a} f_{E}(E_{a}) \big( \mathcal{O}_{i} \big)_{aa} |E_{a} \rangle_{L} | E_{a} \rangle_{R}~.
\ee
The corresponding unnormalized microcanonical Gram matrix is defined by
\be
\mathcal{G}_{ij}^{(mc)}(E)= \left\langle \Psi^{(mc)}_{i} (E) \right| \left. \Psi^{(mc)}_{j} (E) \right\rangle = \sum_{a} f_{E}(E_{a}) \big( \mathcal{O}_{i} \big)_{aa} \big( \mathcal{O}_{j} \big)_{aa}~.
\ee

Although the matrix $\mathcal{G}_{ij}^{(mc)}(E)$ might have a GPI expression, we will compute it via the inverse Laplace transform of the symmetric canonical Gram matrix. To derive this relation, let us consider the non-symmetric canonical Gram matrix. It can be written as
\begin{align}
\mathcal{G}_{ij}^{(c)}(\widetilde{\beta}_{L}, \widetilde{\beta}_{R})= \left\langle \Psi^{(c)}_{i} (\widetilde{\beta}_{L}, \widetilde{\beta}_{R}) \right| \left. \Psi^{(c)}_{j} (\widetilde{\beta}_{L}, \widetilde{\beta}_{R}) \right\rangle \notag \hspace{1.6cm} \\
=  \sum_{a, b} e^{-\widetilde{\beta}_{L} E_{a} -\widetilde{\beta}_{R} E_{b} } \big( \mathcal{O}_{i} \big)_{ba} \big( \mathcal{O}_{j} \big)_{ab} \notag \hspace{1.5cm} \\
= \sum_{a, b} e^{ -\widetilde{\beta}(E_{a}+E_{b}) - \frac{\Delta \widetilde{\beta}}{2} (E_{a}-E_{b}) } \big( \mathcal{O}_{i} \big)_{ba} \big( \mathcal{O}_{j} \big)_{ab} ~,
\end{align}
where we have defined $2\widetilde{\beta}=\widetilde{\beta}_{L} + \widetilde{\beta}_{R}$ and $\Delta \widetilde{\beta}=\widetilde{\beta}_{L} - \widetilde{\beta}_{R}$. If we multiply by $e^{2\widetilde{\beta}E_{a'}}$ and perform the integration with respect to $\widetilde{\beta}$ and $\Delta\widetilde{\beta}$ along the imaginary axis, it becomes
\be
\left(\frac{1}{2\pi i}\right)^2 \int d\widetilde{\beta} \int d \Delta \widetilde{\beta} e^{2\widetilde{\beta}E_{a'}} ~  \mathcal{G}_{ij}^{(c)}(\widetilde{\beta} + \frac{\Delta \widetilde{\beta}}{2}, \widetilde{\beta} - \frac{\Delta \widetilde{\beta}}{2}) = \big( \mathcal{O}_{i} \big)_{a'a'} \big( \mathcal{O}_{j} \big)_{a'a'} ~.
\ee
If $\mathcal{G}_{ij}^{(c)}$ is suitable to a saddle-point approximation for the $\Delta \widetilde{\beta}$ integral at $\Delta\widetilde{\beta}=0$, which is exactly the case for its GPI representation, then the following relation holds
\be
\big( \mathcal{O}_{i} \big)_{a'a'} \big( \mathcal{O}_{j} \big)_{a'a'} \simeq \left(\frac{1}{2\pi i }\right)^2  \int d\widetilde{\beta}  e^{2\widetilde{\beta}E_{a'}} ~  \mathcal{G}_{ij}^{(c)}(\widetilde{\beta}) ~.
\ee
Multiplying by the window function $f_{E}(E_{a'})$ and summing over energy levels, we obtain
\be
\mathcal{G}_{ij}^{(mc)}(E) \simeq \left(\frac{1}{2\pi i}\right)^2 \sum_{a} f_{E}(E_{a}) \int d\widetilde{\beta}  e^{2\widetilde{\beta}E_{a}} ~  \mathcal{G}_{ij}^{(c)}(\widetilde{\beta})~.
\ee
More generally, the $n$-th moment of the microcanonical Gram matrix is related to the canonical one by
\begin{align}
\overline{\mathcal{G}_{i_{1}i_{2}}^{(mc)}(E) \mathcal{G}_{i_{2}i_{3}}^{(mc)}(E) \cdots \mathcal{G}_{i_{n}i_{1}}^{(mc)}(E) } \hspace{8.3cm} \notag \\
\simeq \prod_{p=1}^{n} \left[ \left(\frac{1}{2\pi i}\right)^2 \sum_{a_{p}} f_{E}(E_{a_{p}}) \int d\widetilde{\beta}_{p}  e^{2\widetilde{\beta}_{p} E_{a_{p}}} \right] \overline{\mathcal{G}_{i_{1}i_{2}}^{(c)}(\widetilde{\beta}_{1}) \mathcal{G}_{i_{2}i_{3}}^{(c)}(\widetilde{\beta}_{2}) \cdots \mathcal{G}_{i_{n}i_{1}}^{(c)}(\widetilde{\beta}_{n}) } ~.\label{eq:ILT}
\end{align}

In Sections \ref{sec:heavymass} and \ref{sec:microcanonical}, we use Eq. (\ref{eq:ILT}) to compute the $n$-th moments of the microcanonical Gram matrix from the moments of the canonical Gram matrix obtained in Section \ref{sec:cGm} using the GPI.

\subsection{Schwinger-Dyson equation and
the Hilbert Space Dimension} \label{subsec:dimensionality}
The remaining task is to compute the trace of the resolvent matrix for the microcanonical Gram matrix, $\mathbf{G}^{(mc)}$, and extract the dimension of the flow geometry Hilbert space, $\mathcal{H}_{\textbf{Flow}}(E)$, using the analogue of Eq. (\ref{eq:Disc1}) involving the resolvent. Since we are working with the statistical average of the Gram matrix, we consider the average of the resolvent matrix, which is given by
\be
\overline{R_{ij}(\lambda)} = \left( \overline{ \frac{1}{\lambda \mathbf{1} - \mathbf{G}^{(mc)}}} \right)_{ij} = \frac{1}{\lambda} \delta_{ij} + \sum_{n=1}^{\infty} \frac{1}{\lambda^{n+1}} \left(\overline{(\mathbf{G}^{(mc)})^{n} }\right)_{ij}~.
\ee
In this expression, the $n$-th moment of the normalized microcanonical Gram matrix is given by
\be
\begin{split}
\overline{G_{i_{1}i_{2}}^{(mc)}(E) G_{i_{2}i_{3}}^{(mc)}(E) \cdots G_{i_{n}i_{1}}^{(mc)}(E) }  = \overline{\mathcal{G}_{i_{1}i_{2}}^{(mc)}(E) \mathcal{G}_{i_{2}i_{3}}^{(mc)}(E) \cdots \mathcal{G}_{i_{n}i_{1}}^{(mc)}(E) } ~/ ~ \mathcal{N}_{n} ~ . \\
\left(\mathcal{N}_{n} =  \mathcal{G}_{i_{1}i_{2}}^{(mc)}(E) ~ \mathcal{G}_{i_{2}i_{3}}^{(mc)}(E) \cdots \mathcal{G}_{i_{n}i_{1}}^{(mc)}(E) ~. \right)  \hspace{3cm}  \label{eq:nomalizedGram} 
\end{split}
\ee
Given that these moments have a GPI representation, and both $\Omega$ and $e^{2S}$ are large, the average resolvent $\overline{R(\lambda)}$ satisfies a Schwinger-Dyson equation
\footnote{
This is a sufficient condition but not a necessary condition. The necessary condition is the suppression of non-adjacent cumulants in the cumulant expansion of the moment. 
}
\be
\lambda \overline{R(\lambda)} = \Omega + \sum_{n=1}^{\infty} \left[ \sum_{l_{1}=1}^{\Omega} \sum_{l_{2}=1}^{\Omega} \cdots \sum_{l_{n}=1}^{\Omega} \left.\overline{G_{l_{1}l_{2}}^{(mc)}G_{l_{2}l_{3}}^{(mc)}\cdots G_{l_{n}l_{1}}^{(mc)}} \right|_{conn} \overline{R_{l_{1}l_{1}}(\lambda)}  \cdots \overline{R_{l_{n}l_{n}}(\lambda)}   \right]~.
\label{eq:SD}
\ee
Here $\left.\overline{G_{l_{1}l_{2}}^{(mc)}G_{l_{2}l_{3}}^{(mc)}\cdots G_{l_{n}l_{1}}^{(mc)}} \right|_{conn}$ denotes the $n$-th cumulant, which is the fully connected component in the GPI representation. 

\begin{figure}[t]
\begin{center}
\includegraphics[width=15.cm]{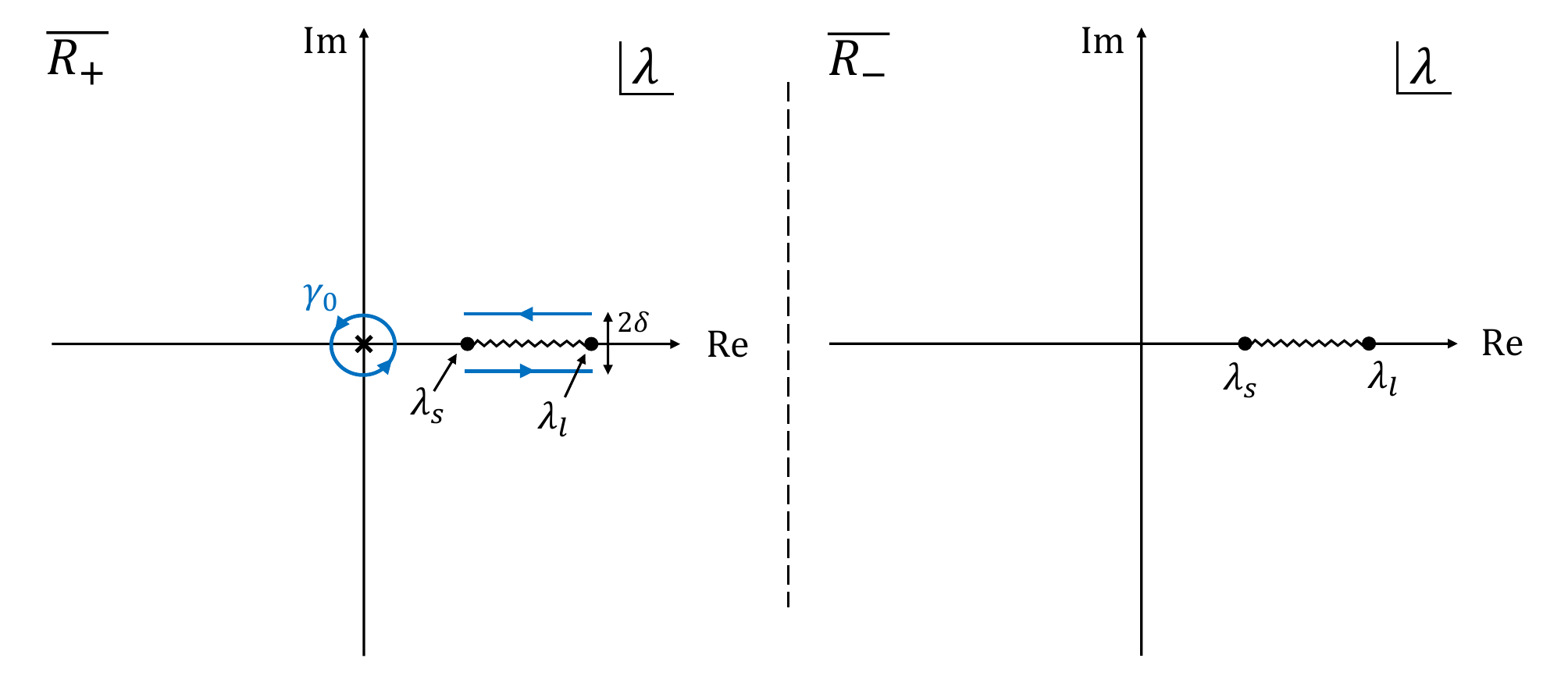} 
	\caption{Riemann surface for $\overline{R(\lambda)}$ when $\Omega > A$. It consists of two complex planes, one for $\overline{R_{+}}$ and the other for $\overline{R_{-}}$. The branch cut can be put on the interval $[\lambda_{s}, \lambda_{l}]$ on the real axis. 
    The integration contour for \eqref{eq:Origin2}, $\gamma_0$, encloses the origin on the first sheet, while the contour for \eqref{eq:Disc2} consists of two straight lines above and below the branch cut.
    }
\label{fig:contour2}
\end{center}
\end{figure}

Crucially, a key simplification occurs if this cumulant is independent of the indices and takes the specific form
\be
\left.\overline{G_{l_{1}l_{2}}^{(mc)}G_{l_{2}l_{3}}^{(mc)}\cdots G_{l_{n}l_{1}}^{(mc)}} \right|_{conn} = A B^n \label{eq:AB}
\ee
for some positive constants $A$ and $B$. In this case, the Schwinger-Dyson equation is simplified to
\be
\lambda \overline{R(\lambda)} = \Omega + \sum_{n=1}^{\infty} A B^{n} \left(\overline{R(\lambda)}\right)^{n} = \Omega + \frac{A B \overline{R(\lambda)}}{1- B \overline{R(\lambda)}}~.
\ee
This is just a quadratic equation for $\overline{R(\lambda)}$, so we can easily solve it;
\be
\overline{R_{\pm}(\lambda)}= \frac{-A B + \lambda +  \Omega B \pm \sqrt{(AB -\lambda - \Omega B)^2 -4 \lambda \Omega B} }{2 \lambda B}~.
\ee
Here, a remarkable point of the random matrix appears. For a usual matrix, its eigenvalues correspond to poles of $R(\lambda)$, as explained around Eq. (\ref{eq:Disc1}). Nevertheless, for a random matrix, a hallmark of random matrix theory is that matrix eigenvalues are also randomly distributed. Thus, the position of poles of $\overline{R(\lambda)}$ is in some sense averaged over and forms a branch cut instead of finite number of poles. Consequently, $\overline{R(\lambda)}$ is defined on two complex planes, as it develops a branch cut $[\lambda_{s}, \lambda_{l}]$ on the real axis (see Fig. \ref{fig:contour2}), where
\begin{align}
\lambda_{s}=B(\sqrt{A}-\sqrt{\Omega})^2 ~, \\
\lambda_{l}=B(\sqrt{A}+\sqrt{\Omega})^2 ~.
\end{align}

When the total number of states $\Omega \neq A$, the numbers of zero and non-zero eigenvalues are given by
\begin{align}
({\rm number ~ of ~ zero ~ eigenvalues}) = \frac{1}{2\pi i} \int_{\gamma_{0}} d\lambda ~\overline{ R(\lambda)} \hspace{1.5cm} \label{eq:Origin2} \\
({\rm number ~ of ~ non}-{\rm zero ~ eigenvalues}) = -\frac{1}{2\pi i}\int^{\lambda_{l}}_{\lambda_{s}} d\lambda ~ {\rm Disc}(\overline{R (\lambda)})~.\label{eq:Disc2} 
\end{align}
Here, $\gamma_{0}$ is a small circle around the origin of the first sheet (\ie the sheet where $\overline{R(\lambda)} = \overline{R_{+}(\lambda)} $), and the second equation is an averaged analogue of (\ref{eq:Disc1}). The discontinuity across the branch cut, ${\rm Disc}(\overline{R(\lambda)})$, is given by
\be
{\rm Disc}(\overline{R(\lambda)}) = \lim_{\delta \to 0} \left( \overline{R_{+}}(\lambda + i \delta) - \overline{R_{+}}(\lambda - i \delta) \right) ~.\label{eq:DefDisc}
\ee
The integration contours for \eqref{eq:Origin2} and \eqref{eq:Disc2} are depicted in Fig.~\ref{fig:contour2}.

Since the total number of eigenvalues is $\Omega$, we can use both expressions \eqref{eq:Origin2} and \eqref{eq:Disc2} to obtain the number of non-zero eigenvalues, \ie, the dimensionality of the Hilbert space, ${\rm dim}(\mathcal{H}_{\mathbf{Flow}})$. Probably, an indirect but relatively easy way is to use \eqref{eq:Origin2}.
When $\Omega > A$, $\overline{R_{+}(\lambda)}$ is approximated as
\be
\overline{R_{+}(\lambda)} \simeq \frac{-A + \Omega  + \sqrt{(A- \Omega)^2}}{2\lambda } = \frac{\Omega - A}{\lambda}~,
\ee
around the origin. Therefore, evaluating (\ref{eq:Origin2}) gives 
\be
({\rm number ~ of ~ zero ~ eigenvalues}) = \Omega- A~,
\ee
and thus the dimensionality
\footnote{
When $\Omega < A$, the pole at the origin of the first sheet disappears. Thus there are no zero-eigenvalues, and ${\rm dim}(\mathcal{H}_{\mathbf{Flow}}) = \Omega$.
}
\be
{\rm dim}(\mathcal{H}_{\mathbf{Flow}}) = A ~ . \label{eq:A}
\ee

Of course, instead of computing the difference between $\Omega$ and the number of zero eigenvalues, we can directly obtain ${\rm dim}(\mathcal{H}_{\mathbf{Flow}})$ by computing (\ref{eq:Disc2}). First, the discontinuity ${\rm Disc}(\overline{R(\lambda)})$ (\ref{eq:DefDisc}) and the dimension ${\rm dim}(\mathcal{H}_{\mathbf{Flow}})$ are given by 
\begin{align}
{\rm Disc}(\overline{R(\lambda)}) = \frac{ i \sqrt{-(AB -\lambda - \Omega B)^2 +4 \lambda \Omega B} }{\lambda B} ~ ,\hspace{1.5cm} \\
\Rightarrow ~~ {\rm dim}(\mathcal{H}_{\mathbf{Flow}}) = - \int_{\lambda_{s}}^{\lambda_{l}} d\lambda \frac{ \sqrt{-(AB -\lambda - \Omega B)^2 +4 \lambda \Omega B} }{2 \pi\lambda B} ~ .
\end{align}
By changing the integration variable to $\theta$ defined by the relation 
\be
\lambda = B(A+\Omega) + 2B\sqrt{A\Omega} \cos\theta ~ ,
\ee
the integration becomes
\footnote{
Here, we used a formula 
\ben
\int^{\pi}_{0}  d\theta \frac{ \sin^2 \theta }{a + b \cos \theta} = \frac{\pi (a-\sqrt{a^2 - b^2})}{b^2} ~~~ (a>|b|) ~ .
\een
}
\begin{align}
 {\rm dim}(\mathcal{H}_{\mathbf{Flow}}) =  \frac{2A\Omega}{\pi} \int^{\pi}_{0}  d\theta \frac{\sin^2 \theta}{A+\Omega + 2\sqrt{A\Omega} \cos \theta } \notag \hspace{2.9cm} \\
 = \frac{2 A \Omega}{\pi} \frac{ \pi (A+ \Omega - \sqrt{(A-\Omega)^2})}{ 4 A \Omega } = \left\{ 
 \begin{array}{ll}
 \Omega & ~~  (A \geq \Omega) \\
 A & ~~ (A \leq \Omega)
 \end{array} ~
. \right.
\end{align}

In Sections \ref{sec:heavymass} and \ref{sec:microcanonical}, we will show that the centaur GPI predicts the cumulant form (\ref{eq:AB}) with 
\be
A\simeq e^{2S_{\rm flow}},~ {\rm and},~ B \simeq e^{-2S_{\rm flow}}~. 
\ee
where $S_{\rm flow}$ is the entropy of the flow geometry. Using the above argument, we conclude that, when the number of states is large, \ie, $\Omega > \exp(2S_{\rm flow})$, yields the final result
\be
{\rm dim}(\mathcal{H}_{\mathbf{Flow}}) \simeq e^{2S_{\rm flow}} ~ .
\ee

\section{Canonical Gram matrix from Gravitational Path Integrals}\label{sec:cGm}
In this section, we calculate the on-shell gravity action for the flow PETS geometries and its generalization to $n$-mouth wormholes. These geometries contribute to the centaur GPI for the statistical average of the canonical Gram matrices 
\be
\overline{\mathcal{G}_{i_{1}i_{2}}^{(c)}(\widetilde{\beta}_{1}) \mathcal{G}_{i_{2}i_{3}}^{(c)}(\widetilde{\beta}_{2}) \cdots \mathcal{G}_{i_{n}i_{1}}^{(c)}(\widetilde{\beta}_{n}) }~. 
\ee

It is important to recall that for $G_{2}>0$ in the canonical ensemble, these flow geometries are not the dominant saddles of the GPI. The dominant saddles are those without a dS region, as discussed in Section~\ref{sec:PETS} and Appendix~\ref{app:thermodynamics}. Thus, the calculation of the canonical Gram matrix at leading order is essentially identical to that in JT gravity. However, the generalization of the flow PETS geometries become the dominant saddles in the microcanonical ensemble, when the energy is specified as negative after performing the inverse Laplace transform.

We begin by computing the average for a single Gram matrix element, $ \overline{\mathcal{G}^{(c)}_{ij} (\widetilde{\beta})}$.
 
 \subsection{Single Gram Matrix Element: $\overline{ \mathcal{G}^{(c)}_{ij} (\widetilde{\beta}) } $}
The GPI computes the ensemble average of the Gram matrix 
$\boldsymbol{\mathcal{G}}^{(c)} (\widetilde{\beta})$. At leading order, this average is diagonal
\be
\overline{ \mathcal{G}^{(c)}_{ij} (\widetilde{\beta}) } = Z^{i}_{1}(2\widetilde{\beta}) \, \delta_{ij}  ~ .
\ee
The diagonal component $Z^{i}_{1}(\widetilde{\beta})$ is precisely the normalization of the canonical PETS and is given by
\be
Z_{1}^{i}(2\widetilde{\beta}) = \int_{\text{conn}} \mathcal{D}\boldsymbol{g} \, \mathcal{D}\Phi \, \mathcal{D} \gamma_{1} ~ e^{-I_{E}^{i}[\boldsymbol{g},\Phi,\gamma_{1}]} ~.\label{eq:GPI1}
\ee
The Euclidean action is
\be
I_{E}^{i}[\boldsymbol{g},\Phi,\gamma_{1}]= I_{E}^{\text{cent}}[\boldsymbol{g},\Phi]+I^{m_{i}}_{E}[\boldsymbol{g},\Phi, \gamma_{1}] ~ ,
\ee
where $I_{E}^{\text{cent}}$ is the action for the centaur geometry and $I^{m_{i}}_{E}$ is the action for a particle of mass $m_i$, defined in Eq.~(\ref{eq:particle}).

We evaluate this path integral by considering its saddle points. The relevant saddle is the Euclidean flow PETS geometry, constructed and analyzed in Subsection~\ref{subsec:flowPETSgeometry} (with its explicit form given in Appendix~\ref{appedix:B}). Although this saddle is subdominant in the canonical ensemble, its on-shell action is well-defined and essential for the subsequent microcanonical analysis. We now proceed to compute this on-shell action.

\subsubsection*{Evaluation of $I^{cent}_{E}$}
The bulk term in the centaur action vanishes on-shell due to equations of motion. The only contribution comes from the boundary term at $r \rightarrow \infty$. The integrand of this boundary term evaluates to
\begin{align}
\left. \frac{-1}{8 \pi G_{2} } \sqrt{h} \Phi \left( K - \frac{1}{\ell} \right)\right|_{r=\infty}
= \frac{ \pi \ell \widetilde{\Phi}_{b}}{4 G_{2} \beta_{i}^2 } ~.
\end{align}
Integrating this over the Euclidean time boundary of length $2\widetilde{\beta}$ and adding the topological term $-\Phi_{0}/4G_{2}$ yields
\be
\left. I^{cent}_{E}\right|_{\rm on-shell} = \frac{\pi \ell \widetilde{\Phi}_{b}}{2 G_{2} \beta_{i}^2 } 
 \widetilde{\beta}  - \frac{\Phi_{0}}{4 G_{2}}~.  \label{eq:onshellcentuaraction}
\ee

\subsubsection*{Evaluation of $I^{m_{i}}_{E}$}

The particle action $I^{m_{i}}_{E}$ consists of a dynamical term and a counterterm. The dynamical term is simplified using the normalization condition for the tangent vector, $g_{\mu\nu}e^{\mu}_{s}e^{\nu}_{s}= f_{\rm flow} \dot{\mathcal{T}}^2 + \frac{1}{f_{\rm flow}}\dot{\mathcal{R}}^2 =1$, and results in 
\be
m_{i}\int_{\gamma_{1}}d\tau \sqrt{h} = m_{i} \int_{-\tau_{\infty}}^{\tau_{\infty}}d\tau = 2m_{i} s_{\infty}~.
\ee
Here $s_{\infty}$ is defined by $\mathcal{R}(s_{\infty})=r_{\infty}$, where $r_{\infty}$ is the location of the AdS boundary, which is set to infinity at the end of the calculation.
\footnote{
Although $r_{\infty}$ did not appear explicitly in the calculation of $I^{cent}_{E}$, we should understand that calculations are performed with the boundary at $r=r_{\infty}$. Then, we will take the limit $r_{\infty} \to \infty$ after completing all calculations.
}
Using the explicit solutions for the particle trajectory from equations (\ref{eq:SolutionLargeMass}) and (\ref{eq:SolutionSmallMass}), we find $s_{\infty}$ has the following asymptotic behavior
\be
s_{\infty} = \left\{
\begin{array}{ll}
\displaystyle \ell \log \left(  \frac{\widetilde{\Phi}_{b} r_{\infty} }{4\pi G_{2} \ell^2} \frac{1}{\sqrt{ m_{i}^2 - \frac{\widetilde{\Phi}_{b}^2 }{4 G_{2}^2 \beta_{i}^2 } } } +\sqrt{ \frac{\widetilde{\Phi}_{b}^2 r_{\infty}^2}{ 
16 \pi^2 G_{2}^2 \ell^4  }\frac{1}{\left( 
m_{i}^2 - \frac{\widetilde{\Phi}_{b}^2 }{4 G_{2}^2 \beta_{i}^2 } \right) } -1 } \right) & ~~~ \displaystyle ( m_{i} > \frac{\widetilde{\Phi}_{b} }{2 G_{2} \beta_{i}} ) \\
 ~ & ~ \\
\displaystyle \ell \log  \left(  \frac{\widetilde{\Phi}_{b} r_{\infty} }{4\pi G_{2} \ell^2} \frac{1}{\sqrt{  \frac{\widetilde{\Phi}_{b}^2 }{4 G_{2}^2 \beta_{i}^2 } - m_{i}^2 } } +\sqrt{ 1+ \frac{\widetilde{\Phi}_{b}^2 r_{\infty}^2}{ 
16 \pi^2 G_{2}^2 \ell^4  }\frac{1}{\left( 
 \frac{\widetilde{\Phi}_{b}^2 }{4 G_{2}^2 \beta_{i}^2 } -m_{i}^2 \right) } } \right) + \frac{\pi}{2}\ell & ~~~ \displaystyle (m_{i} < \frac{\widetilde{\Phi}_{b} }{2 G_{2} \beta_{i}})~.
\end{array}
\right. \label{eq:geodesiclength}
\ee
In both cases, since $s_{\infty}$ diverges in the limit $r_{\infty} \to \infty$, the dynamical term $2 m_{i} s_{\infty}$ is also a divergent quantity. This divergence is precisely canceled by the particle counterterm, defined in Eq.~(\ref{eq:particle})
\footnote{
Note that $\partial \gamma_{1}$ consists of two boundary points. Therefore,
\ben
- \left. m_i \ell \log \left(\frac{\Phi}{2 \pi G_{2} \ell m_{i}}\right) \right|_{\partial \gamma_1} = 2 \times \left\{ -m_i \ell \log \left(\frac{\Phi(r_{\infty})}{2 \pi G_{2} \ell m_{i}}\right) \right\} ~ .
\een
}
\be
- \left. m_i \ell \log \left(\frac{\Phi}{2 \pi G_{2} \ell m_{i}}\right) \right|_{\partial \gamma_1} = -2 m_{i} \ell \log \left( \frac{\widetilde{\Phi}_{b} r_{\infty}}{2 \pi G_{2} \ell^2 m_{i} } \right)~. \label{eq:particlecounter}
\ee
Adding the divergent dynamical term and the counterterm, the divergences cancel exactly. The finite, renormalized on-shell particle action is
\begin{align}
 \left. I^{m_{i}}_{E} \right|_{\rm on-shell} = 2m_{i} \times (\ref{eq:geodesiclength}) + (\ref{eq:particlecounter}) \notag \hspace{6.5cm} \\
 = \left\{
\begin{array}{ll}
\displaystyle - m_{i} \ell \log \left( 
1 - \frac{\widetilde{\Phi}_{b}^2}{4 G_{2}^2 \beta_{i}^2 m_{i}^2} \right) & ~~~ \displaystyle ( m_{i} > \frac{\widetilde{\Phi}_{b} }{2 G_{2} \beta_{i}} ) \\
 ~ & ~ \\
\displaystyle - m_{i} \ell \log \left( 
 \frac{\widetilde{\Phi}_{b}^2}{4 G_{2}^2 \beta_{i}^2 m_{i}^2} -1 \right) + m_{i} \pi \ell & ~~~ \displaystyle (m_{i} < \frac{\widetilde{\Phi}_{b} }{2 G_{2} \beta_{i}})
\end{array}
\right. \label{eq:onshellparticleaction}
\end{align}

For later use, we define a function $\Gamma(m,\beta)$ which represents $ \left. I^{m}_{E} \right|_{\rm on-shell}$, \ie, the contribution from the particle trajectory;
\be
\Gamma(m, \beta)  = \left\{ 
\begin{array}{ll}
\displaystyle - m \ell \log \left( 
1 - \frac{\widetilde{\Phi}_{b}^2}{4 G_{2}^2 \beta^2 m^2} \right) & ~~~ \displaystyle ( m > \frac{\widetilde{\Phi}_{b} }{2 G_{2} \beta} ) ~ ,\\
 ~ & ~ \\
\displaystyle - m \ell \log \left( 
 \frac{\widetilde{\Phi}_{b}^2}{4 G_{2}^2 \beta^2 m^2} -1 \right) + m \pi \ell & ~~~ \displaystyle (m < \frac{\widetilde{\Phi}_{b} }{2 G_{2} \beta}) ~ . \label{eq:Gamma}
\end{array}
\right. 
\ee
Thus, we have $\left. I^{m}_{E} \right|_{\text{on-shell}} = \Gamma(m, \beta)$.

\subsubsection*{Total On-Shell Action $I^{i}_{E}|_{\rm on-shell}$}

The total on-shell action is the sum of the centaur and particle contributions 
\begin{align}
I^{i}_{E}|_{\rm on\text{-}shell} &= \frac{\pi \ell \widetilde{\Phi}_{b}}{2 G_{2} \beta_{i}^2 } 
 \widetilde{\beta} + \Gamma(m_{i}, \beta_{i})  - \frac{\Phi_{0}}{4G_{2}} \notag \\
&= \beta_{i} \left( \frac{\pi \ell \widetilde{\Phi}_{b}}{2 G_{2} \beta_{i}^2 } 
 - \frac{\Phi_{0}}{2G_{2} \beta_{i}} \right) + \frac{\Phi_{0}}{4G_{2}} -  \frac{\pi \ell \widetilde{\Phi}_{b}}{2 G_{2} \beta_{i}^2} \, \Delta \mathcal{T}(m_{i}, \beta_{i})  + \Gamma(m_{i}, \beta_{i})~. \label{eq:oneAction}
\end{align}
The term in parentheses is recognized as twice the free energy of the Euclidean flow geometry at inverse temperature $\beta_{i}$, denoted $F_{\rm flow}(\beta_{i})$ (see Eq. (\ref{eq:AppendixFreeFlow})). Therefore, the action takes the form
\be
I^{i}_{E}|_{\rm on\text{-}shell}  = 2 \beta_{i} F_{\rm flow}(\beta_{i}) + \frac{\Phi_{0}}{4G_{2}} -  \frac{\pi \ell \widetilde{\Phi}_{b}}{2 G_{2} \beta_{i}^2} \, \Delta \mathcal{T}(m_{i}, \beta_{i})  + \Gamma(m_{i}, \beta_{i}) \, . \label{eq:oneAction2}
\ee
Therefore, the on-shell action can be regarded as twice the action of the flow geometry plus corrections due to the presence of the particle (\ie the third and fourth terms), along with a topological contribution (the second term) specific to the two-dimensional dilaton theory. 

\subsubsection*{Saddle Point Energy}
In later sections, we will perform an inverse Laplace transform on $\mathcal{G}_{ij}^{(c)}(\widetilde{\beta})$ integrating the expression
\be 
\mathcal{G}_{ij}(\widetilde{\beta}) \, e^{2E \widetilde{\beta}} \simeq  \sum_i \exp(-I^{i}_{E}|_{\rm on\text{-}shell} + 2E\widetilde{\beta})~,
\ee
over $\widetilde{\beta}$ to obtain the microcanonical Gram matrix. For now we just compute the saddle point using the on-shell action (\ref{eq:oneAction2}). The saddle point of this integral is determined by the condition
\be
\frac{d}{d\widetilde{\beta}} \left( I^{i}_{E}|_{\rm on\text{-}shell} - 2E\widetilde{\beta} \right) = 0~.
\ee
The energy of the saddle is then given by
\be
E = \frac{1}{2} \frac{d \left.I^{i}_{E} \right|_{\rm on\text{-}shell}}{d \widetilde{\beta}} 
= -\frac{\pi \ell \widetilde{\Phi}_{b}}{4 G_{2} \beta_{i}^2} ~ . \label{eq:oneEnergy}
\ee
This is precisely the energy of a flow geometry with inverse temperature $\beta_{i}$ (see Eq. (\ref{eq:AppendixEnergyflow})). This result confirms that the saddle point of the inverse Laplace transform corresponds to the microcanonical ensemble at a specific negative energy $E < 0$, where the flow geometry becomes dominant.

\subsection{Second Moment: $\overline{\mathcal{G}_{ij}^{(c)} \mathcal{G}_{ji}^{(c)}} $} \label{subsec:2moment}
The statistical average of the product $\overline{\mathcal{G}_{ij}^{(c)} (\widetilde{\beta}_{1}) \mathcal{G}_{ji}^{(c)}(\widetilde{\beta}_{2}) } $ (with no summation) has a GPI representation;
\be
\overline{\mathcal{G}_{ij}^{(c)} (\widetilde{\beta}_{1}) \mathcal{G}_{ji}^{(c)}(\widetilde{\beta}_{2}) }  = Z^{i}_{1}(\widetilde{\beta_{1}})Z^{i}_{1}(\widetilde{\beta_{2}}) \delta_{ij} + Z^{ij}_{2} (2\widetilde{\beta}_{1}, 2\widetilde{\beta}_{2}) ~~~~~ ({\rm no ~ sum})~.
\ee
The first term represents the disconnected product of individual normalizations. The second term, $Z_{2}^{ij}$, is the connected component and it is given by
\be
Z_{2}^{ij}(2\widetilde{\beta}_{1}, 2\widetilde{\beta}_{2})
= \int_{conn} \mathcal{D}\boldsymbol{g} \mathcal{D}  \Phi \mathcal{D} \gamma_{1} \mathcal{D} \gamma_{2} ~ e^{-I_{E}^{ij}[\boldsymbol{g}, \Phi, \gamma_{1},\gamma_{2}]}~,
\ee
with the Euclidean action
\be
I_{E}^{ij}[\boldsymbol{g}, \Phi, \gamma_{1}, \gamma_{2}]= I_{E}^{cent}[\boldsymbol{g}, \Phi]+I^{m_{i}}_{E}[\boldsymbol{g}, \Phi, \gamma_{1}] +I^{m_{j}}_{E}[\boldsymbol{g}, \Phi, \gamma_{2}]~.
\ee
For the Schwinger-Dyson equation (\ref{eq:SD}), we require the connected part of the moment
\ben
\left. \overline{\mathcal{G}_{ij}^{(c)}(\widetilde{\beta}_{1}) \mathcal{G}_{ji}^{(c)} (\widetilde{\beta}_{2})} \right|_{conn} \simeq  Z_{2}^{ij}(2\widetilde{\beta}_{1}, 2\widetilde{\beta}_{2}) ~~~~~ ({\rm no ~ sum}) ~ .
\een

\begin{figure}[t]
\begin{center}
\includegraphics[width=13.cm]{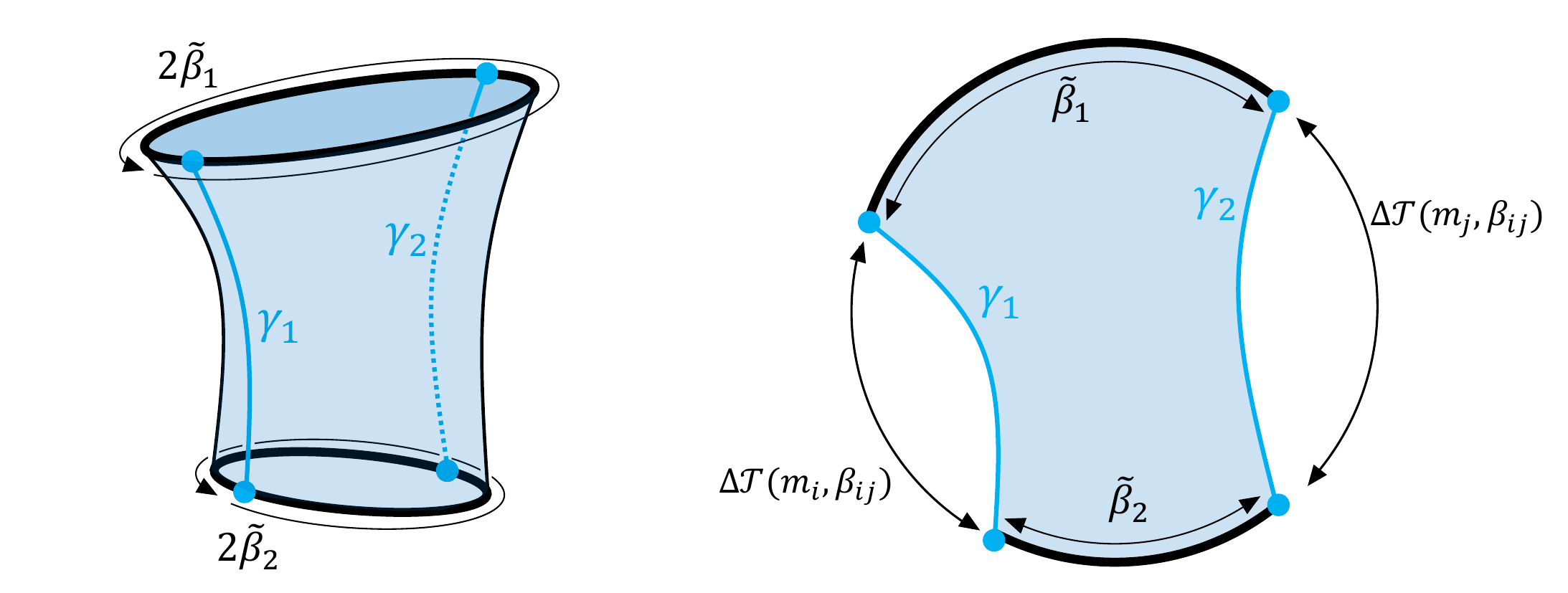} 
	\caption{On-shell geometry of $Z_{2}^{ij}(2\widetilde{\beta}_{1}, 2\widetilde{\beta}_{2})$. The wormhole has two mouths, each with renormalized lengths of $2\widetilde{\beta}_{1}$ and $2\widetilde{\beta}_{2}$. It is simply two copies of the Euclidean flow geometry, with the outsides of $\gamma_{1}$ and $\gamma_{2}$ are cut out and glued together.  }
\label{fig:wormhole}
\end{center}
\end{figure}

The dominant topology for $Z_{2}^{ij}$ is that of a cylinder. Within this topological sector, there exists an on-shell configuration consisting of two copies of the Euclidean flow geometry, each at inverse temperature $\beta_{ij}$, which will be explained shortly.
These geometries are cut along two geodesics (the particle worldlines) and glued together (Fig. \ref{fig:wormhole}). The two arguments of $Z_{2}^{ij}$ represent the renormalized lengths of the two boundaries.  

The inverse temperature, $\beta_{ij}$, of the underlying flow geometry is not equal to $\beta_{i}$ or $\beta_{j}$ from the single-boundary case; it is a new parameter determined by $m_i$, $m_j$, $\widetilde{\beta}_1$, and $\widetilde{\beta}_2$.  In order to derive the relation among these parameters, we consider a flow geometry with an inverse temperature $\beta_{ij}$, and two non-intersecting geodesics, $\gamma_1$ and $\gamma_2$, as shown on the right hand side of Fig. \ref{fig:wormhole}. To prevent two geodesics from intersecting, $\beta_{ij}$ must satisfy the inequality
\be \label{eq:betainequality}
\beta_{ij} > \min \{ \frac{\widetilde{\Phi}_{b}}{2G_{2}m_{i}}, \frac{\widetilde{\Phi}_{b}}{2G_{2}m_{j}} \} ~ .
\ee
If this condition is violated, both geodesics enter the dS region and inevitably intersect each other. 

The relation between $\beta_{ij}$ and the total renormalized boundary length can be derived from the geometry. As we derived in Subsection \ref{subsec:flowPETSgeometry}, the time interval $\Delta \mathcal{T}$ for a particle trajectory depends on its mass and the inverse temperature of the background flow geometry (\ref{eq:TimeDifference}). Using this function, the relation between $\beta_{ij}$ and the total renormalized boundary length $\widetilde{\beta}_{1}+\widetilde{\beta}_{2}$ is written as a consistency condition for the cylinder geometry
\be \label{eq:betaij}
\beta_{ij} = \widetilde{\beta}_{1}+\widetilde{\beta}_{2} + \Delta \mathcal{T}(m_{i}, \beta_{ij}) + \Delta \mathcal{T}(m_{j}, \beta_{ij}) ~ .
\ee
Similarly to the single-particle case (\ref{eq:lengths}), this equation determines $\beta_{ij}$ implicitly and can be solved numerically. 

\begin{figure}[t]
\begin{center}
\includegraphics[width=8.cm]{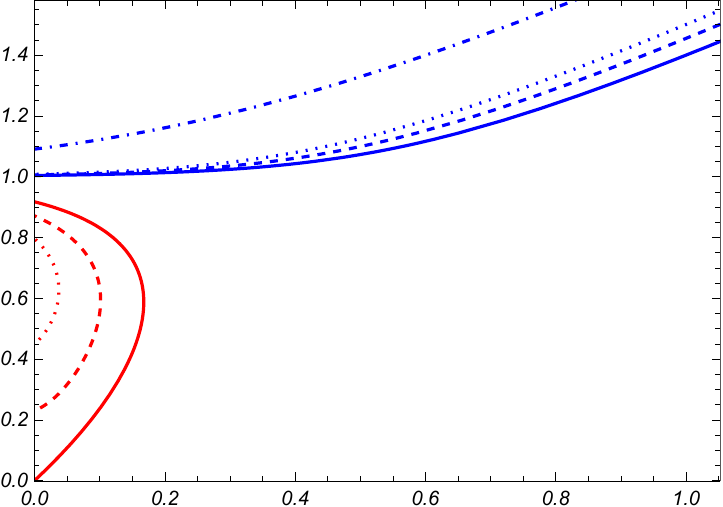}
\put(-145,-15){$m_i (\widetilde\beta_1 +\widetilde\beta_2) $}
\put(-260,70){$m_i \beta_{ij}$}
	\caption{Relations between $\beta_{ij}$ and $\widetilde{\beta}_{1} + \widetilde{\beta}_{2}$ for fixed mass ratio $m_{j}/m_{i}$; $m_{j}/m_{i}\to \infty$ (solid line), $m_{j}/m_{i}=5$ (dashed line), $m_{j}/m_{i} = 2.7$ (dotted line), $m_{j}/m_{i}=1$ (dot-dashed line). The dimensionful quantities are normalized by $m_{i}$, which is always smaller than $m_{j}$. The lower branch disappears when $m_{j}/m_{i} \lesssim 2.325 $. 
    The horizontal and the vertical axes are divided by $2G_{2}\widetilde{\Phi}_{b}$.
    }
\label{fig:betaijfunction}
\end{center}
\end{figure}

Without loss of generality, we assume $m_{i} \leq m_{j}$. Then, from Eq. (\ref{eq:betainequality}), we have $m_{j} > \frac{\widetilde{\Phi}_{b} }{ 2G_{2} \beta_{ij} }$; in other words, the trajectory of the heavier particle ($m_{j}$) always remains outside the dS region. The relationship between $\beta_{ij}$ and the total renormalized length  $\widetilde{\beta}_{1} + \widetilde{\beta}_{2}$, for various mass ratios $m_{i} /m_{j}$ is plotted in Fig.~\ref{fig:betaijfunction}.
In the limit $m_{j}/m_{i} \to \infty$, the curve reduces to the single-particle case shown in the right panel of Fig. \ref{fig:betafunction} (with the identification $\beta_{i} \leftrightarrow \beta_{ij}$ and $\widetilde{\beta} \leftrightarrow \widetilde{\beta}_{1} + \widetilde{\beta}_{2}$). As the mass ratio decreases, the lower (red) branch shrinks and vanishes around the value $m_{j}/m_{i} \simeq 2.325$.

The vertical axis, $\beta_{ij}$, is related to the energy of the saddle by $\sim 1/\sqrt{-E}$. Therefore, for given masses $m_{i}$ and $m_{j}$, there is an energy range (corresponding to the red branch) for which no real Euclidean saddles exist. Specifically, at high energies (small $\beta_{ij}$), real saddles cease to exist when the mass ratio is small. Conversely, at relatively low energy, (large $\beta_{ij}$), real saddles always exist. The left endpoint of the blue branch, $\beta_{ij}(0)$, is restricted to the range
\be 
\frac{2G_{2} m_i}{\widetilde{\Phi}_{b}} \, \beta_{ij}(0) \in \left( \coth(\pi), \coth\left(\frac{\pi}{2}\right) \right) ~ .
\ee
This implies that for energies satisfying
\be
 -E <  \frac{\pi \ell m_{i}^2}{\widetilde{\Phi}_{b}} \, \tanh^2\left(\frac{\pi}{2}\right) ~ , \label{eq:Ebound1}
\ee
there always exists a corresponding real saddle for this two-boundary configuration.

The on-shell action for this saddle is evaluated as
\begin{align}
I^{ij}_{E}|_{\rm on\text{-}shell} &= \frac{\pi \ell \widetilde{\Phi}_{b}}{2 G_{2} \beta_{ij}^2} 
 \left( \widetilde{\beta}_{1} + \widetilde{\beta}_{2} \right) + \Gamma(m_{i}, \beta_{ij})  + \Gamma(m_{j}, \beta_{ij}) \notag \\
&= 2 \beta_{ij} F_{\rm flow} (\beta_{ij}) + 2 \times \frac{\Phi_{0}}{4G_{2}} - \sum_{p=i,j} \frac{\pi \ell \widetilde{\Phi}_{b}}{2 G_{2} \beta_{ij}^2} \, \Delta \mathcal{T} (m_{p}, \beta_{ij}) + \sum_{p=i, j} \Gamma (m_{p}, \beta_{ij}) ~ . \label{eq:onshell2}
\end{align}
The first term represents twice the free energy, the second is a topological contribution from the two boundaries, and the final two terms are corrections from the presence of the particles.

The energy of the saddle, obtained by differentiating with respect to $\widetilde{\beta}_{1}$ or $\widetilde{\beta}_{2}$, is
\be
E= \frac{1}{2}\frac{d \left.I^{ij}_{E} \right|_{\rm on\text{-}shell}}{d \widetilde{\beta}_{1}} 
= \frac{1}{2} \frac{d \left.I^{ij}_{E} \right|_{\rm on\text{-}shell}}{d \widetilde{\beta}_{2}} 
= -\frac{\pi \ell \widetilde{\Phi}_{b}}{4 G_{2} \beta_{ij}^2} ~ .
\ee
This is again the energy of a flow geometry with inverse temperature $\beta_{ij}$ (\ref{eq:AppendixEnergyflow}).

\subsection{$n$-th Moment:  $\overline{\mathcal{G}_{i_{1}i_{2}}^{(c)} \mathcal{G}_{i_{2}i_{3}}^{(c)} \cdots \mathcal{G}_{i_{n}i_{1}}^{(c)}} $}
\label{subsec:nmoment}
\begin{figure}[t]
\begin{center}
\includegraphics[width=13.cm]{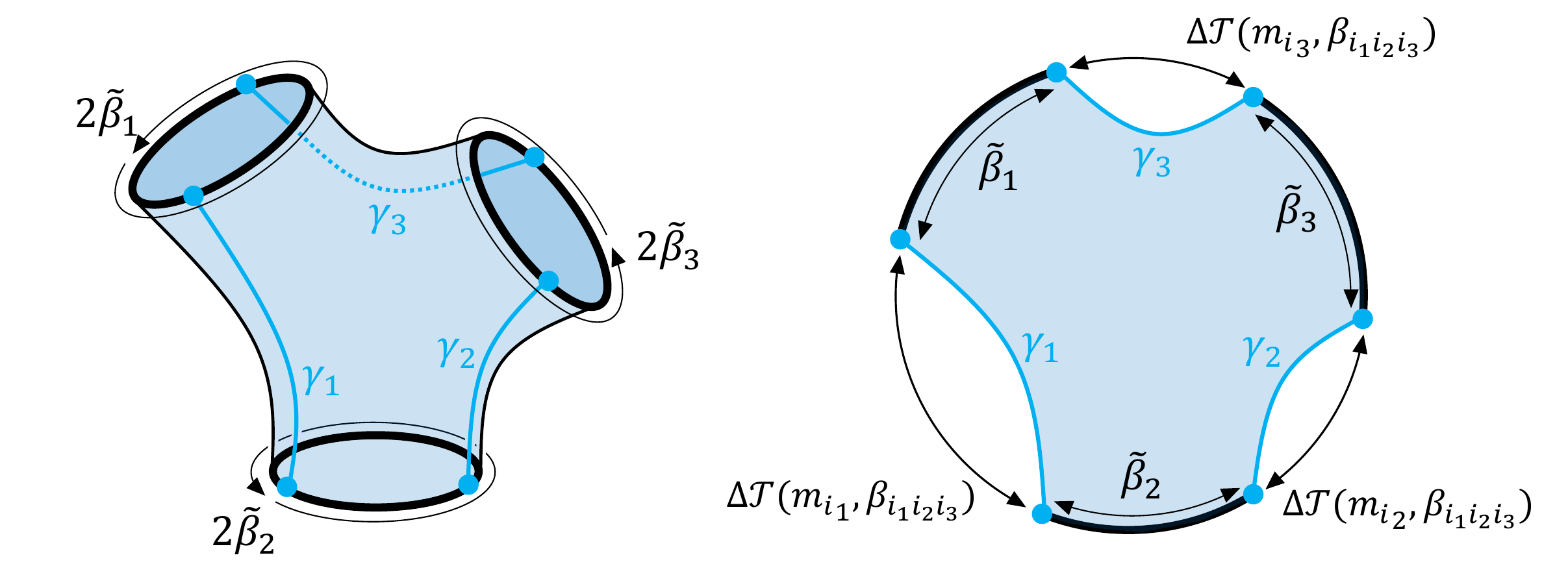} 
	\caption{On-shell geometry of $Z_{3}^{i_{1}i_{2}i_{3}}(2\widetilde{\beta}_{1}, 2\widetilde{\beta}_{2}, 
2\widetilde{\beta}_{3})$. The wormhole has three mouths with renormalized lengths $2\widetilde{\beta}_{1}$, $2\widetilde{\beta}_{2}$ and $2\widetilde{\beta}_{3}$. It is two copies of the Euclidean flow geometry, with the outsides of $\gamma_{1}$, $\gamma_{2}$, and $\gamma_{3}$ are cut out (and glued together).  }
\label{fig:3wormhole}
\end{center}
\end{figure}

Similar to the second moment, the $n$th moment contains both fully connected and (partially) disconnected components. For the Schwinger-Dyson equation, we require only the fully connected component, denoted as $Z_{n}^{i_{1}i_{2} \cdots i_{n}}(2\widetilde{\beta}_{1}, 2\widetilde{\beta}_{2}, \cdots,  2\widetilde{\beta}_{n})$. This is given by the GPI
\begin{align}
Z_{n}^{i_{1}i_{2} \cdots i_{n}}(2\widetilde{\beta}_{1}, 2\widetilde{\beta}_{2}, \cdots,  2\widetilde{\beta}_{n}) 
= \int_{conn} \mathcal{D}\boldsymbol{g} \mathcal{D}  \Phi \prod_{p=1}^{n} \mathcal{D} \gamma_{p} ~ e^{-I_{E}^{i_{1}i_{2}\cdots i_{n}}[\boldsymbol{g}, \Phi, \gamma_{1},\gamma_{2}, \cdots, \gamma_{n}]} ~ ,
\end{align}
with the action
\be
I_{E}^{i_{1}i_{2}\cdots i_{n}}[\boldsymbol{g}, \Phi, \gamma_{1}, \gamma_{2}, \cdots, \gamma_{n}]= I_{E}^{cent}[\boldsymbol{g}, \Phi]+ \sum_{p=1}^{n} I^{m_{p}}_{E}[\boldsymbol{g}, \Phi, \gamma_{p}] ~ .
\ee
This integral has $n$ boundaries, with renormalized lengths $\widetilde{\beta}_1, \ldots, \widetilde{\beta}_n$. 
The dominant topology would be that of a sphere with $n$ holes. One of the on-shell geometries in this topological sector is the one consisting of two copies of the Euclidean flow geometry, from which the exteriors of $n$ geodesics are cut out and then glued together. The case with $n = 3$ is depicted in Fig.~\ref{fig:3wormhole}.

Formally, the discussion parallels the case of the second moment discussed previously (Subsection \ref{subsec:2moment}). Let $\beta_{i_{1}i_{2}\cdots i_{n}}$ be the inverse Hawking temperature of one copy of the flow geometry. To prevent the $n$ geodesics from intersecting, they must satisfy the condition
\be
\beta_{i_{1}i_{2}\cdots i_{n}} > \min \{ \frac{\widetilde{\Phi}_{b} }{2 G_{2} m_{p}} ~ | ~ p=1, \cdots, n \} ~ . \label{eq:nwormholeinnequality}
\ee
The parameter $\beta_{i_{1}\cdots i_{n}}$ is determined implicitly by the consistency condition
\be
\beta_{i_{1}i_{2}\cdots i_{n}} = \sum_{p=1}^{n} \widetilde{\beta}_{p} + \sum_{p=1}^{n} \Delta \mathcal{T}(m_{p}, \beta_{i_{1}i_{2}\cdots i_{n}})~,
\ee
where the time difference function $\Delta \mathcal{T}$ is defined in Eq.~(\ref{eq:TimeDifference}).

Assuming the particle masses are ordered such that $m_{i_1} \leq m_{i_2} \leq \ldots \leq m_{i_n}$, only the trajectory of the lightest particle ($m_{i_1}$) can potentially enter the dS region.  Thus, the relationship between $\beta_{i_{1}\cdots i_{n}}$ and the sum $\sum_{p=1}^{n} \widetilde{\beta}_{p}$ is qualitatively similar to that shown in Fig.~\ref{fig:betaijfunction}, with two branches of solutions.

In this case, the left endpoint of the blue branch, $\beta_{i_{1}\cdots i_{n}}(0)$, is constrained to the range
\be
\frac{2G_{2}}{\widetilde{\Phi}_{b}} m_{i_{1}}\beta_{i_{1}i_{2} \cdots i_{n}}(0) \in \left(\coth(\pi), \coth(\frac{\pi}{n})\right)~.
\ee
It is worth noting that the lower bound is approached when all mass ratios $m_{i_p}/m_{i_1} \to \infty$ for $p=2,\ldots,n$. On the other hand, the upper bound is approached when all masses become equal, \ie, $m_{i_p}/m_{i_1} \to 1$.

This implies that there exists an upper bound for the absolute value of the energy for a given $n$
\be
 -E <  \frac{\pi \ell m_{i}^2}{\widetilde{\Phi}_{b}} \, \tanh^2\left(\frac{\pi}{n}\right) ~ , \label{eq:Ebound2}
\ee
which is the analogue of Eq. (\ref{eq:Ebound1}). This bound becomes more restrictive as $n$ increases. Therefore, for any given energy $E$, there exist a sufficiently large $n$ for which the corresponding real Euclidean saddle does not exist. This potential limitation of the semiclassical expansion will be discussed in later sections.

The on-shell action for the $n$-boundary saddle is a natural generalization of the previous cases
\begin{align}
I_{E}^{i_{1}i_{2}\cdots i_{n}}|_{\rm on\text{-}shell} = \frac{\pi \ell \widetilde{\Phi}_{b}}{2 G_{2} \beta_{i_{1}i_{2}\cdots i_{n}}^2} 
 \sum_{p=1}^{n} \widetilde{\beta}_{p} + \sum_{p=1}^{n}  \Gamma(m_{i_{p}}, \beta_{i_{1}i_{2}\cdots i_{n}})
 \notag \hspace{3.5cm} \\
= 2 \beta_{i_{1}i_{2}\cdots i_{n}} F_{\rm flow} (\beta_{i_{1}i_{2}\cdots i_{n}}) + n \times \frac{\Phi_{0}}{4G_{2}} \notag \hspace{4.6cm} \\
- \sum_{p=1}^{n} \frac{\pi \ell \widetilde{\Phi}_{b}}{2 G_{2} \beta_{i_{1}i_{2}\cdots i_{n}}^2} \, \Delta \mathcal{T} (m_{i_{p}}, \beta_{i_{1}i_{2}\cdots i_{n}}) + \sum_{p=1}^{n} \Gamma (m_{i_{p}}, \beta_{i_{1}i_{2}\cdots i_{n}}) ~ . \label{eq:onshelln}
\end{align}

The energy, obtained by differentiating with respect to any $\widetilde{\beta}_p$, takes a simple form and corresponds to that of a flow geometry with inverse temperature $\beta_{i_{1}i_{2}\cdots i_{n}}$ (\ref{eq:AppendixEnergyflow})
\be
E= \frac{1}{2}\frac{d \left. I^{i_{1}i_{2}\cdots i_{n}}_{E}\right|_{\rm on-shell}}{d \widetilde{\beta}_{p}} =  -\frac{\pi \ell \widetilde{\Phi}_{b}}{4 G_{2} \beta_{i_{1}i_{2}\cdots i_{n}}^2} ~~~ (p=1,2, \cdots, n) ~ . \label{eq:saddleEnergy}
\ee

\section{Heavy-Mass Limit and the Flow ER Bridge} \label{sec:heavymass}
In the works \cite{Balasubramanian:2022gmo, Climent:2024trz}, in order to calculate the moments of the microcanonical Gram matrix and subsequently the dimension of the black hole Hilbert space, the heavy-mass limit is performed
\be
m_{i} \to \infty ~~~ {\rm and} ~~~ |m_{i} - m_{j}| \to \infty ~~~ ({\rm for~} i, j = 1, \cdots, \Omega) ~ . 
\ee
We can also straightforwardly apply this limit to our construction, which drastically simplifies the subsequent computations, closely mirroring the black hole case \cite{Balasubramanian:2022gmo, Climent:2024trz}. As discussed in Section~\ref{sec:PETS}, states with large mass correspond to flow PETS geometries with a flow ER bridge that contains both an AdS and dS regions. Therefore, the flow Hilbert space considered in this section is spanned specifically by states with a flow ER bridge.

\begin{figure}[t]
\begin{center}
\includegraphics[width=13.cm]{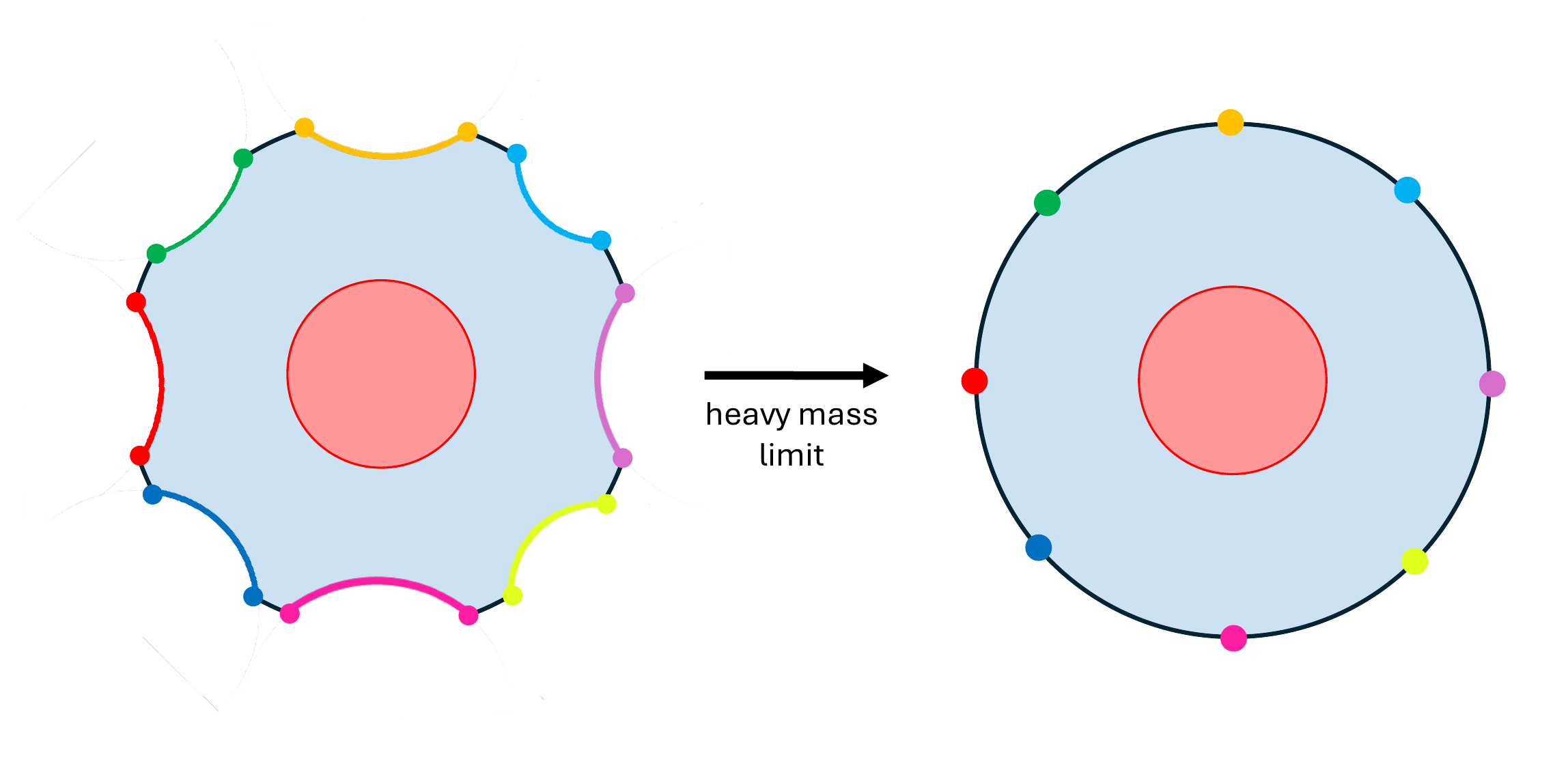} 
\caption{(\textbf{Left}) One copy of the wormhole geometry with particle trajectories of finite masses. 
(\textbf{Right}) In the heavy-mass limit, all trajectories become point-like, and the total boundary length coincides with the inverse temperature of the disc (\ref{eq:betabeta}). 
The two copies of the discs are then glued together at these points.}
\label{fig:heavymasslimit}
\end{center}
\end{figure}

In the heavy-mass limit, the particle trajectories become point-like. The wormhole geometries reduce to two discs joined at $n$ boundary points, as shown in Fig.~\ref{fig:heavymasslimit}. This implies that $\Delta \mathcal{T} \to 0$, and the total renormalized boundary length coincides with the inverse Hawking temperature of the underlying flow geometry, \ie, we have
\be
 \sum_{p=1}^{n} \widetilde{\beta}_{p} = \beta_{i_{1}\cdots i_{n}} ~ . \label{eq:betabeta}
\ee
A significant advantage of this limit is that the point-like nature of the particles ensures the gluing process of two copies of discs is always possible for any value of $\beta_{i_{1}\cdots i_{n}}$ and $n$. In other words, the upper bound for $-E$ in (\ref{eq:Ebound2}) diverges, and we can circumvent the non-existence problem of real wormhole saddles.

Using (\ref{eq:betabeta}) and recalling that our counterterm was chosen so that the on-shell particle contribution vanishes in the point-like limit (\ie, $\Gamma(m_p, \beta) \to 0$),  the general on-shell action (\ref{eq:onshelln}) simplifies to
\begin{align}
I_{E}^{i_{1}\cdots i_{n}}|_{\rm on\text{-}shell} 
= 2 \beta_{i_{1}i_{2}\cdots i_{n}} F_{\rm flow} (\beta_{i_{1}i_{2}\cdots i_{n}}) + n \frac{\Phi_{0}}{4G_{2}} ~ .
\end{align}

We now perform the inverse Laplace transform, as outlined in Subsection~\ref{subsec:mcGram}, to obtain the $n$-th moment of the unnormalized microcanonical Gram matrix. Approximating the window function as $f_{E}(E_{a}) \simeq \varepsilon$ (a constant within the microcanonical window) and setting $E_{a_{p}} = E$, the integrand in (\ref{eq:ILT}) becomes 
\begin{align}
\exp\left( 2 \widetilde{\beta}_{1} E + \cdots + 2 \widetilde{\beta}_{n} E 
 - 2 \beta_{i_{1}\cdots i_{n}} F_{\rm flow} (\beta_{i_{1}\cdots i_{n}}) - n \frac{\Phi_{0}}{4G_{2}}  \right) \notag \\
= \exp\left( 2 \beta_{i_{1}\cdots i_{n}} E - 2 \beta_{i_{1}\cdots i_{n}} F_{\rm flow} (\beta_{i_{1}\cdots i_{n}}) - n \frac{\Phi_{0}}{4G_{2}}  \right) ~ .
\end{align}

Since in the previous Section \ref{subsec:nmoment}, we determined the saddle point expression for $n$ particles yielding the general result (\ref{eq:saddleEnergy}), which implies that $E$ is the flow geometry energy for inverse temperature $\beta_{i_{1}\cdots i_{n}}$, we thus immediately obtain for the $n$-th moment

\begin{align}
\overline{\mathcal{G}_{i_{1}i_{2}}^{(mc)}(E) \mathcal{G}_{i_{2}i_{3}}^{(mc)}(E) \cdots \mathcal{G}_{i_{n}i_{1}}^{(mc)}(E)} 
\simeq \exp\left( 2 S_{\rm flow}(E) - n \frac{\Phi_{0}}{4G_{2}} \right) ~ ,
\end{align}
by applying the saddle-point approximation to the $\widetilde{\beta}_{p}$ integrals and using standard thermodynamic relations.  Here $S_{\rm flow}(E)$ is the entropy of the flow geometry at energy $E$ (see Appendix~\ref{app:thermodynamics}).

The corresponding normalized moment (\ref{eq:nomalizedGram}) is then
\be
\overline{G_{i_{1}i_{2}}^{(mc)}(E) G_{i_{2}i_{3}}^{(mc)}(E) \cdots G_{i_{n}i_{1}}^{(mc)}(E)}  
\simeq \frac{e^{ 2 S_{\rm flow}(E) - n \frac{\Phi_{0}}{4G_{2}}}}{ \left( e^{2  S_{\rm flow}(E) -  \frac{\Phi_{0}}{4G_{2}}}\right)^n } 
= e^{2 S_{\rm flow}(E) (1-n)} ~ .
\ee
This takes the form required by (\ref{eq:AB}) 
\be
A \simeq e^{2S_{\rm flow}} \quad {\rm and}  \quad B \simeq e^{-2S_{\rm flow}}~. 
\ee
Using Eq.~(\ref{eq:A}), the dimension of the flow Hilbert space is
\be
{\rm dim}(\mathcal{H}_{\mathbf{Flow}}) \simeq e^{2S_{\rm flow}} ~ . \label{eq:dimension}
\ee

In summary, by applying the microstate-counting method of \cite{Balasubramanian:2022gmo, Climent:2024trz} in the heavy-mass limit to the flow geometry in centaur gravity, we have shown that the flow Hilbert space can be spanned by a family of states with a flow ER bridge, and that its dimensionality is given by the exponential of the horizon entropy (\ref{eq:dimension}), saturating the Bekenstein-Hawking bound.

\section{Finite-mass and dS ER bridge} \label{sec:microcanonical}
The main goal of this paper is to investigate the dS horizon entropy using the flow geometry, which contains a dS region with its cosmological horizon located in the deep interior. In the previous section, we succeeded in reproducing the dimensionality of the horizon Hilbert space by applying the black hole microstate–counting method. However, all the corresponding microstates in the family that spans the Hilbert space contain a {\it centaur} ER bridge, which includes an AdS portion. These interior structures differ from those of pure dS spacetime and are specific to flow geometries. To bring the interior structure closer to that of pure dS spacetime, we now focus on states created by injecting small-mass particles. The semiclassical geometry of these states has a dS ER bridge, meaning that the interior behind the horizon is purely dS and contains no AdS region. In this section, we explore this possibility by extending the microstate-counting method \cite{Balasubramanian:2022gmo, Climent:2024trz} to the finite-mass regime.

We will show that the dS entropy can also be reproduced in the finite-mass case, albeit under certain assumptions. The derivation requires careful treatment of the inverse Laplace transform, which we detail in this section. 

\subsection{Inverse Laplace Transform of $\mathcal{G}_{ij}^{(c)}(\widetilde{\beta}) $ }
To obtain the microcanonical Gram matrix element, $\mathcal{G}_{ij}^{(mc)}(E)$, we apply an ILT to its canonical counterpart, which results in the expression 
\begin{align}\label{eq:ILT1}
\mathcal{G}_{ii}^{(mc)}(E) \simeq \left(\frac{1}{2\pi i}\right)^2 \sum_{a} f_{E}(E_{a}) \int d\widetilde{\beta}  e^{2\widetilde{\beta}E_{a}} ~  \mathcal{G}_{ii}^{(c)}(\widetilde{\beta}) \notag \\
= \left(\frac{1}{2\pi i}\right)^2 \sum_{a} f_{E}(E_{a}) \int d\widetilde{\beta} e^{2\widetilde{\beta}E_{a}} ~ Z_{1}^{i}(2\widetilde{\beta})~.
\end{align} 
As mentioned in Section \ref{sec:cGm}, the dominant saddle of the GPI, $Z_{1}^{i}(2\widetilde{\beta})$, is a black hole PETS geometry, which do not have dS regions. Thus $\mathcal{G}_{ij}^{(c)}(\widetilde{\beta})$ (or $Z_{1}^{i}(2\widetilde{\beta})$) itself is essentially the same as that of JT gravity. However, a key difference from JT gravity is that $Z_{1}^{i}(2\widetilde{\beta})$ also includes subdominant saddles with negative energy, corresponding to the flow PETS geometries. In the above ILT (\ref{eq:ILT1}), when the energy is specified to be negative ($E<0$), the flow PETS geometry becomes the dominant saddle in the $\widetilde{\beta}$ integral.

\begin{figure}[t]
\begin{center}
\includegraphics[width=9.cm]{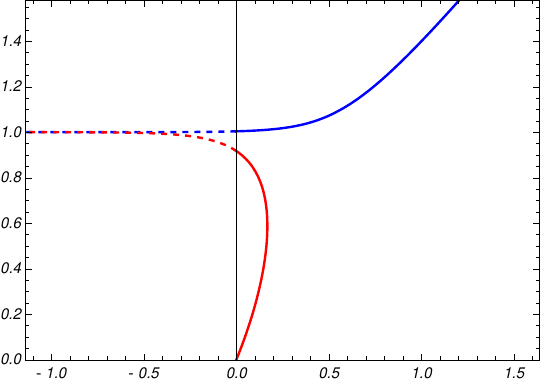} 
\put(-295,80){$ \displaystyle  \frac{m_{i}}{\sqrt{-E}}$}
\put(-160,-15){$2 G_2 m_i\widetilde\beta /\widetilde\Phi_b$}
\caption{The same plot as Fig. \ref{fig:betafunction} but extended to negative $\widetilde{\beta}$. The vertical axis represents the inverse of $\sqrt{-E}$. (The vertical axis is divided by $\sqrt{ \frac{\pi \ell G_{2}}{\widetilde{\Phi}_{b}} }$.) The blue curve represents the case where $m_{i} > \sqrt{ -\frac{\widetilde{\Phi}_{b} E}{ \pi \ell G_{2} } }$ and the red curve represents the case where $m_{i} < \sqrt{ -\frac{\widetilde{\Phi}_{b} E}{ \pi \ell G_{2} } }$. The dashed curves for negative $\widetilde{\beta}$  asymptotically approach $1$.  }
\label{fig:betafunctionNegative}
\end{center}
\end{figure}

Using equations (\ref{eq:oneAction}) and (\ref{eq:oneEnergy}), Eq. (\ref{eq:ILT1}) becomes
\small \begin{align}
\mathcal{G}_{ii}^{(mc)}(E) \hspace{14.8cm} \notag \\ \propto \left\{
\begin{array}{l}
\displaystyle \exp\left[ 4E\left( \sqrt{- \frac{\pi \ell \widetilde{\Phi}_{b}}{4 G_{2} E}}  - \Delta \mathcal{T}^{(mc)}(m_{i}, E)  \right)
 + m_{i} \ell \log \left( 
1  + \frac{\widetilde{\Phi}_{b} E}{\pi \ell G_{2} m_{i}^2} \right) + \frac{\Phi_{0}}{4G_{2}} \right]  \\ 
\displaystyle \hspace{12cm} ( m_{i} > \frac{\widetilde{\Phi}_{b} }{2 G_{2} \beta_{i}} ) \\
 ~  \\
\displaystyle \exp\left[  4E\left( \sqrt{- \frac{\pi \ell \widetilde{\Phi}_{b}}{4 G_{2} E}}  - \Delta \mathcal{T}^{(mc)}(m_{i}, E)  \right) + m_{i} \ell \log \left( 
 - \frac{\widetilde{\Phi}_{b} E}{\pi \ell G_{2} m_{i}^2} -1 \right) - m_{i} \pi \ell + \frac{\Phi_{0}}{4G_{2}} \right] \\
\displaystyle \hspace{12cm}  (m_{i} < \frac{\widetilde{\Phi}_{b} }{2 G_{2} \beta_{i}}) 
\end{array}
\right.~. \label{eq:microGram1}
\end{align}
where the time difference function $\Delta \mathcal{T}^{(mc)}(m_{i}, E) $ is given by
\begin{align}
\Delta \mathcal{T}^{(mc)}(m_{i}, E) =  \Delta \mathcal{T}(m_{i}, \beta) \Big|_{\beta=\sqrt{ - \frac{\pi \ell \widetilde{\Phi}_{b}}{4 G_{2} E} } } \hspace{7.cm} \notag\\
= \left\{
\begin{array}{ll}
\displaystyle \frac{1}{2} \sqrt{ -\frac{\ell \widetilde{\Phi}_{b} }{\pi G_{2} E} } {\rm arctanh} \left[ \frac{1}{m_{i}}\sqrt{ - \frac{\widetilde{\Phi}_{b} E}{\pi \ell G_{2}} } \right]  & ~~~ \displaystyle ( m_{i} > \sqrt{ -\frac{\widetilde{\Phi}_{b} E}{ \pi \ell G_{2} } } ) \\
 ~ & ~ \\
\displaystyle \frac{1}{2} \sqrt{ -\frac{\ell \widetilde{\Phi}_{b} }{\pi G_{2} E} } {\rm arctanh} \left[ m_{i} \sqrt{ -\frac{\pi \ell G_{2} }{\widetilde{\Phi}_{b} E} }  \right] + \frac{1}{4} \sqrt{ -\frac{\pi \ell \widetilde{\Phi}_{b} }{G_{2} E} } & ~~~ \displaystyle (m_{i} < \sqrt{ -\frac{\widetilde{\Phi}_{b} E}{ \pi \ell G_{2} } } )
\end{array}
\right. 
\end{align}
Note that, as we mentioned in Subsection \ref{sec:ERstates}, in the energy range
\ben
\tanh^2 (\pi) <\frac{\widetilde{\Phi}_{b}}{\pi \ell G_{2}} \frac{(-E)}{m_{i}^2} < \frac{1}{\tanh^2(\frac{\pi}{2})} ~ ,
\een
real Euclidean geometries may not exist via the gluing procedure introduced in Subsection \ref{subsec:flowPETSgeometry}. The most straightforward approach, for a given energy $-E$, is to choose the family of masses outside this problematic range
\ben
\tanh\left( \frac{\pi}{2} \right) \sqrt{ \frac{-E \widetilde{\Phi}_{b}}{ \pi \ell G_{2} } } < m_{i} < \frac{1}{\tanh(\pi)} \sqrt{ \frac{-E \widetilde{\Phi}_{b}}{ \pi \ell G_{2} } } ~ .
\een
Alternatively, one could interpret geometries in this range as having analytically continued negative boundary length. The relation between $-E$ and $\widetilde{\beta}$, in this case, is shown in Fig. \ref{fig:betafunctionNegative}. Note that while we do not need this assumption for the single-boundary case, a similar consideration will be necessary for higher moments in the finite-mass analysis.

For later convenience, let us define the following functions;
\begin{align}
\mathfrak{b}(E) = \sqrt{ -\frac{\pi \ell \widetilde{\Phi}_{b}}{4 G_{2} E}} ~ , \hspace{8.1cm} \\
\Gamma^{(mc)}(m, E) = \left\{
\begin{array}{l}
\displaystyle - m \ell \log \left( 
1  + \frac{\widetilde{\Phi}_{b} E}{\pi \ell G_{2} m^2} \right)  ~~~ \hspace{1.45cm}  ( m > \sqrt{ -\frac{\widetilde{\Phi}_{b} E}{ \pi \ell G_{2} } } ) \\
 ~  \\
\displaystyle - m \ell \log \left( 
- \frac{\widetilde{\Phi}_{b} E}{\pi \ell G_{2} m^2} -1 \right) -m \pi \ell ~~~ ( m < \sqrt{ -\frac{\widetilde{\Phi}_{b} E}{ \pi \ell G_{2} } } )
\end{array}
\right. ~ .
\end{align}
Here, $\mathfrak{b}$ represents the inverse Hawking temperature of the flow geometry with energy $E<0$, while $\Gamma^{(mc)}$ is the microcanonical version of $\Gamma(m,\beta)$ (\ref{eq:Gamma}),
\ben
\Gamma^{(mc)}(m,E) = \Gamma(m,\beta)\Big|_{\beta=\sqrt{ - \frac{\pi \ell \widetilde{\Phi}_{b}}{4 G_{2} E} } }~,
\een
representing the contribution from a particle to the `microcanonical action'. Using these functions, we can rewrite (\ref{eq:microGram1}) as
\be
\mathcal{G}_{ii}^{(mc)}(E) \propto \exp \left[ 4E\mathfrak{b}(E)-4E\Delta \mathcal{T}^{(mc)}(m_{i}, E) + \Gamma^{(mc)}(m_{i}, E) + \frac{\Phi_{0}}{4G_{2}} \right]~. \label{eq:microGram2}
\ee

\subsection{Inverse Laplace Transform of $ \overline{
\mathcal{G}_{ij}^{(c)}(\widetilde{\beta}_{1})
\mathcal{G}_{ji}^{(c)}(\widetilde{\beta}_{2})} $ }
\begin{figure}[t]
\begin{center}
\includegraphics[width=9.cm]{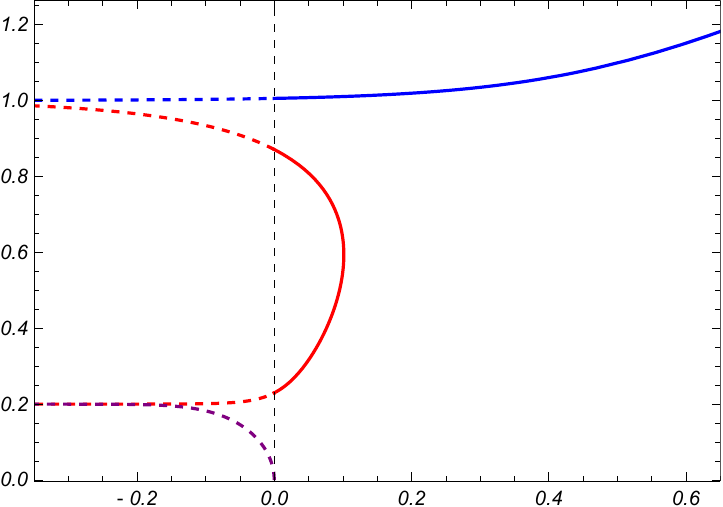} 
\put(-295,80){$ \displaystyle \frac{m_{i}}{\sqrt{-E}}$}
\put(-180,-15){$2 G_2 m_i (\widetilde\beta_1 + \widetilde\beta_2) /\widetilde\Phi_b$}
	\caption{Same plot as Fig. \ref{fig:betaijfunction} ($m_{j}/m_{i} = 5$), but extended to negative $\widetilde{\beta}_{1} + \widetilde{\beta}_{2}$ with a newly added (purple-colored) branch. The vertical axis is divided by $\sqrt{ \frac{\pi \ell G_{2}}{\widetilde{\Phi}_{b}} }$. Both the blue and red curves represent the case where $m_{j} > \sqrt{ -\frac{\widetilde{\Phi}_{b} E}{ \pi \ell G_{2} } }$, with $m_{i} > \sqrt{ -\frac{\widetilde{\Phi}_{b} E}{ \pi \ell G_{2} } } $ for the blue curve and $m_{i} < \sqrt{ -\frac{\widetilde{\Phi}_{b} E}{ \pi \ell G_{2} } } $ for the red curve. The purple curve represents the case where both $m_i$ and $m_j$ are less than $\sqrt{ -\frac{\widetilde{\Phi}_{b} E}{ \pi \ell G_{2} } }$. Therefore, the purple branch does not represent a real Euclidean saddle.  }    
\label{fig:betaijfunctionNegative}
\end{center}
\end{figure}

We now compute the second moment of the microcanonical Gram matrix, by performing an inverse Laplace transform, which is given by 
\begin{align}
\overline{\mathcal{G}_{ij}^{(mc)}(E) \mathcal{G}_{ji}^{(mc)}(E) } \notag \hspace{10.5cm} \\
\simeq \left(\frac{1}{2\pi i}\right)^4 \sum_{a, b} f_{E}(E_{a}) f_{E}(E_{b}) \int d\widetilde{\beta}_{1} \int d\widetilde{\beta}_{2}   e^{2\widetilde{\beta}_{1}E_{a} + 2\widetilde{\beta}_{2}E_{b}} ~  \overline{\mathcal{G}_{ij}^{(c)}(\widetilde{\beta}_{1}) \mathcal{G}_{ji}^{(c)}(\widetilde{\beta}_{2}) } \notag \hspace{0.cm}\\
=  \left(\frac{1}{2\pi i}\right)^4 \sum_{a, b} f_{E}(E_{a}) f_{E}(E_{b}) \int d\widetilde{\beta}_{1} \int d\widetilde{\beta}_{2}   e^{2\widetilde{\beta}_{1}E_{a} + 2\widetilde{\beta}_{2}E_{b}} ~  Z_{2}^{ij}(2\widetilde{\beta}_{1}, 2\widetilde{\beta}_{2}) ~.\hspace{0.5cm}
\end{align} 
When the width of the energy window is sufficiently small, $\varepsilon$, the sum over energy levels, $E_{a}$ and $E_{b}$, may be approximated by setting $E_{a}=E_{b}=E$ and multiplying by a factor $\varepsilon^2$, yielding
\begin{align}
\overline{\mathcal{G}_{ij}^{(mc)}(E) \mathcal{G}_{ji}^{(mc)}(E)} \simeq \left(\frac{1}{2\pi i}\right)^4 \varepsilon^2  \int d\widetilde{\beta}_{1} \int d\widetilde{\beta}_{2} e^{2\widetilde{\beta}_{1}E + 2\widetilde{\beta}_{2}E} ~  Z_{2}^{ij}(2\widetilde{\beta}_{1}, 2\widetilde{\beta}_{2})  \notag \\
= \left(\frac{1}{2\pi i}\right)^4 \varepsilon^2  \int d\widetilde{\beta}_{1} \int d\widetilde{\beta}_{2} e^{2\widetilde{\beta}_{1}E + 2\widetilde{\beta}_{2}E} ~  e^{-I_{E}^{ij}|_{\rm on-shell}} \hspace{0.4cm}
\end{align}
The saddle point geometry for the $\widetilde{\beta}_{1}$ and $\widetilde{\beta}_{2}$ integrals is identical to that derived in Subsection \ref{subsec:2moment}. Consequently, we find 
\begin{align}
\overline{\mathcal{G}_{ij}^{(mc)}(E) \mathcal{G}_{ji}^{(mc)}(E)} \notag \hspace{11.8cm} \\
\propto \exp \left[ 4E\mathfrak{b}(E)-4E\Delta \mathcal{T}^{(mc)}(m_{i}, E) -4E\Delta \mathcal{T}^{(mc)}(m_{j}, E) + \Gamma(m_{i}, E) + \Gamma(m_{j}, E) \right] \label{eq:micro2nd}
\end{align}
However, this result requires careful interpretation, as the saddle point does not always correspond to a real Euclidean geometry. First, as established in Subsection \ref{subsec:2moment}, the red curve (see Fig. \ref{fig:betaijfunctionNegative})  does not lie in the region $\widetilde{\beta}_{1}+\widetilde{\beta}_{2}>0$ when the mass ratio is below the critical value
\be
\frac{m_{j}}{m_{i}} \leq \frac{3+\cosh(\pi) + 2\sqrt{2(\cosh(\pi)+1)}  }{\cosh(\pi)-1} \simeq 2.34 ~ .
\ee
In order to obtain the result (\ref{eq:micro2nd}) for the entire energy range, $E<0$, we must therefore assume that such non-real Euclidean geometries are included in the GPI for $\widetilde{\beta}_{1}+\widetilde{\beta}_{2}<0$, even if there is no corresponding Euclidean geometry with $\widetilde{\beta}_{1}+\widetilde{\beta}_{2}>0$. 

In addition, these two types of saddles alone cannot cover the whole energy range for the higher moments. We must further assume that geometries with multiple particles entering the dS region are also included in the GPI, even though they are not real Euclidean saddles either. In the case of $n=2$, we consider geometries with two particles entering the dS region. The relation between $\widetilde{\beta}_{1} + \widetilde{\beta}_{2}$ and $E$ is then formally given by
\begin{align}
\widetilde{\beta}_{1}+\widetilde{\beta}_{2} = \mathfrak{b}(E) - \Delta \mathcal{T}^{(mc)}(m_{i},E) - \Delta \mathcal{T}^{(mc)}(m_{j},E) \notag \hspace{4.8cm} \\
= -\frac{\mathfrak{b}(E)}{\pi} \left[ {\rm arctanh}\left( m_{i} \sqrt{- \frac{\pi \ell G_{2}}{ \widetilde{\Phi}_{b} E } } \right) + {\rm arctanh}\left( m_{j} \sqrt{- \frac{\pi \ell G_{2}}{\widetilde{\Phi}_{b} E} } \right)\right] < 0 ~,\\
\left( {\rm when ~} m_{i}, m_{j}<\sqrt{- \frac{\widetilde{\Phi}_{b} E}{\pi \ell G_{2}} } ~.\right) \notag 
\end{align}
The complete relation between $-E$ and $\widetilde{\beta}_{1}+\widetilde{\beta}_{2}$ is shown in Fig. \ref{fig:betaijfunctionNegative}, which reveals three distinct branches in total. 

\subsection{Inverse Laplace Transform of the $n$-th Moment}
We now generalize the previous calculation to the $n$-th moment of the microcanonical Gram matrix. The quantity, $\overline{\mathcal{G}_{i_{1}i_{2}}^{(mc)}(E) \mathcal{G}_{i_{2}i_{3}}^{(mc)}(E) \cdots \mathcal{G}_{i_{n}i_{1}}^{(mc)}(E) }$, is obtained by applying the ILT to its canonical counterpart. Using the same narrow-window approximation  ($E_{a_p} = E$, yielding a factor of $\varepsilon^n$), we find
\begin{align}
\overline{\mathcal{G}_{i_{1}i_{2}}^{(mc)}(E) \mathcal{G}_{i_{2}i_{3}}^{(mc)}(E) \cdots \mathcal{G}_{i_{n}i_{1}}^{(mc)}(E) } \hspace{8.4cm} \notag \\
\simeq \prod_{p=1}^{n} \left[ \left(\frac{1}{2\pi i}\right)^2 \sum_{a_{p}} f_{E}(E_{a_{p}}) \int d\widetilde{\beta}_{p}  e^{2\widetilde{\beta}_{p} E_{a_{p}}} \right] \overline{\mathcal{G}_{i_{1}i_{2}}^{(c)}(\widetilde{\beta}_{1}) \mathcal{G}_{i_{2}i_{3}}^{(c)}(\widetilde{\beta}_{2}) \cdots \mathcal{G}_{i_{n}i_{1}}^{(c)}(\widetilde{\beta}_{n}) } \hspace{0.1cm} \notag \\
= \left(\frac{1}{2\pi i}\right)^{2n} \varepsilon^{n} \left( \prod_{p=1}^{n} \int d\widetilde{\beta}_{p} e^{2\widetilde{\beta}_{p} E} \right) e^{-\left.  I_{E}^{i_{1}i_{2}\cdots i_{n}}\right|_{\rm on-shell} } \hspace{4.2cm} \notag\\
\propto \exp [ 4E \mathfrak{b}(E) - \sum_{p=1}^{n}4E \Delta \mathcal{T}^{(mc)}(m_{i_{p}}, E) + \sum_{p=1}^{n} \Gamma(m_{i_{p}}, E) + \frac{\Phi_{0}}{4 G_{2}}(2-n) ] 
\end{align}
As in the lower moments, this derivation assumes the inclusion of non-Eulcidean saddles in the GPI. The situation is now richer: there exist $n+1$ types of saddles depending on how many particles enter the dS region. When $k$ particles enter, the relation between the boundary length $\sum_{p=1}^{n} \widetilde{\beta}_{p}$ and the energy $-E$ is given by
\begin{align}
\sum_{p=1}^{n} \widetilde{\beta}_{p} = \mathfrak{b}(E) \left( 1- \frac{k}{2} \right) \hspace{9.6cm} \notag \\ 
-\frac{\mathfrak{b}(E)}{\pi} \left[ \sum_{p=1}^{k} {\rm arctanh}\left( m_{i_{p}} \sqrt{- \frac{\pi \ell G_{2}}{ \widetilde{\Phi}_{b} E } } \right) + \sum_{p=k+1}^{n} {\rm arctanh}\left( \frac{1}{ m_{i_{p}} } \sqrt{- \frac{\widetilde{\Phi}_{b} E}{\pi \ell G_{2}} } \right)\right]
\end{align}

An important subtlety arises for large $n$ and a broad mass spectrum. When the ratio between the smallest mass $m_1$ and the largest mass $m_\Omega$ is sufficiently large, the positive $\sum_{p=1}^{n} \widetilde{\beta}_{p}$ region for saddles with one or more particles entering the dS region may disappear for generic particle combinations and large $n$. Therefore, those saddles may not be interpreted as analytic continuations of real Euclidean geometries. The only exception is the saddle where no particles enter the dS region. This saddle always possesses a region with positive $\sum_{p=1}^{n} \widetilde{\beta}_{p}$, and its negative counterpart can be interpreted as its analytic continuation. Therefore, for energies satisfying
\be
-E < \frac{\pi \ell G_{2}}{\widetilde{\Phi}_{b}} m_{1}^2 ~,
\ee
they can be relatively safely included in the GPI, assuming the validity of the analytic continuation. For the general case, the status of saddles that cannot be related to a real Euclidean geometry via analytic continuation remains formally unclear.  In the subsequent analysis, we will proceed under the assumption that the gravitational path integral includes all such saddles, regardless of their direct connection to real Euclidean geometries.

\subsection{Dimension of the Flow Geometry Hilbert Space}
Using the results from the previous sections, we can now compute the normalized version of microcanonical Gram matrix and extract the Hilbert space dimension. For the second moment, using equations (\ref{eq:microGram2}) and (\ref{eq:micro2nd}), we find
\be
\begin{split}
\overline{G^{(mc)}_{ij}(E) G^{(mc)}_{ji}(E) } =  \overline{\mathcal{G}^{(mc)}_{ij}(E) \mathcal{G}^{(mc)}_{ji}(E) } ~ / ~ \mathcal{N}_{2} \simeq e^{-4E\mathfrak{b}(E) - \frac{\Phi_{0}}{4G_{2}}\times 2} = e^{-2S_{\rm flow}}~, \\
\left( \mathcal{N}_{2} = \mathcal{G}^{(mc)}_{ii}(E) ~ \mathcal{G}^{(mc)}_{jj}(E) ~ . \right)
\end{split}
\ee
Crucially, the particle contributions from $\Delta \mathcal{T}$ and $\Gamma$ in both the numerator and denominator cancel out, leaving a result that is simply the inverse of the density of states. 

This pattern generalizes to the $n$-th moment
\be
\begin{split}
\overline{G_{i_{1}i_{2}}^{(mc)}(E) G_{i_{2}i_{3}}^{(mc)}(E) \cdots G_{i_{n}i_{1}}^{(mc)}(E) }  =  \overline{\mathcal{G}_{i_{1}i_{2}}^{(mc)}(E) \mathcal{G}_{i_{2}i_{3}}^{(mc)}(E) \cdots \mathcal{G}_{i_{n}i_{1}}^{(mc)}(E) } ~ / ~ \mathcal{N}_{n} \hspace{0.5cm}\\
\simeq e^{ \left( - 4 E \mathfrak{b}(E) - \frac{\Phi_{0}}{4G_{2}} \times 2 \right) (n-1) } = e^{-2S_{\rm flow} (n-1)}~, \\
\left( \mathcal{N}_{n}  = \mathcal{G}_{i_{1}i_{2}}^{(mc)}(E) ~ \mathcal{G}_{i_{2}i_{3}}^{(mc)}(E) \cdots \mathcal{G}_{i_{n}i_{1}}^{(mc)}(E) ~ . \right)
\end{split}
\ee

This takes the form given in (\ref{eq:AB}) with 
\be
A \simeq e^{2S_{\rm flow}}~~ {\rm and}~~ B \simeq e^{-2S_{\rm flow}}~.
\ee

Therefore, using Eq. (\ref{eq:A}), the dimension of the flow geometry Hilbert space is given by
\be
{\rm dim}(\mathcal{H}_{\mathbf{Flow}}) \simeq e^{2S_{\rm flow}} ~ .
\ee

\section{Discussion} \label{sec:Discuss}
This work was motivated by the challenge of understanding the microscopic origin of the de Sitter horizon entropy. A powerful method for constructing and counting black hole microstates exists, but its application to dS space is obstructed by the absence of a timelike (AdS) boundary, which is required to prepare states and inject matter that expand the horizon interior.

In this paper, we have addressed this problem by embedding a dS region within an asymptotically AdS spacetime, so that the dS horizon remains in causal contact with the boundary. While such setup is prohibited in higher spacetime dimensions with positive-energy matter, it becomes feasible in two-dimensional dilaton gravity with a V-shaped potential. The resulting geometry is the so-called \textit{flow geometry}, and the corresponding theory is called \textit{centaur} gravity.

We demonstrate that the construction method for horizon microstates can be successfully applied within this framework, yielding microstates with a long ER bridge behind the dS horizon. Because the theory admits both AdS and dS regions, the interior structure depends critically on the mass of the injected matter. For sufficiently heavy masses, the bridge is a \textit{flow ER bridge} containing both AdS and dS regions. For light masses, it becomes a \textit{dS ER bridge}, which consists purely of a dS region.

A straightforward application of the counting method requires taking the large-mass limit, in which all resulting microstates belong to the family containing flow ER bridges. This procedure works without obstacles. We show that the constructed corresponding microstates span a Hilbert space of dimension $\exp(2S_{\rm flow})$. This constitutes one of the concrete results of this paper.  

While our explicit analysis utilized centaur geometries—which interpolate from an asymptotic AdS$_2$ boundary to a dS$_2$ interior—the framework we developed is applicable to a broader class of flow geometries. The crucial element enabling our microstate construction and state counting is the asymptotic AdS$_2$ region, which guarantees the existence of a long wormhole in a suitable parameter regime and provides a tractable holographic anchor for the gravity calculation.

Nevertheless, from the perspective of understanding the dS horizon and its intrinsic microstates, this result is somewhat unsatisfactory: the interior geometry in this construction necessarily includes an AdS region, which may not exist in higher-dimensional dS spacetimes. A more faithful description of dS microstates should exclude the AdS region, \ie, it should rely on microstates with dS ER bridges, realized by light particles.  To this end, we have extended the counting method to families with finite masses, and, under certain assumptions, we succeed in reproducing the horizon Hilbert space dimension $\exp(2S_{\rm flow})$ solely from microstates associated with dS ER bridges.  

In summary, we have succeeded in constructing microstates of the dS horizon and in providing an explanation of its entropy in two dimensions, by embedding the dS spacetime into an AdS spacetime. Nevertheless, we should be cautious about the precise implications of this result, and further study is required. In what follows, we outline key subtleties involved and possible directions for future investigation.

\subsection{Microcanonical Universal Overlaps} \label{subsec:microUniOver}
In Section \ref{sec:microcanonical}, we argued that, under certain assumptions, the $n$-th moment of the microcanonical Gram matrix has the form
\ben
\overline{G_{i_{1}i_{2}}^{(mc)}(E) G_{i_{2}i_{3}}^{(mc)}(E) \cdots G_{i_{n}i_{1}}^{(mc)}(E) }  \simeq e^{-2S_{\rm flow} (n-1)}~,
\een
which is independent of the particle masses. In fact, this universal form of the overlaps also holds in black hole cases, \ie, not only for black holes in centaur or JT gravity but also for those in higher-dimensional gravity. We now revisit the rationale for this formula, explaining the underlying assumptions from a slightly different, more geometric perspective.
 
The $n$-th moment of the Gram matrix is given by the ratio of an unnormalized version of PETS to the product of single-state normalizations. Both terms admit GPI expressions and thus have geometric representations (for appropriate choices of particle or shell masses with given $n$ and $E$). 
\begin{figure}[t]
\begin{center}
\includegraphics[width=15.cm]{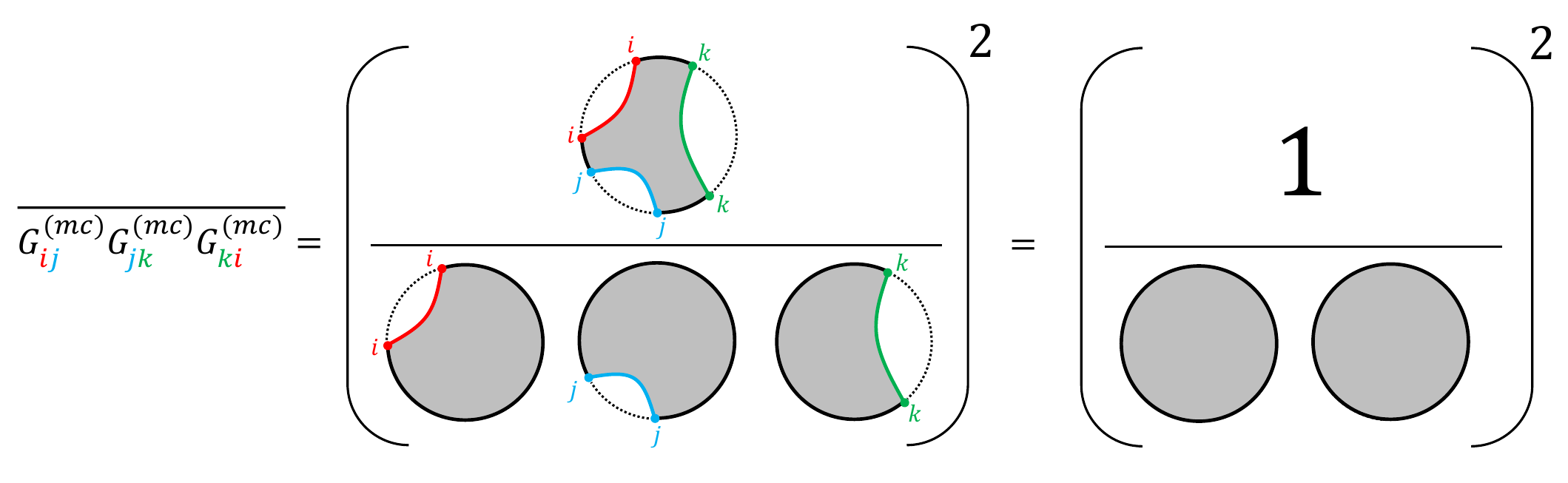} 
	\caption{Geometrical representation of the 3rd moment of the microcanonical Gram matrix (for an appropriate set of particles or shells such that the 3-wormhole geometry in the numerator exists). The numerator represents half of the 3-wormhole, while the denominator represents half of the product of the Euclidean PETS geometries. Since the background geometries are the same, the contributions from the particles or shells, as well as the geometric defects they introduce, are identical and cancel out. In addition, the remaining disc in the numerator and one disc in the denominator also cancel. Recalling that one disc contributes $e^{S}$ in the microcanonical ensemble, the final result is  $\left( \frac{1}{e^{2S}}\right)^2=e^{-4S}$.}   
\label{fig:UniOverlap}
\end{center}
\end{figure}
The numerator can be expressed as an $n$-boundary wormhole partition function and the denominator as (the product of) discs partition functions. Crucially, both can be decomposed into contributions from two fundamental discs, with defects introduced by particles or shells at the junctions. This decomposition for the $n=3$ case is illustrated in Fig.~\ref{fig:UniOverlap}.

The important point is that in the microcanonical ensemble the background disc geometry is the same both in the numerator and denominator, in contrast to the case of canonical ensembles. Both terms correspond to the same Euclidean geometry with energy $E$, whose contribution is $e^{S}$. 
Since the background geometries are the identical, the particle (or shell) trajectories are also identical. Consequently, the contributions from the particles, as well as some of the full discs, cancel exactly in the ratio, yielding the simple result
\be
\begin{gathered}
\overline{G_{i_{1}i_{2}}^{(mc)}(E) G_{i_{2}i_{3}}^{(mc)}(E) \cdots G_{i_{n}i_{1}}^{(mc)}(E) }  \simeq e^{-2S(n-1)} ~ .\\
\hspace{5.5cm} ({\rm for ~ sufficiently ~ large ~ masses})
\end{gathered} \label{eq:UniOverlap1}
\ee
At this stage, we have made no additional assumptions, and expression (\ref{eq:UniOverlap1}) is valid only for masses for which the $n$-wormhole geometry exists as a real Euclidean saddle. 

\begin{figure}[t]
\begin{center}
\includegraphics[width=15.cm]{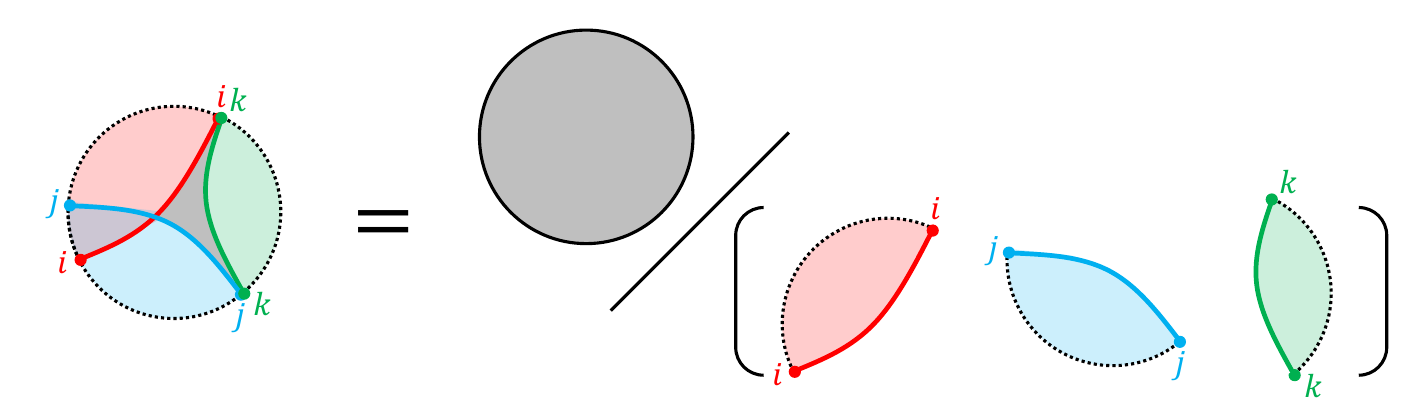} 
	\caption{One disc with particles or shells of small masses in the numerator of the 3rd moment of the Gram matrix. In this case, we may not be able to construct a 3-wormhole via the gluing procedure. However, under the `microcanonical universal overlap' assumption, we regard this simply as a disc with particles or shells, whose semiclassical GPI is given by the full disc divided by the contributions from the particles or shells, as well as the defects enclosed by their trajectories and the boundary. In other words, we analytically continue the GPI to the small-mass region.}    
\label{fig:UniOverlap2}
\end{center}
\end{figure}

The situation becomes more subtle when the masses are not sufficiently heavy. In that case, the particle or shell trajectories on a single copy of the disc would cross each other, making the actual gluing into a smooth wormhole geometry impossible. Our core assumption is that we can disregard the global gluing constraints and simply consider the $n$-wormhole as two copies of a disc with particle insertions. In this case, the contributions can be decomposed into those from the full disc, shell trajectories, and the regions enclosed by a trajectory and the boundary, as illustrated in Fig.~\ref{fig:UniOverlap2}. 
If a region is enclosed multiple times, we also count its contribution multiple times. This is equivalent to analytically continuing the wormhole on-shell action into the small-mass regime where a real Euclidean solution does not exist. Under this assumption, the cancellation of geometric factors persists, and we recover the universal result:
\be
\overline{G_{i_{1}i_{2}}^{(mc)}(E) G_{i_{2}i_{3}}^{(mc)}(E) \cdots G_{i_{n}i_{1}}^{(mc)}(E) }  \simeq e^{-2S(n-1)} ~ .
\label{eq:UniOverlap2}
\ee

This microcanonical universality contrasts
the canonical ensemble, where universal overlaps typically require a large-mass limit to enforce cancellation of the particle contributions.
In the microcanonical ensemble, this kind of cancellation occurs automatically, once the energy is fixed (without further large-mass limit), provided one accepts the analytic continuation.

We do not attempt a rigorous mathematical justification for this procedure. However, we find it physically reasonable because it self-consistently yields the correct Hilbert space dimension $\text{dim}(\mathcal{H}) \simeq  \exp({S(E)})$ for finite masses—a result for which we have no physical reason to expect a deviation.

\subsection{The dS ER bridge} \label{subsec:dSER} 
If the microcanonical universal overlap discussed previously holds, then applying it to a family of microstates with small masses in centaur gravity implies that the flow-geometry Hilbert space can be spanned by states with a dS ER bridge. One might be tempted to interpret this structure as analogous to an ER bridge associated with interior of a dS horizon. On the other hand, we know from the theorem of Gao and Wald  \cite{Gao:2000ga} that such a long ER bridge cannot occur in higher-dimensional de Sitter spacetimes, \ie, dS spacetimes can become taller but not fatter under perturbations satisfying the null energy condition. Why does this apparent contradiction arise?

\begin{figure}[t]
\begin{center}
\includegraphics[width=6.45cm]{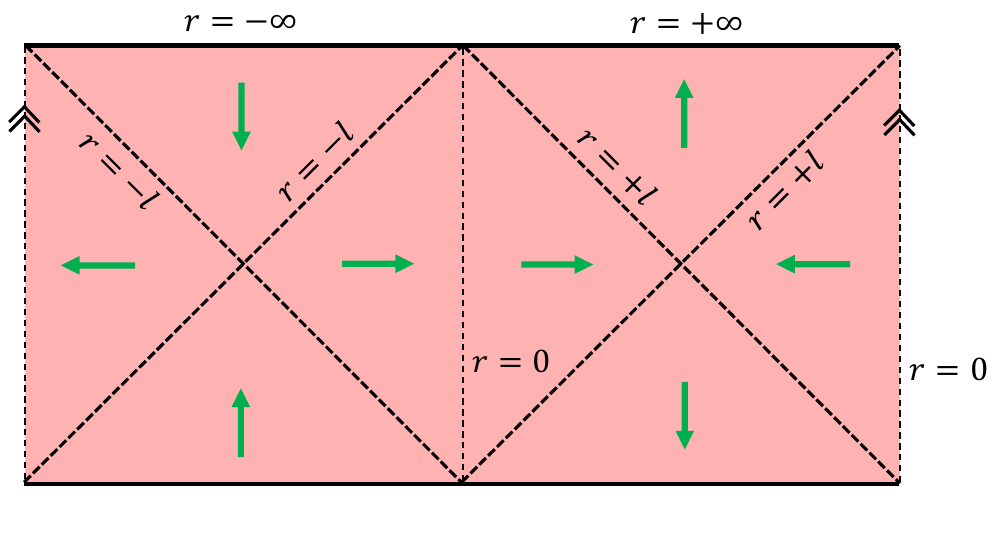}  ~~ \includegraphics[width=8.cm]{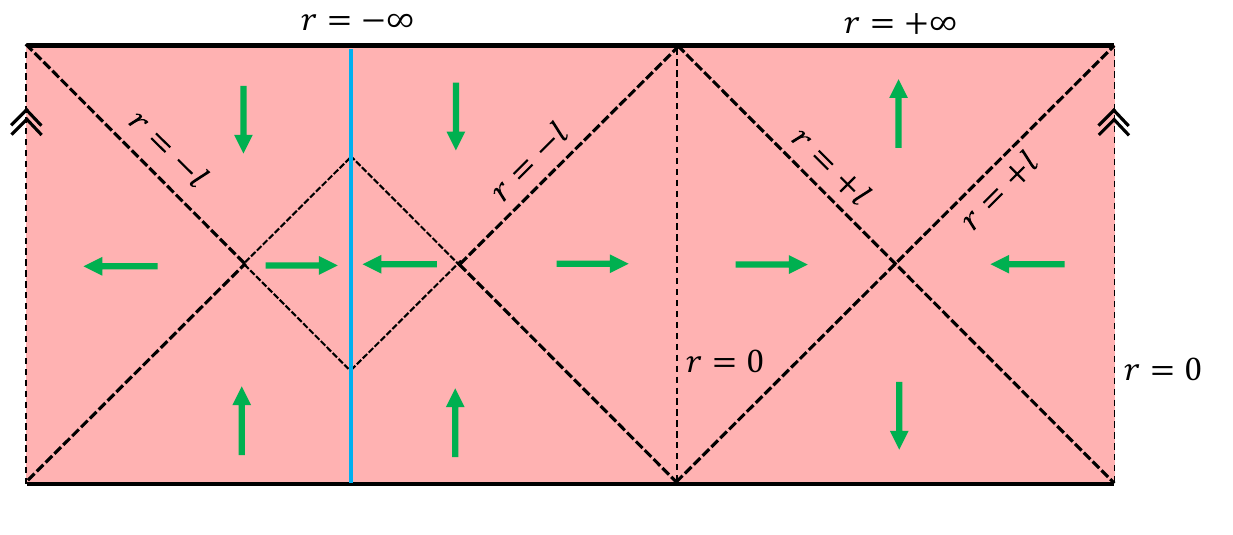} 
	\caption{(\textbf{Left}) dS$_{2}$ in dS JT gravity. The dilaton solution is given by $\Phi(r) = \widetilde{\Phi}_{b} \frac{r}{\ell}$, and the green arrows indicate the increasing direction of the dilaton $\Phi$. (\textbf{Right}) A fatter dS$_{2}$ in dS JT  gravity. When a heavy particle is located in the region $r \in (-\ell,0)$ of the static patch (e.g., in the leftmost patch), it may backreact on the geometry and make the wormhole between the central diamond and the leftmost static patch longer.
}   
\label{fig:dS2}
\end{center}
\end{figure}

The resolution lies in dimensionality: the Gao–Wald theorems do not apply to two-dimensional spacetimes.
\footnote{
Although the theorems and corollary in their paper do not assume a specific spacetime dimension, they are not (at least straightforwardly) applicable to two-dimensional spacetime. This is because they require the null generic condition, which states that every null geodesic has at least one point where the particular contraction between the Riemann tensor and its (null) tangent vector does not vanish. However, this condition does not hold for general two-dimensional geometries, since the Riemann tensor is always written as
$ R_{\mu\nu\alpha\beta} = \tfrac{1}{2} R \,(g_{\mu\alpha} g_{\nu \beta} - g_{\mu\beta} g_{\nu\alpha})$ 
and its contraction with null vectors always vanishes.
}
Therefore, they do not forbid fat dS spacetimes, or long ER bridges behind dS horizons, in two dimensions. Indeed, using the particle solution \eqref{eq:SolutionSmallMass} (Wick-rotated to Lorentzian signature) for the dS region and performing the gluing procedure of Subsection \ref{subsec:flowPETSgeometry}, we can explicitly construct a fatter dS$_{2}$ geometry in dS JT gravity without any exotic matter (Fig.~\ref{fig:dS2}).

In this case, it is important to place the particle in the static patch region $r \in (-\ell, 0)$, where the dilaton increases toward the pole. By contrast, if we put the particle in the region $r \in (0, +\ell)$, the dS spacetime becomes taller, analogous to the higher dimensional case. This demonstrates that in two dimensions, the gravitational response to matter differs between the regions $r \in (-\ell, 0)$ and $r \in (0, +\ell)$. The dS region in the flow geometry corresponds to the former, which partly explains why we can construct a long ER bridge in the flow geometry.

Another important point we should mention is the following. Although we have shown that the flow Hilbert space can be spanned solely by microstates with a dS ER bridge, we cannot exclude microstates with a flow ER bridge from the Hilbert space. These later states also reside in $\cal{H}_{\bf Flow}$. This is evident from the non-vanishing overlap between a microstate with a dS ER bridge and one with a flow ER bridge. The fact that the Hilbert space dimension computed in Sections \ref{sec:heavymass} and \ref{sec:microcanonical} is identical confirms that both types of microstates belong to the same Hilbert space.

This implies that the interior structure of the horizon microstates is strongly influenced by the UV behavior of the theory. In other words, since centaur gravity can be regarded as an IR-modified version of JT gravity, the interior structure of horizon microstates somehow inevitably contains its UV part, namely an AdS region in the ER bridge. Therefore, if we wish to exclude the ER bridge containing the AdS region from the Hilbert space completely, we must also modify the UV behavior of the theory. One way to achieve this is to consider dS JT gravity with the boundary placed at $r=0$, restricting to the region $r<0$. In this case, the flow ER bridge can be excluded, but the holographic dual becomes much more obscure.

\subsection{Algebraic approach}
Recent years have seen significant progress in understanding the algebra of diffeomorphism-invariant observables in systems with a large number of degrees of freedom. A surprising insight is that gravitational effects can lead to better defined algebras of type II\footnote{ For completeness, the classification is as follows: \textbf{Type I} describes conventional quantum mechanics with finite trace, \textbf{Type II$_\infty$} characterizes infinite systems where traces exist but are unbounded, \textbf{Type II$_1$} represents systems where all operators admit a finite trace, and \textbf{Type III} captures quantum field theory where no trace exists.}. As a consequence, an algebraic definition of density matrix and entropy becomes possible even in the absence of a factorized Hilbert space. This novel approach has found fruitful applications from black hole physics to cosmology. 

In this work, we have analyzed aspects of the microcanonical thermofield double for flow geometries,
\be\label{eq:TFDmicro}
\left|\Psi^{(mc)}_{\rm TFD}(E) \right\rangle = \sum_{a} f_{E}(E_{a}) |E_{a}\rangle_{L} | E_{a}\rangle_{R}~,
\ee
and used the gravitational path integral to compute the corresponding microcanonical Gram matrix and the dimensionality of the Hilbert space spanned by flow PETS. A key result is that microstates with a dS ER bridge—and their corresponding gravitational solutions— have enough information to span the flow Hilbert space of dimension $\exp(2 S_{\bf Flow})$. 

The mircrocanonical version of such a state has been analyzed in the context of von Neumann algebras for black holes in AdS \cite{Chandrasekaran:2022eqq}. A natural subsequent question is to determine the underlying entanglement structure of the analogous state for flow geometries. It would be very interesting to investigate the algebraic properties of the new class of microstates presented in this paper: the flow PETS geometries. This research direction would extend the canonical approach studied in \cite{Aguilar-Gutierrez:2023odp}. We expect that the Murray–von Neumann algebra (type II$_1$) would play a central role, as it does in the canonical case. Within this algebraic framework, the dS ER bridge constructed in Section \ref{sec:microcanonical} could shed new light on the encoding of de Sitter physics.

More broadly, our framework, which incorporates wormhole solutions, provides a platform to analyze the role of spacetime wormholes in the algebraic structure of flow PETS geometries. This opens the possibility to investigate how the boundary algebra of observables is affected by topology change and wormhole corrections. In particular, it would be interesting to study the role of baby universes in the algebra of observables \cite{Liu:2025cml, Kudler-Flam:2025cki}.

Finally, ongoing work builds on the flow PETS geometries and wormhole constructions developed here to investigate the closed universes that arise from them. Unlike previously studied closed universes with crunching interiors (due to a negative cosmological constant) \cite{Usatyuk:2024mzs, Usatyuk:2024isz}, these baby universes exhibit novel features, including the possibility of interiors with positive cosmological constant, offering a potentially new perspective on de Sitter entropy and cosmology. We hope to report on these developments in future work \cite{ourpaper}.

\section*{Acknowledgments}
It is a pleasure to thank Ana Climent, Bart{\l}omiej Czech, Roberto Emparan, Dami\'an A. Galante, Xu-Yao Hu, Viktor Jahnke, Martin Sasieta, Herman Verlinde, and Edward Witten for illuminating discussions. RE thanks the Shanghai Institute for Mathematics and Interdisciplinary Sciences (SIMIS) for the warm hospitality during the final stage of this work. We both thank the Institute of Cosmos Sciences of the University of Barcelona (ICCUB)  where this work was initiated. RE thanks the organizers of the workshop ``Quantum Gravity, Holography and Quantum Information'' held at IIF-UFRN Natal, the conference ``Gravity, Geometry, and Operator Algebras'' held at SIMIS and the workshop ``Black Holes, Quantum Chaos, and Quantum Information'' held at YITP. RE is supported by the Dushi Zhuanxiang Fellowship and acknowledges a Shuimu Scholarship as part of the Shuimu Tsinghua Scholar Program. SM is supported in part by the National Science and Technology Council (No. 111-2112-M-259-016-MY3).

\appendix

\section{Thermodynamics} \label{app:thermodynamics}

\subsection{Canonical Ensemble}
The thermodynamic properties of centaur gravities of $G_{2}>0$ and $G_{2}<0$ are formally similar within the canonical ensemble. Approximating the partition function by the dominant saddle point geometry, free energy, energy, and entropy are given by \cite{Anninos:2017hhn}
\begin{align}
F(T) = - \frac{\pi \ell}{4 |G_{2}|}\widetilde{\Phi}_{b} T^2 - \frac{\Phi_{0}}{4 |G_{2}|} T~, \label{eq:AppendixFree} \\
E(T) = \frac{\pi \ell}{4 |G_{2}|} \widetilde{\Phi}_{b} T^2~,  \hspace{2.1cm} \label{eq:AppendixEnergy} \\
S(T) = \frac{\pi \ell}{2 |G_{2}|} \widetilde{\Phi}_{b} T
 + \frac{\Phi_{0}}{4 |G_{2}|}~.  \hspace{0.8cm} \label{eq:AppendixEntropy}
\end{align}
Notably, these expressions are completely identical to those of JT gravity at the zero-loop level.
\footnote{
Note that although the dominant saddle contribution is exactly the same as in JT gravity, there also exists a linearly unstable subdominant saddle in centaur gravity, as shown in Fig.~\ref{fig:AppendixFree}. Therefore, we may have a non-perturbative instability similar to that of hot flat space~\cite{Gross:1982cv}, implying that the canonical ensemble of centaur gravity could be ill-defined. We do not make any definite statement about this possible instability or ill-definedness here, and proceed with the formal computation as if there were no such problem. Note also that in the main part of this paper we focus on microcanonical ensembles, so this subtlety does not affect the main results.
}
The fundamental distinction between centaur gravity and JT gravity lies not in the local thermodynamic formulas, but in the number of saddle points and the dominant saddle point. 

As it is well known, JT gravity has only one single saddle point, Euclidean BH geometry;
\begin{align}
\underline{ {\rm Euclidean ~ BH} } ~~~~~~  ds^2 = f_{\rm BH}(r) d\tau^2 + \frac{1}{f_{\rm BH}(r)}dr^2 , ~~~~ \Phi = \widetilde{\Phi}_{b} \frac{r}{\ell} ~ , \hspace{3cm} \label{eq:AppendixBH} \\
f_{\rm BH}(r)= -(2\pi \ell T)^2 + \frac{r^2}{\ell^2}, ~~ \tau \in [0, 1/T], ~~ r\in(2\pi \ell^2 T, \infty) ~ ,
\end{align}
where the parameter $T$ represents the temperature of the system.
\footnote{
In this parametrization of the geometry, we can see that the horizon is located at $r=2\pi \ell^2 T$. And the smoothness condition at the horizon leads to the periodicity of Euclidean time circle as
\ben
({\rm periodicity}) = \left. \frac{f_{BH}'}{4\pi}\right|_{r=2\pi \ell^2 T} = \frac{1}{T} ~ .
\een
}

On the other hand, centaur gravity admits two types of saddle point geometries. The first is the Euclidean BH geometry stated above. The second is the Eulclidean flow geometry, which is given by
\begin{align}
\underline{ {\rm Euclidean ~ flow} } ~~~~~~ ds^2 = f_{\rm flow}(r) d\tau^2 + \frac{1}{f_{\rm flow}(r)}dr^2 , ~~~~ \Phi = \widetilde{\Phi}_{b} \frac{r}{\ell} ~ , \hspace{2.8cm}  \label{eq:AppendixFlow}\\
f_{\rm flow}(r)= (2\pi \ell T)^2 + \frac{r}{|r|} \frac{r^2}{\ell^2}, ~~ \tau \in [0, 1/T], ~~ r\in(-2\pi \ell^2 T, \infty) ~ ,
\end{align}
where again, the parameter $T$ represents the temperature of the system. As we explained in the main text, the region of negative $r<0$ corresponds to dS space, while positive $r>0$ corresponds to AdS space. As we will explain shortly, the dominant saddle depends on the sign on $G_{2}$; when $G_{2}>0$ Euclidean BH saddles dominate and when $G_{2}<0$ Euclidean flow saddles dominate. Before going into the details, we summarize these differences in Table \ref{tab:JTandCentaur}.
\begin{table}[h]
\centering
\begin{tabular}{c|c|c|c}
 ~ & number of saddles & saddle & dominant saddle \\
 \hline \hline 
 JT & 1 & BH & BH \\
 \hline
 centaur ($G_{2}>0$) & 2 & BH, flow & BH \\
 \hline
 centaur ($G_{2}<0$) & 2 & BH, flow & flow
\end{tabular}
    \caption{The differences among JT gravity, centaur gravity$(G_{2}>0)$ and centaur gravity$(G_{2}<0)$ for canonical ensembles.  }
\label{tab:JTandCentaur}
\end{table}
\subsubsection*{Partition Function and Free Energy}
The Euclidean action for centaur gravity is given by 
\begin{align}
    I_E=\frac{-1}{16\pi G_2}\int_{\mathcal{M}} \rmd^2x\sqrt{g} \left( R\Phi+ \frac{2}{\ell^2} \frac{\Phi}{|\Phi|} \Phi \right)+\frac{-1}{8\pi G_2}\int_{\partial\mathcal{M}}\rmd \tau\sqrt{h}\,\Phi K \hspace{1cm} \notag \\
   +\frac{1}{8\pi G_2}\int_{\partial\mathcal{M}}\rmd \tau\sqrt{h}\,\frac{\Phi}{\ell} - \frac{\Phi_{0}}{4 |G_{2}|}\chi ~ . \label{eq:AppendixAction}
\end{align}
Except for the last term, the sign of $G_{2}$ only changes the overall sign, but does not affect the equation of motion, nor the saddle point solutions. By solving the equation of motion, we see that both the Euclidean BH geometry (\ref{eq:AppendixBH}) and the Euclidean flow geometry (\ref{eq:AppendixFlow}) are valid solutions. Note that only for the last term, the absolute value of $G_{2}$ is taken. This is chosen to suppress higher-genus contributions for both signs of $G_{2}$. 

\begin{figure}[t]
\begin{center}
\includegraphics[width=7.cm]{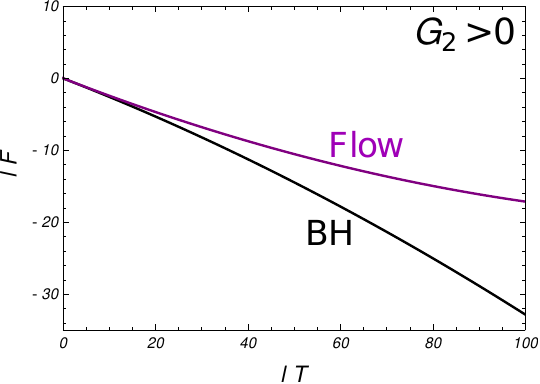}
~~~~ \includegraphics[width=7.cm]{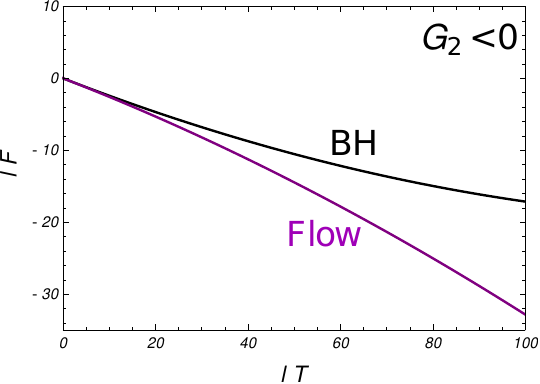}
	\caption{Behaviors of free energies of BH and flow saddles in centaur gravity. We choose $\widetilde{\Phi}_{b}=0.001, \Phi_{0}=1, |G_{2}|=1$. (left) The case of $G_{2}=1$. The dominant saddle is Euclidean BH geometry. (right) The case of $G_{2}=-1$. The dominant saddle is Euclidean flow geometry. }
\label{fig:AppendixFree}
\end{center}
\end{figure}

Substituting (\ref{eq:AppendixBH}) and (\ref{eq:AppendixFlow}) in the action (\ref{eq:AppendixAction}), we can obtain their respective free energies;
\begin{align}
F_{\rm BH} = - \frac{\pi \ell}{4 G_{2}} \widetilde{\Phi}_{b} T^2 - \frac{\Phi_{0}}{4|G_{2}|}T ~ , \label{eq:AppendixFreeBH} \\
F_{\rm flow} = \frac{\pi \ell}{4 G_{2}} \widetilde{\Phi}_{b} T^2 - \frac{\Phi_{0}}{4|G_{2}|} T ~ . \label{eq:AppendixFreeFlow}
\end{align}
The behavior of these free energies is illustrated in Fig. \ref{fig:AppendixFree}.
When $G_{2}>0$, the dominant saddle is the Euclidean BH geometry, whereas for $G_{2}<0$, the Euclidean flow geometry dominates.
\footnote{
These results illustrate the general statement, shown in \cite{Witten:2020ert}, that in dilaton gravity theories satisfying the {\it asymptotically JT condition}, a thermodynamically unstable solution always appears together with a thermodynamically stable solution that has lower free energy. More precisely, only the $G_{2} > 0$ centaur gravity satisfies this condition, but a straightforward variant of the argument can also be applied to the $G_{2} < 0$ centaur gravity.
}
The physical free energy of the system is given by $F = \min\{ F_{\rm BH}, F_{\rm flow}\}$,  which leads to the result in (\ref{eq:AppendixFree}). Using the standard thermodynamic relations, we also obtain the energy (\ref{eq:AppendixEnergy}) and the entropy (\ref{eq:AppendixEntropy}).

\subsection{Microcanonical Ensemble}
\begin{figure}[t]
\begin{center}
\includegraphics[width=7.cm]{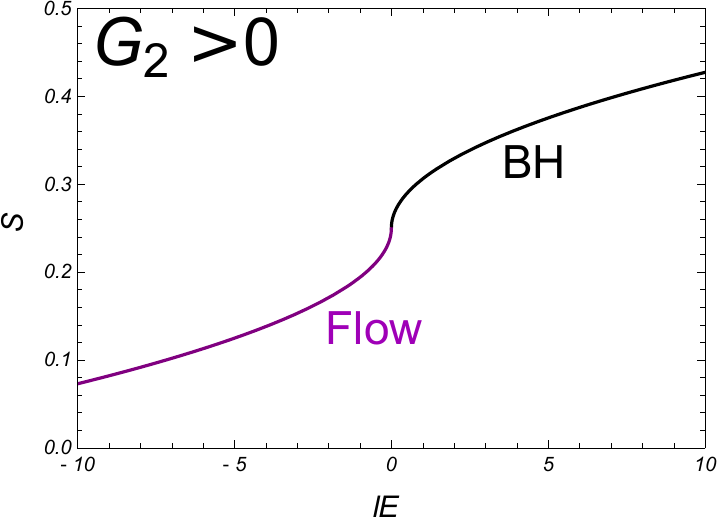}
~~~~ \includegraphics[width=7.cm]{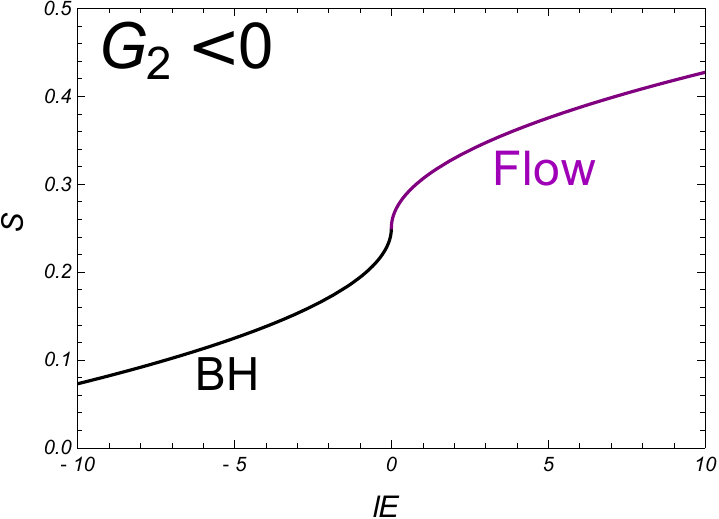}
	\caption{Behaviors of entropies of BH and flow saddles in centaur gravity. We choose $\widetilde{\Phi}_{b}=0.001, \Phi_{0}=1, |G_{2}|=1$. (left) The case of $G_{2}=1$.  (right) The case of $G_{2}=-1$. }
\label{fig:AppendixEntropy}
\end{center}
\end{figure}

While the flow geometry (for $G_{2}>0$) or BH geometry (for $G_{2}<0$) may not show up as a stable equilibrium states in the canonical ensemble, they can appear as physical states when we consider microcanonical ensembles. Firstly, using the expression of free energies (\ref{eq:AppendixFreeBH}), and (\ref{eq:AppendixFreeFlow}), we obtain their corresponding entropies and energies;
\begin{align}
S_{\rm BH}= \frac{\pi \ell}{2G_{2}} \widetilde{\Phi}_{b} T + \frac{\Phi_{0}}{4|G_{2}|}, ~~~~~~~ E_{\rm BH} = \frac{\pi \ell}{4 G_{2}} \widetilde{\Phi_{b}} T^2 ~ , ~~~ \\
S_{\rm flow} = -\frac{\pi \ell}{2G_{2}} \widetilde{\Phi}_{b} T + \frac{\Phi_{0}}{4|G_{2}|}, ~~~~ E_{\rm flow} = -\frac{\pi \ell}{4 G_{2}} \widetilde{\Phi_{b}} T^2 ~ .\label{eq:AppendixEnergyflow} 
\end{align}
Equivalently, we can express the entropy as the function of energy $E$;
\begin{align}
S_{\rm BH}(E)= \frac{\Phi_{0}}{4|G_{2}|} + \sign(G_{2}) \sqrt{\pi \ell \widetilde{\Phi}_{b}} \sqrt{\frac{E}{G_{2}}} ~ , ~~~\\
S_{\rm flow}(E) = \frac{\Phi_{0}}{4|G_{2}|} - \sign(G_{2}) \sqrt{\pi \ell \widetilde{\Phi}_{b}} \sqrt{-\frac{E}{G_{2}}} ~ .
\end{align}
The domains of definition for these saddles are
\begin{align*}
{\rm BH ~ saddle}: ~~ E \in (0, \infty) {\rm ~ for ~} G_{2}>0, ~~~~~ E\in(-\infty, 0) {\rm ~ for ~} G_{2}<0 ~ , \\
{\rm Flow ~ saddle}: ~~ E \in ( - \infty, 0) {\rm ~ for ~} G_{2}>0, ~~~~~ E\in(0, \infty) {\rm ~ for ~} G_{2}<0 ~ .
\end{align*}
The entropy versus energy behavior is plotted in Fig. \ref{fig:AppendixEntropy}. 
This analysis confirms that both geometries can represent physcal states when we consider the microcanonical ensemble. In the main text, we specifically focus on the negative energy region for the $G_{2}>0$ case, which corresponds to the flow geometry saddle.

\section{Gram matrix and counting microstates} \label{AppGram}
This paper constructs families of states that share the same exterior geometry $\cal{M}$ but differ in their interior structure. We interpret these as microstates of the flow geometry, dubbed \emph{centaur microstates}. In the main text, we consider the Hilbert space spanned by the microstates and compute its dimensionality using the Gram matrix obtained from the GPI, which may exhibit random matrix behavior. Here, we review the standard relation (i.e., in the case where the Gram matrix does not exhibit randomness) between an arbitrary family of quantum states and the dimensionality of the Hilbert space they span.

\begin{figure}[t]
\begin{center}
\includegraphics[width=7.cm]{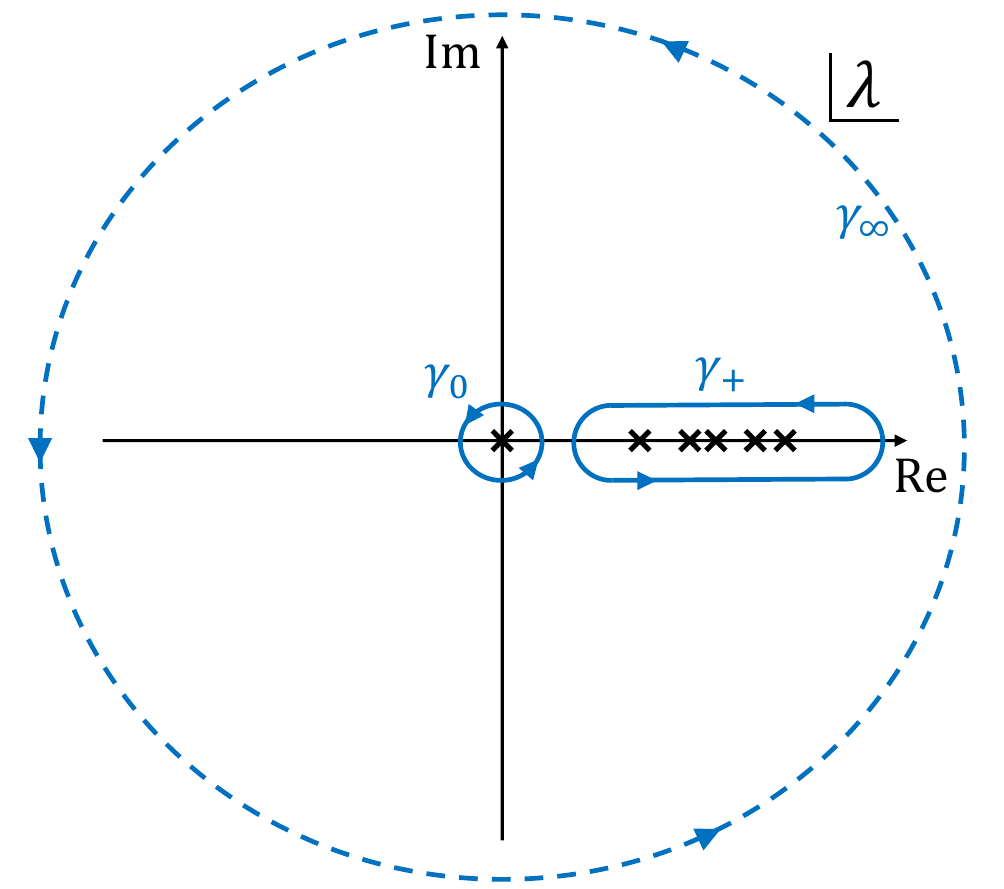} 
	\caption{Contour integration paths involved in the integrals for the resolvent $R(\lambda)$ in the complex $\lambda$-plane (\ref{eq:Disc1}). The poles of $R(\lambda)$ on the real axis correspond to the eigenvalues of the Gram matrix $\mathbf{G}$. The contour $\gamma_{0}$ which encircles the origin gives $\Omega - {\rm dim}(\mathcal{H}_{\mathbf{F}})$. The contour integral of $\gamma_{+}$, which encircles the positive region of the real axis, gives $ {\rm dim}(\mathcal{H}_{\mathbf{F}})$. These contours can be deformed to the contour at infinity $\gamma_{\infty}$, and its contour integral gives $\Omega$.}
\label{fig:contour1}
\end{center}
\end{figure}

Consider a Hilbert space $\cal{H}$ and a family of quantum states $\mathbf{F} = \{\ket{\psi_{i}} \in \mathcal{H} \mid i=1, \cdots, \Omega \}$. Let the subspace spanned by $\mathbf{F}$ be $\mathcal{H}_{\mathbf{F}} := {\rm Span}(\mathbf{F})$.\footnote{Throughout this paper, $\Omega$ denotes the size of a family of states. We therefore suppress the explicit $\Omega$-dependence of subspaces.} In general, these states are not orthogonal. Consequently, the dimension of $\mathcal{H}_{\mathbf{F}}$ does not necessarily coincide with $\Omega$; it rather satisfies
\be
{\rm dim}(\mathcal{H}_{\mathbf{F}}) \leq \Omega~.
\ee
Our focus is on computing the dimension of the Hilbert space spanned by the family $\mathbf{F}$. The key object for determining ${\rm dim}(\mathcal{H}_{\mathbf{F}})$ is the Gram matrix $\mathbf{G}$, an $\Omega \times \Omega$ matrix of overlaps defined by
\be
G_{ij} = \braket{\Psi_{i}}{\Psi_{j}}~, \quad i,j=1,\cdots,\Omega~.
\ee
The rank of $\mathbf{G}$ is precisely equal to ${\rm dim}(\mathcal{H}_{\mathbf{F}})$. To extract this rank, we study its resolvent. The resolvent matrix, $\mathbf{R}(\lambda)$, is defined by
\be
R_{ij}(\lambda) := \left( \frac{1}{\lambda \mathbf{1} - \mathbf{G}} \right)_{ij} = \frac{1}{\lambda} \delta_{ij} + \sum_{n=1}^{\infty} \frac{1}{\lambda^{n+1}} (\mathbf{G}^n)_{ij}~,
\ee
where $\mathbf{1}$ is the  $\Omega \times \Omega$ identity matrix. Let $\lambda_{a} ~ ({\rm for}~ a=1,\cdots,\Omega)$ be the eigenvalues of $\mathbf{G}$. The trace of the resolvent, $R(\lambda) := {\rm Tr~} \mathbf{R}(\lambda)$, is then given by
\be
R(\lambda) = \sum_{a=1}^{\Omega} \frac{1}{\lambda - \lambda_{a}}~.
\ee
Since $\mathbf{G}$ is a positive semi-definite matrix, its eigenvalues are non-negative. The number of zero eigenvalues is $\Omega - {\rm dim}(\mathcal{H}_{\mathbf{F}})$. Denoting positive eigenvalues by $\lambda_{A}>0 ~ ({\rm for~}A=1,\cdots, {\rm dim}(\mathcal{H}_{\mathbf{F}}))$, we can rewrite the trace as
\be
R(\lambda) = \frac{\Omega - {\rm dim}(\mathcal{H}_{\mathbf{F}}) }{\lambda} + \sum_{A=1}^{{\rm dim}(\mathcal{H}_{\mathbf{F}}) } \frac{1}{\lambda - \lambda_{A}}~.
\ee
The dimension ${\rm dim} (\mathcal{H}_{\mathbf{F}})$ can be isolated by computing a contour integral of $R(\lambda)$ in the complex $\lambda$-plane. By computing the integral around a contour $\gamma_+$ that encircles all non-zero eigenvalues, or equivalently, by computing the integral around a contour $\gamma_0$ that encircles the origin, which encloses the pole from all zero eigenvalues (see Fig. \ref{fig:contour1}), we obtain
\be\label{eq:Disc1}
{\rm dim}(\mathcal{H}_{\mathbf{F}}) = \frac{1}{2\pi i} \int_{\gamma_{+}} d\lambda ~ R(\lambda) 
= \Omega - \frac{1}{2\pi i} \int_{\gamma_{0}} d\lambda ~ R(\lambda)~. 
\ee
These equations indicate that the rank of $\mathbf{G}$, \ie ${\rm dim}(\mathcal{H}_{\mathbf{F}})$, can be computed directly from the resolvent $R(\lambda)$ via a contour integral, without explicitly diagonalizing $\mathbf{G}$.

When the number of states $\Omega$ is infinite, the resolvent must take continuous integral representation in the complex plane \cite{Coleman} 
\be
\label{eq:Rintegral}
R(z) = \int d \lambda \frac{\rho(\lambda)}{z-\lambda}~,
\ee
where $\rho(\lambda)$ is a normalized eigenvalue density with support on a finite interval $\cal{C}$. In this continuum limit, the resolvent develops a branch cut in the complex plane. The density $\rho(\lambda)$ can be recovered from (\ref{eq:Rintegral}) by computing the discontinuity of $R(z)$ across the cut $\cal{C}$ (\ie the residue at $\lambda = z$),
\be
\rho(\lambda) = \lim_{\delta \to 0}\frac{1}{2\pi i}(R(\lambda + i\delta) - R(\lambda - i\delta))~.
\ee
Finally, the Hilbert space dimension spanned by $\mathbf{F}$ is given by the integral
\be
{{\rm dim}(\cal{H}_{\bf F})} = \int_{\cal{C}} d \lambda~\rho(\lambda)~. 
\ee

In the main text, we consider the case where $\Omega$ is finite. However, due to its random nature, the resolvent must also be expressed in an integral representation, as in \eqref{eq:Rintegral}.

\section{Particle Trajectories}
\label{appedix:B}
This appendix presents the solution to the junction condition~(\ref{eq:junction}), which exhibits different behaviors depending on the particle mass~$m$. In Appendix~\ref{B1}, we provide the solution for a heavy mass, $m > \frac{\widetilde{\Phi}_{b}}{2G_{2}\beta}$, where the trajectory remains in the AdS region. In Appendix~\ref{B2}, we provide the solution for a light mass, $m < \frac{\widetilde{\Phi}_{b}}{2G_{2}\beta}$, where the trajectory enters the dS region.

\subsection{Heavy Mass: $\displaystyle m > \frac{\widetilde{\Phi}_{b} }{2G_{2}\beta } $} \label{B1}
When the mass is sufficiently large, the particle trajectory stays entirely in the AdS region ($\mathcal{R}(s) > 0$). In this regime, Eq.~\eqref{eq:geodesicf} can be written as
\be\label{eq:eomparticle}
\left( \frac{2\pi \ell}{\beta} \right)^2 + \frac{\mathcal{R}(s)^2}{\ell^2} - \dot{\mathcal{R}}(s)^2 = \frac{16\pi^2 G_{2}^2 \ell^2 m^2}{\widetilde{\Phi}_{b}^2} ~ .
\ee
It is straightforward to verify that
\be
\mathcal{R}(s)=\frac{4\pi G_{2} \ell^2 }{\widetilde{\Phi}_{b}} \sqrt{m^2-\frac{\widetilde{\Phi}_{b}^2}{4 G_{2}^2 \beta^2}} \cosh\left( \frac{s}{\ell} \right), ~~~ s\in (-\infty, \infty)~ , \label{eq:SolutionLargeMass}
\ee
solves Eq. (\ref{eq:eomparticle}) and satisfies the boundary condition $\mathcal{R}(\pm \infty) = \infty$. The exact expression for the time coordinate $\mathcal{T}(s)$ is obtained by using Eq.~\eqref{eq:ParticleConstraint}:
\be
\mathcal{T}(s)= \frac{\beta}{2\pi } {\rm arctanh} \left[  \frac{\widetilde{\Phi}_{b} }{2 G_{2} m\beta} \tanh\left( \frac{s}{\ell} \right)\right] , ~~~ s\in (-\infty, \infty) ~ . \label{eq:centaurTlargemass}
\ee

\subsection{Light Mass: $\displaystyle m < \frac{\widetilde{\Phi}_{b} }{2G_{2}\beta } $} \label{B2}
In the small-mass case, the particle enters the dS region, that is, $\mathcal{R}(s)$ becomes negative for some interval range of $s$. Eq.~\eqref{eq:geodesicf} then takes different forms in each region
\ben
\left( \frac{2\pi \ell}{\beta} \right)^2 + \frac{\mathcal{R}(s)^2}{\ell^2} - \dot{\mathcal{R}}(s)^2 = \frac{16\pi^2 G_{2}^2 \ell^2 m^2}{\widetilde{\Phi}_{b}^2}, \qquad ({\rm for}~\mathcal{R}(s) > 0)~,
\een
\ben
\left( \frac{2\pi \ell}{\beta} \right)^2 - \frac{\mathcal{R}(s)^2}{\ell^2} - \dot{\mathcal{R}}(s)^2 = \frac{16\pi^2 G_{2}^2 \ell^2 m^2}{\widetilde{\Phi}_{b}^2}, \qquad ({\rm for}~\mathcal{R}(s) < 0)~.
\een
It is straightforward to verify that
\be \label{eq:SolutionSmallMass}
\mathcal{R}(s)=
\left\{
\begin{array}{l}
\displaystyle 
\frac{4\pi G_{2} \ell^2 }{\widetilde{\Phi}_{b}} \sqrt{\frac{\widetilde{\Phi}_{b}^2 }{4 G_{2}^2 \beta^2}-m^2} \sinh\!\left( \frac{s}{\ell} -\frac{\pi}{2} \right), \quad s\in \left[ \frac{\pi}{2}\ell, \infty \right),  \\[1.2em]
\displaystyle -\frac{4\pi G_{2} \ell^2 }{\widetilde{\Phi}_{b}} \sqrt{\frac{\widetilde{\Phi}_{b}^2 }{4 G_{2}^2 \beta_{i}^2 }-m^2 } \cos\!\left( \frac{s}{\ell} \right), \quad s\in \left[ -\frac{\pi}{2}\ell, \frac{\pi}{2}\ell \right], \\[1.2em]
\displaystyle \frac{4\pi G_{2} \ell^2 }{\widetilde{\Phi}_{b}} \sqrt{\frac{\widetilde{\Phi}_{b}^2 }{4 G_{2}^2 \beta^2 }-m^2} \sinh\!\left( -\frac{s}{\ell} - \frac{\pi}{2} \right), \quad s\in \left(-\infty, -\frac{\pi}{2}\ell \right],
\end{array}
\right.
\ee
solves the equations of motion, and the corresponding time coordinate $\mathcal{T}(s)$ is given by
\be
\mathcal{T}(s)=
\left\{
\begin{array}{l}
\displaystyle 
\frac{\beta}{2\pi } {\rm arctanh}\!\left[ \frac{2 G_{2} m\beta }{\widetilde{\Phi}_{b} } \tanh\!\left( \frac{s}{\ell} - \frac{\pi}{2} \right)\right] + \frac{\beta}{4}, \quad s\in \left[ \frac{\pi}{2}\ell, \infty \right), \\[1.2em]
\displaystyle \frac{\beta}{2\pi } {\rm arctan}\!\left[ \frac{\widetilde{\Phi}_{b} }{2 G_{2} m \beta} \tan\!\left( \frac{s}{\ell} \right)\right], \quad s\in \left[ -\frac{\pi}{2}\ell, \frac{\pi}{2}\ell \right], \label{eq:centaurTsmallmass} \\[1.2em]
\displaystyle \frac{\beta}{2\pi} {\rm arctanh}\!\left[  \frac{2 G_{2} m \beta}{\widetilde{\Phi}_{b} } \tanh\!\left( \frac{s}{\ell} + \frac{\pi}{2} \right)\right] - \frac{\beta}{4}, \quad s\in \left(-\infty, -\frac{\pi}{2}\ell \right].
\end{array}
\right.
\ee

\section{$E>0$: AdS Black Hole Geometry}
The gravitational path integral of centaur gravity also includes the AdS BH geometry in addition to the flow geometry. In this Appendix, we briefly discuss the AdS BH case. The computation and the results presented here are exactly the same as those in JT gravity.

We define the black hole Hilbert space as
\begin{align}
\mathbf{BH}(E)  &= \left\{ \left|\Psi_{i}^{(mc)}(E) \right\rangle \in \mathcal{H} ~ \big| ~ i=1,\cdots, \Omega \right\} ~ ,\\
\mathcal{H}_{\mathbf{BH}}(E) &= {\rm Span}\!\left(\mathbf{BH}(E)\right) ~ ,
\end{align}
where $\mathcal{H}_{\mathbf{BH}}(E)$ is spanned by a family of microcanonical PETS states $\left|\Psi_{i}^{(mc)}(E)\right\rangle$ with positive energy $E>0$.

\begin{figure}[t]
\begin{center}
\includegraphics[width=4.9cm]{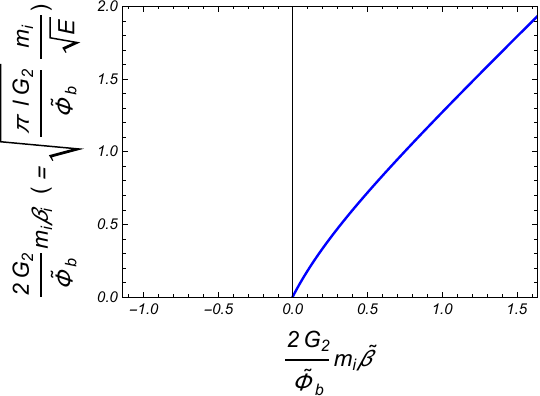}  \includegraphics[width=4.9cm]{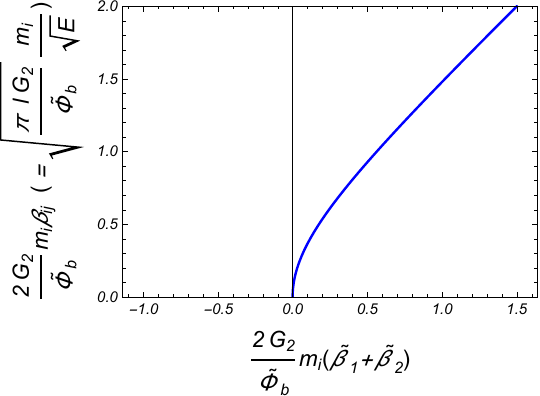}  \includegraphics[width=4.9cm]{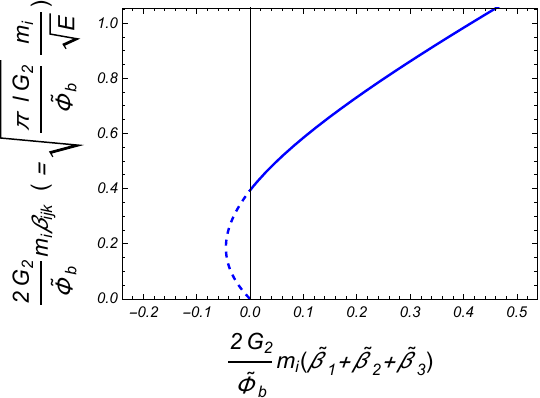}
	\caption{Relation between the total boundary length ($\sum_{p=1}^{n}\widetilde{\beta}_{p}$) and the effective inverse temperature ($\beta^{BH}_{i_{1}i_{2}\cdots i_{n}}$). (\textbf{Left}) The $n=1$ case.  (\textbf{Middle}) The $n=2$ case with $m_{j}/m_{i} = 1.5$. (\textbf{Right}) The $n=3$ case with $m_{j}/m_{i} = 1.5$ and $m_{k}/m_{i} = 2$. \\
    As shown in (\textbf{Left}) and (\textbf{Middle}), $\beta^{BH}_{i_{1}i_{2}\cdots i_{n}}$ is a monotonically increasing function for $n=1$ and $2$. When $n\geq 3$, the total renormalized boundary lenght becomes negative for small $\beta^{BH}_{i_{1}i_{2}\cdots i_{n}}$, or high $E$ region. }
\label{fig:betafunctionBH}
\end{center}
\end{figure}

AdS PETS geometries and their wormhole generalizations can be similarly constructed by injecting particles from the Euclidean boundary. In this case, the trajectory of a particle with mass $m$ in an AdS BH geometry with inverse temperature $\beta$ is given by
\begin{align}
\mathcal{R}_{\rm BH}(s) &= \frac{4\pi G_{2}\ell^2}{\widetilde{\Phi}_{b}} 
\sqrt{m^2 + \frac{\widetilde{\Phi}_{b}^2}{4 G_{2}^2 \beta^2}} 
\cosh\!\left( \frac{s}{\ell} \right) ~ ,\\
\mathcal{T}_{\rm BH}(s) &= \frac{\beta}{2\pi} 
\arctan\!\left[ \frac{\widetilde{\Phi}_{b}}{2 G_{2} m \beta} 
\tanh\!\left( \frac{s}{\ell} \right)\right] ~ .
\end{align}
The time difference $\Delta \mathcal{T}_{BH} = \mathcal{T}_{BH}(\infty) - \mathcal{T}_{BH}(-\infty)$ is therefore given by
\be\label{eq:TBH}
\Delta \mathcal{T}_{BH}(m, \beta) = \frac{\beta}{\pi} \arctan \left[ \frac{\widetilde{\Phi}_{b}}{2 G_{2} m \beta} \right] ~ .
\ee
For an $n$-mouth wormhole, the relation between the total boundary length ($\sum_{p=1}^{n}\widetilde{\beta}_{p}$) and the effective inverse temperature ($\beta^{BH}_{i_{1}i_{2}\cdots i_{n}}$) is given in terms of the time difference (\ref{eq:TBH}), 
\be
\sum_{p=1}^{n}\widetilde{\beta}_{p}= \beta^{BH}_{i_{1}i_{2}\cdots i_{n}} - \sum_{p=1}^{n} \Delta \mathcal{T}_{BH}(m_{i_{p}},\beta_{i_{1}i_{2}\cdots i_{n}}) ~ .
\ee
Examples for $n=1$, $n=2$, and $n=3$ with specific mass choices are shown in Fig. \ref{fig:betafunctionBH}, where the vertical axes represent $\sim \frac{1}{\sqrt{E}}$.

The energy of the wormhole saddles are given by
\be
E = \frac{1}{2} \frac{\partial \left. I_{E}^{i_{1}i_{2}\cdots i_{n}} \right|_{\rm on-shell(BH)} }{\partial \widetilde{\beta}_{p} } = \frac{\pi \ell \widetilde{\Phi}_{b}}{4 G_{2} (\beta^{BH}_{i_{1}i_{2}\cdots i_{n} })^2 } ~~~ (p=1,2,\cdots, n) ~,
\ee 
where, $ \left. I_{E}^{i_{1}i_{2}\cdots i_{n}} \right|_{\rm on-shell(BH)} $ represents the on-shell action for the AdS BH wormhole geometry:
\begin{align}
\left. I_{E}^{i_{1}i_{2}\cdots i_{n}} \right|_{\rm on-shell(BH)} = - \frac{ \pi \ell \widetilde{\Phi}_{b} }{ 2 G_{2} (\beta^{BH}_{i_{1}i_{2} \cdots i_{n} })^2 } \sum_{p=1}^{n} \widetilde{\beta}_{p} - \frac{\Phi_{0}}{4 G_{2}} (2-n) \notag \hspace{3cm} \\
 - \sum_{p=1}^{n} m_{i_{p}} \ell \log \left( 1 + \frac{ \widetilde{\Phi}_{b}^2 }{ 4 G_{2} m_{i_{p}}^2 (\beta^{BH}_{i_{1}i_{2} \cdots i_{n} })^2 } \right) ~ .
\end{align}
As shown in Fig \ref{fig:betafunctionBH}, regions with negative $\sum_{p=1}^{n}\widetilde{\beta}_{p}$ appear for $n \geq 3$.
\footnote{
This may be easily understood because the range of $\Delta \mathcal{T}_{BH}(\beta,m)$ as a function of $m$ is $(0,\frac{\beta}{2})$ so, at least 3 particles are needed in order for $\beta^{BH}_{i_{1}i_{2}\cdots i_{n}} - \sum_{p=1}^{n} \Delta \mathcal{T}_{BH}(m_{i_{p}},\beta^{BH}_{i_{1}i_{2}\cdots i_{n}})$ to be negative.
}

The range of $\beta^{BH}_{i_{1}i_{2}\cdots i_{n}}$ depends on the masses. For a given $n$, the upper bound is achieved when all masses are equal ($m_{i_{1}} = m_{i_{2}} = \cdots = m_{i_{n}}$) and given by
\be
\frac{2 G_{2}}{\widetilde{\Phi}_{b}} m_{i_{1}} \beta^{BH}_{i_{1}i_{2}\cdots i_{n}} \in \left(0, \cot\left( \frac{\pi}{n} \right) \right) ~ .
\ee
This bound increases with $n$, so for finite masses and generic mass combinations at large $n$, the use of non-real Euclidean saddles is unavoidable. 

Following the same aproach as in the main text, and assuming non-Eulcidean saddles are also included in the GPI, the inverse Laplace transform of the $n$-th moment of the unnormalized canonical Gram matrix yields
\begin{align}
\overline{\mathcal{G}_{i_{1}i_{2}}^{(mc)}(E) \mathcal{G}_{i_{2}i_{3}}^{(mc)}(E) \cdots \mathcal{G}_{i_{n}i_{1}}^{(mc)}(E) } \hspace{8.8cm} \notag \\
\propto \exp [ 4E \mathfrak{b}_{BH}(E) - \sum_{p=1}^{n}4E \Delta \mathcal{T}^{(mc)}_{BH}(m_{i_{p}}, E) - \sum_{p=1}^{n} \Gamma_{BH}(m_{i_{p}}, E) + \frac{\Phi_{0}}{4 G_{2}}(2-n) ] \\
(E>0) ~ ,\notag 
\end{align}
for $E>0$, where we define the following functions:
\begin{align}
\mathfrak{b}_{BH}(E) = \sqrt{ \frac{\pi \ell \widetilde{\Phi}_{b}}{4 G_{2} E} } ~ , \hspace{4.1cm} \\
\Delta \mathcal{T}_{BH}^{(mc)}(m, E) = \Delta \mathcal{T}_{BH}(m, \mathfrak{b}_{BH}(E)) ~\hspace{2.4cm}  \notag \\
= \frac{1}{2} \sqrt{ \frac{\ell \widetilde{\Phi}_{b}}{\pi G_{2} E} } \arctan \left[ \frac{1}{m} \sqrt{ \frac{\widetilde{\Phi}_{b} E}{ \pi \ell G_{2}   } }  \right] ~ , \\
\Gamma_{BH}(m, E) = - m \ell \log \left( 1 + \frac{\widetilde{\Phi}_{b} E}{\pi \ell G_{2} m^2} \right) ~ .\hspace{1.1cm}
\end{align}
Then, we can compute its normalized version as
\begin{align}
\overline{G_{i_{1}i_{2}}^{(mc)}(E) G_{i_{2}i_{3}}^{(mc)}(E) \cdots G_{i_{n}i_{1}}^{(mc)}(E) }  
\simeq e^{ \left( - 4 E \mathfrak{b}_{BH}(E) - \frac{\Phi_{0}}{4G_{2}} \times 2 \right) (n-1) } = e^{-2S_{\rm BH} (n-1)} ~ ,
\end{align}
which takes the form of (\ref{eq:AB}) with $A \simeq e^{2S_{\rm BH}}$ and $B \simeq e^{-2S_{\rm BH}}$. Using Eq. (\ref{eq:A}), we obtain 
\be
{\rm dim}(\mathcal{H}_{\mathbf{BH}}) \simeq e^{2S_{\rm BH}} ~ .
\ee

\bibliographystyle{JHEP}
\bibliography{references.bib}
\end{document}